\documentclass[12pt]{article}
\usepackage[latin9]{inputenc}
\usepackage{geometry}
\usepackage{amsmath}
\usepackage{amssymb}
\usepackage{stmaryrd}
\usepackage{setspace}
\usepackage[authoryear]{natbib}
\usepackage[unicode=true,
 bookmarks=false,
 breaklinks=false,pdfborder={0 0 1},backref=false,colorlinks=false]
 {hyperref}
\hypersetup{
 colorlinks,linkcolor=blue,citecolor=blue}

\makeatletter
\usepackage{amsfonts}
\usepackage{amsthm}
\usepackage{lscape}
\usepackage{appendix}
\usepackage{dcolumn}
\usepackage{rotating}
\usepackage{comment}
\usepackage{color}

\usepackage{diagbox}
\usepackage[flushleft]{threeparttable}
\usepackage{subfigure}
\usepackage{pgf}
\usepackage{tikz}\usepackage{caption}
\usepackage{tkz-tab}
\usepackage{tikz-3dplot}
\usetikzlibrary{calc}
\usetikzlibrary{arrows, automata}

\usepackage{changepage}
\usepackage{booktabs}
\usepackage{graphicx}
\usepackage{titletoc}
\usepackage{enumitem}

\usepackage{xr}
\newcolumntype{d}[1]{D{.}{.}{#1}}
\newcolumntype{t}[1]{D{,}{,}{#1}}
\newcolumntype{i}[1]{D{.}{}{#1}}
\newtheorem{theorem}{Theorem}[section]

\newtheorem{algorithm}{Algorithm}[section]
\newtheorem{assumption}{Assumption}[section]

\newtheorem{example}{Example}
\newtheorem{lemma}{Lemma}[section]

\newtheorem{proposition}{Proposition}[section]
\newtheorem{remark}{Remark}[section]

\theoremstyle{plain}

\newtheorem{innercustomas}{Assumption}
\newenvironment{myas}[1]
  {\renewcommand\theinnercustomas{#1}\innercustomas}
  {\endinnercustomas}

\numberwithin{equation}{section}

\usepackage{xcolor}
\newcommand{\indep}{\perp\!\!\!\!\perp}
\newcommand{\Y}{\mathcal{Y}}
\newcommand{\D}{\mathcal{D}}
\newcommand{\iD}{\mathring{\D}}
\newcommand{\Ep}{\mathrm{E}}
\newcommand{\V}{\mathcal{V}}

\makeatother

\begin{document}
\title{Estimating Causal Effects of Discrete and Continuous Treatments with Binary Instruments\thanks{The authors respectively represent MIT, BU, U of Bristol, and U of Michigan. W\"uthrich is also affiliated with CESifo. For helpful comments, we are grateful to the editor Keisuke Hirano, three anonymous referees, Yingying Dong, Dalia Ghanem, Jinyong Hahn, Stella Hong, Desire Kedagni, Michael Knaus, Eric Mbakop, Ulrich Mueller, Whitney Newey, Jack Porter, Andres Santos, Ed Vytlacil, Martin Weidner, Daniel Wilhelm, and participants in seminars at WashU, Indiana, MSU, Yale, Zhejiang, HKUST, UCSD, Northwestern, Stanford, University of Washington, University of Wisconsin-Madison, LMU, Mannheim, Bonn, Erasmus, Amsterdam, UC3M, Aarhus, Cantabria, and participants at the Asian Meeting of the Econometric Society, the Bristol Econometric Study Group, Munich Econometrics Workshop, KU Leuven Summer Event, California Conference, UCLA Econometrics Mini Conference and IAAE 2026. We also appreciate Matt Hong, Meng Hsuan Hsieh, ChatGPT, Claude, and Refine for excellent research assistance. This research was supported in part through computational resources and services provided by Advanced Research Computing at the University of Michigan, Ann Arbor. W\"uthrich gratefully acknowledges funding from a University of Michigan start-up package from the William Haber Economics Faculty Fund.  The usual disclaimer applies.}}
\author{Victor Chernozhukov \hspace{20pt} Iv\'an Fern\'andez-Val \\ 
Sukjin Han \hspace{20pt} Kaspar W\"uthrich}

\date{\today}

\maketitle
\vspace{-0.2cm}

\begin{abstract}
We propose an instrumental variable framework for identifying and estimating causal effects of discrete and continuous treatments with binary instruments. The basis of our approach is a local copula representation of the joint distribution of the potential outcomes and unobservables determining treatment assignment. This representation allows us to introduce an identifying assumption, so-called \emph{constant local dependence}, that restricts the local dependence of the copula with respect to the treatment propensity. We show that constant local dependence identifies treatment effects for the entire population and other subpopulations such as the treated.  The identification results are constructive and lead to practical estimation and inference procedures based on distribution regression. An application to estimating the effect of sleep on well-being uncovers interesting patterns of heterogeneity.

\vspace{0.1in}


\end{abstract}

\newpage 
\section{Introduction\label{sec:Introduction}}

Endogeneity and heterogeneity are key challenges in causal inference. Endogeneity arises because most treatments and policies of interest are the result of decisions made by economic agents. Unobserved heterogeneity also arises naturally as many of the agents' characteristics are unobserved to the researcher. Accounting for endogeneity and heterogeneity in treatment effects estimation is crucial to answer policy questions, such as how to allocate social resources and
combat inequalities. This paper  proposes a flexible instrumental variable (IV) modeling framework for identifying heterogeneous treatment effects under endogeneity, which yields practical estimation and inference procedures.

Without additional assumptions, IV strategies cannot point-identify meaningful treatment effects in the presence of heterogeneity. The literature has proposed different solutions to deal with this challenge  that exhibit trade-offs between adding structure to the treatment assignment mechanism and potential outcomes. One line of research has restricted the structure
and heterogeneity of the potential outcomes while allowing for flexible 
treatment assignment mechanisms (e.g., \citet{chernozhukov2005iv}
with binary treatments and \citet{newey2003instrumental} with continuous
treatments). Another line of research has shown the usefulness of restricting the treatment assignment 
while being flexible regarding how the potential outcomes are formed (e.g.,
\citet{imbens1994identification} with binary treatments and \citet{imbens2009identification}
with continuous treatments).

We explore an intermediate route that imposes structure on the relationship between the treatment assignment and the potential outcomes to achieve point identification of treatment effects. The basis of this approach is a
local Gaussian representation of the copula capturing the dependence between the potential outcomes and the unobservable determinants of treatment assignment. This representation is fully \textit{nonparametric}, that is, it does not require that potential outcomes and treatment unobservables are jointly or marginally Gaussian. 
Indeed, the bivariate Gaussian structure always holds locally by treating the correlation parameter as an implicit function that equates the bivariate Gaussian copula with the copula of the potential outcome and selection unobservables. We use this representation to introduce an assumption that has not
been previously considered for identification of treatment effects. This assumption, so-called
\emph{constant local dependence} (CLD), restricts the local dependence of the copula
 with respect to treatment propensity, and thus it restricts the form of endogeneity.  We provide a range of statistical and economic interpretations of CLD, which clarify the empirical content of the assumption.

We show that,
even with a binary IV, CLD identifies quantile
and average treatment effects (QTE and ATE) of binary and ordered discrete 
treatments and quantile and average structural functions (QSF and ASF) of continuous treatments for the entire population.   The results for ordered discrete and continuous treatments have particular empirical relevance, as previous identification results for global treatment parameters are scarce and typically rely on rich instrument variation; see the literature review below.\footnote{There is a wide range of empirical studies estimating the effects of continuous and ordered endogenous variables using IVs with limited variation. Prominent examples include work on intergenerational mobility \citep[e.g.,][]{black2011recent}, studies of the returns to schooling \citep[e.g.,][]{angrist1995twostage}, analyses of the impact of air pollution \citep[e.g.,][]{chay2005does}, demand analyses with discrete IVs \citep[e.g.,][]{angrist2000interpretation}, and studies where discrete experimental treatments are used as IVs for ordered discrete and continuous treatments \citep[e.g.,][]{bessone2021economic}.}
As a byproduct, we also identify the dependence function, which captures the direction and magnitude of endogenous selection and may be of interest in applications. When covariates are available, we impose CLD conditional on them, allowing for an additional source of heterogeneity in our model.

The CLD-based identification strategy is constructive and leads to practical semiparametric estimation procedures based on distribution regression for both discrete and continuous treatments.\footnote{We refer to these estimators as ``semiparametric'' because distribution regression models involve function-valued parameters.} We establish the asymptotic Gaussianity of the  estimators of the potential outcome distributions, dependence function, and functionals such as the unconditional QSF and QTE. We show that the bootstrap is valid for estimating the limiting laws and provide an explicit algorithm for constructing uniform confidence bands.

The proposed method adds to the IV toolkit by expanding the directions of modeling trade-offs in identifying treatment effects. We allow for richer patterns of effect heterogeneity, compared to \citet{chernozhukov2005iv} and \citet{newey2003instrumental}, and more heterogeneity in the treatment assignment, compared to \citet{imbens1994identification}  and \citet{imbens2009identification}, while imposing more restrictions on the dependence structure (i.e., the form of endogeneity), as we detail below.

We apply our method to estimate the distributional effects of sleep on well-being. In this case,  sleep time is treated as a continuous treatment. We
use data from the experimental analysis by \citet{bessone2021economic}, who studied the effects of randomized interventions to increase the sleep time of low-income adults in India. A simple two-stage least squares analysis suggests that sleep has moderate average effects on well-being. Using our method, we document interesting patterns of heterogeneity across the distributions of sleep time and well-being that are missed by average effects. For example, the quantile treatment effects of increasing sleep on well-being at the lower tail, which are particularly policy-relevant, are larger than the corresponding average effect. The estimates are pointwise significant in the tails, but the uniform confidence band includes zero at all quantiles, which is unsurprising given the small sample size. Our estimates also suggest greater heterogeneity than estimates obtained under conditional exogeneity or based on the IV quantile regression model \citep{chernozhukov2005iv,chernozhukov2006instrumental}.

\vspace{8pt}

\noindent\textbf{Related Literature.} The literature on identification of heterogeneous treatment effects with endogeneity is vast. We focus the review on approaches that do not impose parametric distributional assumptions to achieve point identification.

A first strand of the literature imposed assumptions on the generation of the potential outcomes, such as rank invariance (RI) and rank similarity (RS). RI imposes that the outcome equation is strictly monotonic in a scalar unobservable that is the same for all potential outcomes. This convenient assumption allows for identification of QTE and ATE with discrete and continuous treatments \citep{chernozhukov2005iv}. Additive error structures which imply RI can also be used to identify the ASF with continuous treatments \citep{newey2003instrumental,blundell2007semi}.\footnote{\cite{ai2003efficient} and \cite{chen2009efficient, chen2012estimation, chen2015sieve} considered a general framework that nests these models.} \citet{chernozhukov2005iv}'s RS is slightly weaker as it does not necessarily restrict the unobservable to be the same across potential outcomes, although both produce the same testable restriction \citep{chernozhukov2013quantile}. \citet{vuong2017counterfactual} further showed that RI and strict monotonicity identify individual treatment effects under suitable regularity conditions. This literature remains flexible about the treatment selection process.

The proposed CLD assumption and RS or RI are non-nested: the former concerns the dependence between potential outcomes and selection unobservables across instrument levels, whereas the latter restricts this dependence across \emph{treatment levels}. Indeed, we show that RS can be viewed as another form of constant local dependence. However, our CLD allows for more general patterns of treatment effect heterogeneity. Also, approaches for point identification based on RI and RS rely on strict monotonicity of the outcome equation, which can only hold for continuous outcomes, and on full rank and completeness conditions that rule out, for example, ordered discrete and continuous treatments with a binary instrument. When treatments are continuous, they typically lead to ill-posedness. By contrast, the proposed method  does not suffer from ill-posedness and accommodates discrete and mixed discrete-continuous outcomes and ordered and continuous treatments with a binary instrument.

A second strand focuses on assumptions imposed on the treatment assignment. \citet{imbens1994identification} and \citet{heckman2005structural} combined a scalar unobservable in the treatment assignment with monotonicity of potential treatments in a binary instrument to identify local average and marginal effects of binary treatments. \citet{abadie2002instrumental} and \citet{carneiro2009estimating} extended this approach to local quantile effects, and \citet{vytlacil2006ordered} to ordered discrete treatments. \citet{newey1999nonparametric} and \citet{imbens2009identification} imposed strict monotonicity of the treatment selection equation in a scalar unobservable and large support of the instrument (ruling out discrete instruments) to identify global effects using a control variable approach, which only applies to continuous treatments. \citet{newey2021control} avoided the large support requirement of \citet{imbens2009identification} by assuming a parametric conditional expectation of the treatment that enables extrapolation outside the instrument support. Compared to this strand, CLD restricts the relationship between potential outcomes and treatment assignment, but identifies global effects of discrete and continuous treatments with discrete instruments, without relying on a scalar unobservable in the assignment mechanism.

There are also approaches that combine or modify the assumptions of the previous strands. \citet{chesher2003identification} showed identification of the quantile effect of continuous treatments on continuous outcomes with continuous instruments assuming strict monotonicity of the outcome and treatment selection equations in scalar unobservables. Under similar assumptions, \citet{dhaultfoeuille2015identification} and \citet{torgovitsky2015identification} found that quantile effects can be identified with discrete instruments, and \citet{torgovitsky2017minimum} proposed a two-step minimum distance estimator for the finite-dimensional parameter of the outcome function. Again, these restrictions are not nested with CLD and, for continuous treatments, our identification strategy is constructive, yielding straightforward plug-in estimators. There are also partial identification solutions imposing less structure on the treatment assignment and potential outcomes (e.g., \citet{manski1990nonparametric} and \citet{balke1997bounds} for earlier references, and \citet{chesher2020generalized} for a more recent survey).

The copula has been previously employed in econometrics for identification and estimation. For example, \citet{chen2006efficient} used a parametric copula to achieve efficient estimation in a class of multivariate distributions.\footnote{In time series, \citet{chen2006estimation,chen2006estimation2}, \citet{beare2010copulas}, \citet{chen2021efficient,chen2022copula}, and \citet{fan2023estimation}, among others, used copulas to model temporal dependence.} In a semiparametric triangular model with binary dependent variables, \citet{han2017identification} introduced a class of single-parameter copulas for the dependence between the unobservables and established a condition---satisfied by many well-known copulas, including the Gaussian---under which the parameters are identified. With a binary treatment and binary outcome, the current paper's framework is relevant to \citet{han2017identification}. However, while they assumed a parametric copula, we assume CLD, of which their Gaussian copula is a special case corresponding to a local dependence function that is constant everywhere.  \citet{han2019estimation} developed sieve estimation and inference methods based on \citet{han2017identification}, and \citet{han2023semiparametric} extended them to semiparametric models for dynamic treatment effects. \citet{mourifie2025layered} used copulas as a channel to impose assumptions in characterizing identified sets for marginal treatment effects.

\citet{arellano2017quantile} and \citet{chernozhukov2025distribution} studied IV identification of selection models using assumptions on the copula between the latent outcome and selection unobservable: the former assumed a real analytic copula and required continuous instruments, while the latter used the local Gaussian representation and copula exclusion, a special case of CLD, with a binary instrument. Therefore, our framework with a binary treatment is related to their setup but differs in several dimensions. First, our setting requires two-way sample selection due to the switching of treatment status. Second, we introduce a general threshold-crossing selection model with a multivariate unobservable, allowing for rich selection patterns. Third, these features make the local representations and constant local dependence assumptions specific to treatment status and the value of the IV. More importantly, the use of local Gaussian representation and CLD for ordered and continuous treatments is completely new to this paper, with the two cases requiring distinct identification strategies and proof techniques.

Finally, in the difference-in-differences setup, \citet{athey2006identification} showed that average and quantile treatment effects on the treated (and the untreated) can be identified when the unobservable determinant of the untreated potential outcome is independent of time within groups. \citet{ghanem2023evaluating} provided general identification results under a time invariance assumption on the copula between the potential outcomes and the group indicator, which they show is equivalent to the assumptions in \citet{athey2006identification} with continuous outcomes. When interpreted through the lens of a local Gaussian representation, their approach restricts the time dependence of the local dependence parameter.

\vspace{8pt}
\noindent\textbf{Organization of the Paper.} Section  \ref{ssec:Preliminaries}
introduces the key variables and parameters of interest, Section \ref{ssec:LGR} states the local
Gaussian representation, and Section \ref{subsec:Assumptions}
posits the main identifying assumptions that will be used throughout
the analyses. We devote Sections \ref{subsec:Models-with-Binary}--\ref{subsec:Models-with-Continuous}
to the identification analyses with binary, ordered discrete, and continuous
treatments, respectively. Section \ref{sec:Discussions-on-Copula}
discusses CLD in further detail, and Section \ref{sec:Estimation-and-inference}
deals with estimation and inference. Section \ref{sec:Applications}
provides an empirical illustration. Section \ref{sec:Conclusions} presents concluding remarks. The Supplemental Appendix (SA) contains the proofs and additional results \citep{chernozhukov2026supplemental}.

\vspace{8pt}
\noindent\textbf{Notation.}
For scalar random variables (r.v.s) $X$ and $Y$ and a possibly multivariate r.v. $Z$, $F_{X,Y \mid Z}$ denotes the joint distribution of $(X,Y)$ conditional on $Z$, $F_{X \mid Z}$ denotes the (marginal) distribution of $X$ conditional on $Z$, and $F_{Z}$ denotes the marginal (joint) distribution of $Z$. We use calligraphic letters for support sets of r.v.s (e.g., $\mathcal{Z}$ denotes the support of $Z$). The symbol $\indep$ denotes (stochastic) independence. The interior of the set $\D$ is denoted by $\iD$.

\section{Setup and Assumptions\label{sec:Models-and-Identification}}

\subsection{Preliminaries\label{ssec:Preliminaries}}

Let $Y\in\mathcal{Y} \subseteq \mathbb{R}$ denote the scalar outcome and $D\in\mathcal{D}  \subseteq \mathbb{R}$ denote the scalar
treatment. We consider binary, discrete ordered, and continuous treatments with
$\mathcal{D}=\{0,1\}$, $\mathcal{D}=\{0, 1, \ldots,K\}$, and $\mathcal{D}$ equal to an uncountable set,
respectively. The outcome is not restricted; it can be continuous, discrete or mixed continuous-discrete. Let $Z\in\{0,1\}$ be the binary IV. We focus on a
binary $Z$ as the most challenging case; the analysis readily
extends to multi-valued  or multivariate $Z$. Let $Y_{d}$ denote the potential
outcome given $d\in\mathcal{D}$ and $D_{z}$ the potential treatment
given $z\in\{0,1\}$. They are related to the observed outcome and
treatment through $Y=Y_{D}$ and $D=D_{Z}$.\footnote{When $D$ is continuous, we require that $Y_d$ is suitably measurable \citep{hirano2004propensity}.} Let $X \in\mathcal{X} \subseteq \mathbb{R}^{d_x}$ be a vector of covariates. All the assumptions and analyses are conditional on $X$, which we suppress to lighten the notation. We make it explicit when we discuss estimation and inference in Section \ref{sec:Estimation-and-inference}.

We consider a general treatment assignment equation: 
\begin{align}
D_{z} & =h(z,V_{z}), \ \ z \in \{0,1\},\label{eq:gen_sel}
\end{align}
where $v \mapsto h(z,v)$ is weakly increasing, and we normalize $V_{z} \sim U(0,1)$. We provide examples of the function $h$ for each type of treatment below. The variable $V_z$ can be interpreted as an unobserved treatment propensity ranking, in the sense that high values of $V_z$ are associated with high treatment intensity. By allowing for a different unobservable $V_{z}$ at each value of $z$, we essentially
permit different propensity rankings for each value of $z$ and therefore $D$ to be a function of the \emph{vector} of unobservables $(V_{0},V_{1})$.

We are interested in identifying the distribution of $Y_d$, $F_{Y_{d}}$,
for $d\in\mathcal{D}$, and functionals of $F_{Y_{d}}$, such
as QSFs and ASFs. Thus, 
by using appropriate operators:
\begin{align*}
QSF_{\tau}(d)  := Q_{Y_{d}}(\tau)=\mathcal{Q}_{\tau}(F_{Y_{d}}),\ \
ASF(d)  := \Ep[Y_{d}]=\mathcal{E}(F_{Y_{d}}),
\end{align*}
where $\mathcal{Q}_{\tau}(F):=\inf\{y\in\mathcal{Y}:F(y)\ge\tau\}$
and $\mathcal{E}(F):=\int y \mathrm{d}F(y)$. QTE and ATE
 can be expressed as $\{QSF_{\tau}(d)-QSF_{\tau}(d')\}/(d-d')$ and $\{ASF(d)-ASF(d')\}/(d-d')$, $d,d' \in \D$ with $d' \neq d$, and for continuous treatments we can also consider
$\partial QSF_{\tau}(d)/\partial d$
and $\partial ASF(d)/\partial d$, $d \in \D$. When the treatment is binary, we may also be interested in the distribution of $Y_d$ in subpopulations such as the treated, $F_{Y_{d} \mid D}(\cdot \mid 1)$, and untreated, $F_{Y_{d} \mid D}(\cdot \mid 0)$,
for $d\in\{0,1\}$,  and functionals of these distributions. More generally, we may be interested in these objects for subpopulations defined by values of the covariates in $X$.

\subsection{Local Gaussian Representation\label{ssec:LGR}}

Treatment endogeneity can be captured by the joint distribution of
the potential outcome and unobservable treatment propensity ranking in equation \eqref{eq:gen_sel}. We use  the local Gaussian representation (LGR) to represent this joint distribution. This representation is the basis of our identification
and estimation strategies. Throughout the paper, let $C(u_{1},u_{2};\rho)$
denote the Gaussian copula with correlation coefficient $\rho$, that is,
$C(u_{1},u_{2};\rho) := \Phi_2(\Phi^{-1}(u_1), \Phi^{-1}(u_2); \rho)$
where $\Phi_2(\cdot, \cdot; \rho)$ is the standard bivariate Gaussian distribution with parameter $\rho$ and $\Phi$ is the standard univariate Gaussian distribution.

The following lemma shows that the distribution of any bivariate r.v. has a local Gaussian representation \citep{anjos2005representation,kolev2006copulas, chernozhukov2025distribution}.
\begin{lemma}[LGR] \label{lem:LGR}
For any r.v.s $Y$ and $V$, the joint distribution of $Y$ and $V$  admits the following representation: for $(y,v)$ such that $0 <  F_{Y}(y), F_{V}(v) < 1$,
\begin{align*}
F_{Y,V}(y,v) \equiv C(F_{Y}(y),F_{V}(v);\rho_{Y,V}(y,v)),
\end{align*}
 where $\rho_{Y,V}(y,v)$
is the unique solution in $\rho$ to 
$F_{Y,V}(y,v) = C(F_{Y}(y),F_{V}(v);\rho)
$.\footnote{For completeness, we set $\rho_{Y,V}(y,v):=0$ when one of the marginals is equal to $0$ or $1$.}
\end{lemma}
In the lemma, the LGR is fully nonparametric and is not imposing Gaussianity on the joint or marginal distributions. The distribution is equated to the Gaussian copula (with the nonparametric marginal distributions) by adjusting the value of the correlation coefficient $\rho$ for each evaluation point $(y,v)$. Note that the solution $\rho_{Y,V}(y,v)$ depends on both the dependence structure  and  marginals. Lemma A.1 in \citet{chernozhukov2025supplement} shows that $\rho_{Y,V}(y,v)$ has the same sign as $\operatorname{Cov}(1\{Y\le y\},1\{V\le v\})$. 

Gaussianity in the local representation in Lemma \ref{lem:LGR} is convenient for introducing identifying assumptions, interpreting these assumptions using joint normality as a benchmark, and developing  estimators. However, Gaussianity is not essential for the local representation, and other comprehensive single-parameter copulas could be used; see SA \ref{app:local_rep_copula}.

\subsection{Assumptions\label{subsec:Assumptions}}

We maintain the following assumptions.\footnote{We impose the assumptions for all values of $d\in \D$. If the assumptions only hold for a subset of values $\D_{A}\subset \D$, then our analysis applied to $d\in \D_A$ yields identification of $F_{Y_d}$, $d\in \D_A$, and of its functionals.}
\begin{myas}{EX}[Independence]\label{as:EX} For
$d \in \D$ and $z\in\{0,1\}$, $Z\indep (Y_{d},V_{z})$.\end{myas}

\begin{myas}{REL}[Relevance]\label{as:REL} (i) $Z\in\{0,1\}$; (ii) $0 < \Pr[Z=1] < 1$; and (iii) if $\mathcal D=\{0,1\}$,  $\Pr[D=1\mid Z=1] \neq \Pr[D=1\mid Z=0]$ and $0 < \Pr[D=1\mid Z=z] < 1$, $z\in\{0,1\}$;  if $\mathcal D= \{0, 1, ...,K\}$, $F_{D \mid Z}(d \mid 0) > F_{D \mid Z}(d \mid 1)$, $d\in\mathcal D \setminus \{K\}$,\footnote{The direction of dominance is up to relabeling the levels of $Z$; see related discussions in SA \ref{app:REL(iii)}.} and $\Pr[D=d\mid Z=z] > 0$, $(z,d) \in \{0,1\}\times \mathcal D$; and if $\mathcal D$ is uncountable, $F_{D \mid Z}(d \mid 1) \neq F_{D \mid Z}(d \mid 0)$ and $0 < F_{D \mid Z}(d \mid z) < 1$, $(z,d) \in \{0,1\}\times \iD$.\end{myas}

\ref{as:EX} is a standard exogeneity condition in IV strategies that imposes that the distribution of $Y_d$ and $V_z$ conditional on $Z$ is the same as the (unconditional) distribution of $Y_d$ and $V_z$. It is weaker than $Z\indep (\{Y_{d}\}_{d\in\mathcal{D}},V_0,V_1)$ because it does allow the relationship between $Y_d$ and $Y_{d'}$, $d,d' \in \D$, and between $V_0$ and $V_1$ to depend on $Z$. Also, in \ref{as:EX}, a standard exclusion restriction is implicit in the notation:
$Y_{d}=Y_{d,z}$ almost surely (a.s.), where $Y_{d,z}$ is the potential outcome given
$(d,z)$. \ref{as:REL}(ii)--(iii) are the usual IV relevance and  non-degeneracy conditions. \ref{as:REL}(iii) for $\mathcal D=\{0,1\}$ can be formulated as a special case of \ref{as:REL}(iii) for $\mathcal D= \{0,1,...,K\}$ with $K=1$, but we state it separately for clarity.

Let $\pi_d(z) := F_{D \mid Z}(d \mid z)$.  We refer to $\pi_d(z)$ as the $d$-treatment threshold induced by $Z=z$ as it gives the marginal value of the treatment propensity ranking $V_z$ for which the treatment changes from $D=d$ to the next level when $Z=z$. For example,  when $\D = \{0,1\}$, the treatment changes from $0$ to $1$ when $V_z$ passes the threshold $\pi_0(z) = F_{D \mid Z}(0 \mid z) = \Pr[D=0 \mid Z=z]$.  We make the following assumption about the local dependence parameter of the LGR of $(Y_{d},V_{z})$:
\begin{myas}{CLD}[Constant Local Dependence]\label{as:CLD} Consider the following condition:
\begin{equation}\label{eq:CLD}
   \rho_{Y_d,V_0}(y,\pi_d(0)) = \rho_{Y_d,V_1}(y,\pi_d(1)) =: \rho_{Y_d}(y) \in (-1,1).
\end{equation}
For $y \in \Y$, (i) if  $\D = \{0,1\}$, \eqref{eq:CLD} holds for $d=0$ and
\begin{align}\label{eq:CLD(i)}
   \rho_{Y_{1},V_0}(y,\pi_{0}(0)) &= \rho_{Y_{1},V_1}(y,\pi_{0}(1)) =: \rho_{Y_1}(y) \in (-1,1),
\end{align}
(ii) if $\D = \{0,\ldots, K\}$ with $K\ge2$, \eqref{eq:CLD} holds for all $d \in \{0,\ldots,K-1\}$ and
\begin{align}
   &\rho_{Y_{K},V_0}(y,\pi_{K-1}(0)) = \rho_{Y_{K},V_1}(y,\pi_{K-1}(1)) =: \rho_{Y_K}(y) \in (-1,1),\label{eq:CLD(ii)-1}\\
   &\rho_{Y_d,V_z}(y,\pi_d(z)) = \rho_{Y_d,V_z}(y,\pi_{d-1}(z)), \quad d \in \{1,\ldots, K-1 \}, \quad z \in \{0,1\},\label{eq:CLD(ii)-2}
\end{align}
(iii) if $\D$ is  uncountable, \eqref{eq:CLD} holds, $v\mapsto F_{Y_d,V_z}(y,v)$ is differentiable,\footnote{\label{fn:diff} Differentiability of $v\mapsto\rho_{Y_{d},V_{z}}(y,v)$
follows by the differentiability of $v\mapsto F_{Y_d,V_z}(y,v)$ and $v\mapsto C(\cdot,v;\cdot)$ and the implicit function
theorem.}  and
\begin{align}\label{eq:CLD(iii)}
   &\left. \frac{\partial \rho_{Y_d,V_z}(y,v)}{\partial v}\right|_{v = \pi_d(z)} = 0, \quad d \in \iD.
\end{align}
\end{myas}

\ref{as:CLD} is a high-level condition that imposes a shape restriction on the dependence between $Y_d$ and $V_z$. In particular, it imposes that the local dependence of $Y_d$ and $V_z$ is constant at the $d$-treatment thresholds induced by the two levels of the instrument when $D$ is binary and also around these thresholds when $D$ is discrete ordered or continuous. A sufficient condition for \ref{as:CLD} is that $\rho_{Y_d,V_0}(y,v) = \rho_{Y_d,V_1}(y,v) = \rho_{Y_d}(y)$ for all $v \in \V$, where $\V \subseteq (0,1)$ is an open interval containing the set of possible values for $\pi_d(z)$, $z \in \{0,1\}$ (with $\V=(0,1)$ this condition is equivalent to a Gaussian copula, see SA \ref{app:cld_gauss}). Under selection rank invariance (SRI) (i.e., $V_0=V_1=V$ a.s.), \ref{as:CLD} holds trivially when the copula of $(Y_d,V)$ is Gaussian because $\rho_{Y_d,V}(y,v)$ is constant everywhere. Moreover, we show in SA \ref{app:LATE_Monotonicity} that \ref{as:CLD} does not require SRI and in SA \ref{app:cld&indep} that it can be compatible with dependence between $Y_d$ and $V_z$ (i.e., endogeneity) even with parametric data generating processes (DGPs).

Section \ref{sec:Discussions-on-Copula} provides further discussions of the statistical and economic content of \ref{as:CLD} and a comparison to identification under RS. We note here that restrictions on the shape of the copula parameter, such as \ref{as:CLD}, are inherently specific to the copula chosen for the local representation. Moreover, for the LGR to represent a valid distribution function, the local dependence parameter must satisfy certain restrictions \citep[see, e.g.,][]{nelsen2006introduction}. These restrictions could, in principle, be exploited to develop a specification test for  \ref{as:CLD}.\footnote{For example, in the binary treatment case with $d=0$, these restrictions are that $y \mapsto C(F_{Y_0}(y),\pi_0(z);\rho_{Y_0}(y))$ and $y \mapsto F_{Y_0}(y) - C(F_{Y_0}(y),\pi_0(z);\rho_{Y_0}(y))$ be nondecreasing on $\Y$ for $z\in\{0,1\}$.}

\begin{figure}[h!]
	\begin{center}
     \caption{Examples of Distributions that Satisfy \ref{as:CLD}}\label{fig:ci}
  		\includegraphics[width=\textwidth]{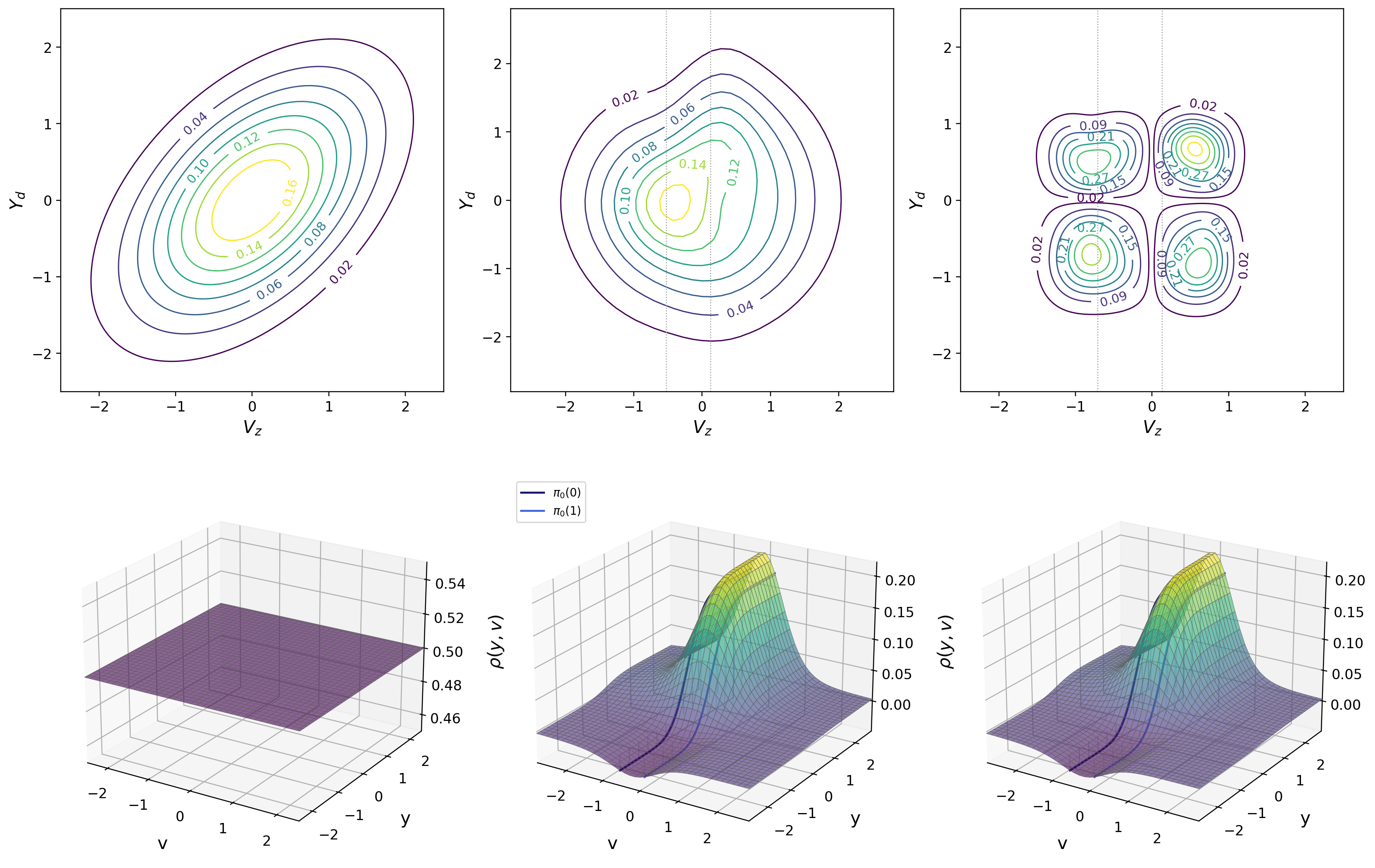}
        (a)\hspace{5cm}(b)\hspace{4.9cm}(c)
	\end{center}

\vspace{0.1cm}
\footnotesize{\textit{Notes:} Panel (a): Bivariate Gaussian distribution (i.e., Gaussian copula and univariate Gaussian marginals) with correlation 0.5; Panel (b): a two-channel-deviation (non-Gaussian) copula with $\rho_{Y_{d},V_0}(y,\pi_0(0)) = \rho_{Y_{d},V_1}(y,\pi_0(1)) = \rho_{Y_{d}}(y)$ for all $y$, with $\rho_{Y_{d},V_z}(y,v)$ non-constant for all other $(y,v)$, and univariate Gaussian marginals (see Section \ref{ssec:Roy}); Panel (c): the copula in (b) and non-Gaussian marginals $\Phi(a(\text{sgn}(u) + 1(u>0)/2)u^2+b)$ for both $Y_d$ and $V_z$, where $a$ and $b$ are calibrated such that the distribution has zero mean and unit variance. The upper figures display contour plots of the joint distribution and the lower
figures display the corresponding local dependence functions. All the figures are drawn in terms of $\Phi^{-1}$-transformed $V_z$ for readability.}	
\end{figure}

Figure \ref{fig:ci} shows examples of joint distributions of $(Y_d,V_z)$ that satisfy \ref{as:CLD}. Panel (a) corresponds to a bivariate Gaussian distribution (i.e., a Gaussian copula and univariate Gaussian marginals); Panel (b) corresponds to a non-Gaussian copula and standard univariate Gaussian marginals that we introduce later in Section \ref{ssec:Roy}; and Panel (c) corresponds to the copula in (b) coupled with non-Gaussian marginals. The upper figures display contour plots of the joint distribution, and the lower figures display the corresponding local dependence functions. These examples showcase that \ref{as:CLD} is compatible with very diverse shapes for the joint distribution, including multimodality and asymmetry.\footnote{We refer to \citet{wasserman-nonparanormal} and \citet{chernozhukov2025distribution} for more examples.} Moreover, even though the Gaussian copula does not have tail dependence, \ref{as:CLD} allows for it because the local dependence parameter can change with the level of $Y_d$.

\section{Identification Analysis\label{identification-analyses}}

\subsection{Binary Treatment\label{subsec:Models-with-Binary}}

We start by considering the identification of the causal effects of
a binary treatment $D\in\mathcal{D}=\{0,1\}$. To reflect this, we
consider a treatment selection equation 
\begin{equation}\label{eq:gen_sel_binary}
    D_z=h(z,V_{z})=1\{V_{z} > \pi_0(z)\}, \ \ z \in \{0,1\},
\end{equation}
with  $\Pr[D=0 \mid Z=z] = \Pr[D_z = 0 \mid Z=z] = \Pr[V_{z}\leq \pi_0(z)] = \pi_0(z)$
by \ref{as:EX} and the normalization $V_{z} \sim U(0,1)$.
Note that $V_{1}$ and $V_{0}$ are two potentially distinct unobservables that
do not restrict the relationship between $D_0$ and $D_1$.\footnote{This is in contrast to the LATE monotonicity
assumption \citep{imbens1994identification}, which imposes either $D_1 \geq D_0$ or $D_0 \geq D_1$, a.s., which corresponds to $V_{1}=V_{0}$, a.s. 
\citep{vytlacil2002independence}. See SA \ref{app:LATE_Monotonicity} for further discussions.}

For the identification analysis, consider
\begin{align}
\Pr[Y \leq y, D = 0 \mid Z=z] & =\Pr[Y_{0}\le y,D_{z} = 0 \mid Z=z] \notag =\Pr[Y_{0}\le y,V_{z} \leq \pi_0(z) \mid Z=z]  
\notag \\
&  =C(F_{Y_{0}}(y),\pi_0(z);\rho_{Y_{0},V_{z}}(y,\pi_0(z))) , \quad (y,z) \in\mathcal{Y}\times\{0,1\},\label{eq:sys_Y1}
\end{align}
where the second equality uses equation \eqref{eq:gen_sel_binary}, and the last equality
follows from \ref{as:EX} and Lemma \ref{lem:LGR}. For each $y \in \mathcal{Y}$, this is a system of two equations in three unknowns, $F_{Y_{0}}(y)$, $\rho_{Y_{0},V_{0}}(y,\pi_0(0))$ and $\rho_{Y_{0},V_{1}}(y,\pi_0(1))$. \ref{as:CLD}(i) (\eqref{eq:CLD} in particular) reduces the number of unknowns to two: $(F_{Y_{0}}(y)$, $\rho_{Y_{0}}(y))$. The following theorem shows that the nonlinear system of equations \eqref{eq:sys_Y1} has a unique solution under \ref{as:CLD}(i). This result follows from a global univalence theorem of \cite{gale1965jacobian}, because the Jacobian of the system of equations is a P-matrix under \ref{as:REL}(iii), the map defined by the system is differentiable (as the copula is differentiable) and the parameter space $(0,1)\times (-1,1)$ for $(F_{Y_{0}}(y), \rho_{Y_{0}}(y))$ is open and rectangular.\footnote{We discuss identification using other local representations in SA \ref{app:local_rep_copula}.} A similar argument shows that the distribution of $Y_1$ and the local dependence parameter of the LGR of $Y_1$ and $V_z$ are identified under \eqref{eq:CLD(i)} in \ref{as:CLD}(i).\footnote{The  system \eqref{eq:sys_Y1} becomes $\Pr[Y \leq y, D = 1 \mid Z=z] = C(F_{Y_{1}}(y),1-\pi_0(z);-\rho_{Y_{1},V_{z}}(y,\pi_0(z)))$, $(y,z) \in\mathcal{Y}\times\{0,1\}$.} Let $\bar \Y_d := \{y \in \mathbb{R} : 0 < F_{Y_d}(y) < 1\}$.

\begin{theorem}[Identification for Binary Treatment]\label{thm:ID_binary}Suppose $D_z\in\{0,1\}$ satisfies
\eqref{eq:gen_sel_binary} for $z \in \{0,1\}$. Under  \ref{as:EX}, \ref{as:REL},
and \ref{as:CLD}(i), the functions $y \mapsto F_{Y_{d}}(y)$ and $y \mapsto \rho_{Y_{d}}(y)$ are identified on $y\in \bar \Y_d$, 
for $d\in\{0,1\}$.\footnote{The set $\bar \Y_d$ is identified because $F_{Y_d}(y) = 1$ if and only if $\Pr(Y \leq y \mid D= d, Z=z) = 1,$ $z \in \{0,1\},$ and $F_{Y_d}(y) = 0$ if and only if $\Pr(Y \leq y \mid  D= d, Z=z) = 0,$ $z \in \{0,1\}$.} 
\end{theorem}

The proof of Theorem \ref{thm:ID_binary} is contained in SA \ref{app:proof_ID_binary}. It is interesting to compare Theorem \ref{thm:ID_binary} to \cite{chernozhukov2005iv}, who also establish identification of $F_{Y_{d}}$ when $D$ and $Z$ are binary. Compared to them, we do not impose RS and thus, as we show in SA \ref{app:Rank-Similarity-Restricts}, allow for richer effect heterogeneity. As a trade-off, we restrict the form of endogeneity by imposing \ref{as:CLD}. For a more detailed comparison, see Section \ref{ssec:RS}.

\begin{remark}[Identification of $F_{Y_1 \mid D}$ and $F_{Y_0 \mid D}$]\label{remark:tt} We focus on identification for the untreated, $D=0$; identification for the treated, $D=1$, follows by a similar argument. The distribution of $Y_0$ is trivially identified from $F_{Y_0 \mid D}(y \mid 0) = F_{Y \mid D}(y \mid 0)$. Identification of the distribution of $Y_1$ follows from 
$F_{Y_{1} \mid D}(y \mid 0)=(F_{Y_{1}}(y)-\Pr[D=1] F_{Y \mid D}(y\mid 1))/\Pr[D=0]$,  where $F_{Y_{1}}(y)$ is identified by Theorem \ref{thm:ID_binary}. \qed\end{remark}

\begin{remark}[One-sided non-compliance]\label{rem:random_ITT}Under one-sided non-compliance, where $D=1$ whenever $Z=1$ (i.e., $\Pr[D=1 \mid Z=1] =1$),  \ref{as:REL}(iii) is violated. In this case
$F_{Y_{0}}(y)$ is no longer identified because one of the equations in \eqref{eq:sys_Y1} becomes
uninformative as $\pi_0(1)=0$. We can still identify $F_{Y_1}(y)=F_{Y\mid Z}(y\mid 1)$. Therefore, we can also identify $F_{Y_1\mid D}(y\mid d)$, $d\in \{0,1\}$, as in  Remark \ref{remark:tt}; and $F_{Y_{0}\mid D}(y\mid 0)=F_{Y \mid D}(y \mid 0)$. %We can still identify $F_{Y_{d}\mid D}(y\mid 0)$ for $d=0,1$ using the same analysis of Remark \ref{remark:tt} for $F_{Y_{1}\mid D}(y\mid 0)$, because $F_{Y_{1}}(y)=F_{Y \mid Z}(y \mid 1)$; and $F_{Y_{0}\mid D}(y\mid 0)=F_{Y \mid Z}(y \mid 0)$. 
\qed\end{remark}

\begin{remark}[Overidentification with non-binary instruments]\label{rem:overID}
    It is clear from the identification analysis that, when the IV takes more than two values (or equivalently, when there are multiple binary IVs), the resulting system of equations produces overidentifying restrictions. These restrictions can be used to test the model specification, and \ref{as:CLD} in particular. For example, if we have multiple binary IVs, we can combine the first-order conditions of the estimators for each IV in Section \ref{sec:Estimation-and-inference} through a GMM approach and use the standard $J$-statistic to test for overidentification. Multi-valued IVs can also be used to make the model more flexible by allowing the local dependence parameter to partially vary  with the value of $V_z$ as \ref{as:CLD} only needs to be satisfied for two values. Although we shall not repeat it, this remark also applies to the subsequent cases of discrete ordered and continuous $D$. \qed
\end{remark}

\begin{remark}[Relaxing \ref{as:CLD}] Here, we outline how one can relax \ref{as:CLD} and analyze the impact of failures of \ref{as:CLD} on the identification of $F_{Y_0}$ and $F_{Y_1}$. We focus on $F_{Y_0}$; the analysis for $F_{Y_1}$ is symmetric. Consider the following relaxation of \ref{as:CLD}(i)
\begin{equation}
|\rho_{Y_0,V_0}(y,\pi_0(0)) - \rho_{Y_0,V_1}(y,\pi_0(1))|\le r_0 \label{eq:relaxation_CLD}
\end{equation}
for some $0<r_0\le 2$. Under this relaxation, we can build on \citet[][Appendix A.4]{chernozhukov2025supplement} to characterize the identified set for $F_{Y_0}(y)$ as a function of $r_0$. To describe the resulting identified set, consider the following system of equations:
\begin{align}
\Pr[Y \leq y, D = 0 \mid Z=z] 
&=C(F_0,\pi_0(z);\rho_{0,z}) , \quad z \in\{0,1\}.\label{eq:sys_partial_ID}
\end{align}
For $z\in\{0,1\}$, denote by $S_{0,z}$ the set of all solutions to  \eqref{eq:sys_partial_ID}. Then, \citet[][Theorem A.1]{chernozhukov2025supplement} implies that the identification region for $F_{Y_0}(y)$ is 
$\operatorname{Proj}_1 R_0(r_0)$ where $\operatorname{Proj}_1$ is the projection operator with respect to the first coordinate and 
$$
R_0(r_0):=\{(F_0,\rho_{0,0},\rho_{0,1}):(F_0,\rho_{0,z})\in S_{0,z}, z\in \{0,1\}\}\cap\{(F_0,\rho_{0,0},\rho_{0,1}):|\rho_{0,0}-\rho_{0,1}|\le r_0\}.$$ 
In practice, $\operatorname{Proj}_1 R_0(r_0)$ needs to be computed numerically. By varying $r_0$, one can perform sensitivity and breakdown analyses. \qed \end{remark}

\subsection{Ordered Discrete Treatment\label{subsec:Models-with-Ordered}}

We consider identification of the causal effects of a multi-valued
ordered treatment $D\in\mathcal{D}=\{0,\dots,K\}$ using a binary
instrument $Z\in\{0,1\}$. We assume a threshold-crossing model for the treatment selection equation, which can be viewed as a natural
extension of model \eqref{eq:gen_sel_binary} from two to multiple treatment levels, 
\begin{equation}
D_{z}= h(z,V_z) = \sum_{d \in \D} d \ 1\{ \pi_{d-1}(z) < V_{z}\le\pi_{d}(z) \}= \begin{cases}
0, & \pi_{-1}(z) < V_{z}\le\pi_{0}(z)\\
1, & \pi_{0}(z)<V_{z}\le\pi_{1}(z)\\
\vdots & \vdots\\
K, & \pi_{K-1}(z)<V_{z}\le \pi_K(z)
\end{cases},\label{eq:sel_ordered}
\end{equation}
where $\pi_{-1}(z) = 0$ and $\pi_K(z) =1$. 
Equation \eqref{eq:sel_ordered}  generalizes the model in Section 7.2 of \citet{heckman2007chapter71} by allowing for two unobservables $(V_0,V_1)$ and a different impact of the instrument on the different
cutoffs; see SA \ref{subsec:HV07}. It is not fully general, however, as it imposes that the unobservables are the same for all the treatment levels.\footnote{\citet{cunha2007identification} showed that ordered choice models with multivariate unobservables are point identified under strong assumptions.}

Under the normalization $V_{z} \sim U(0,1)$ and \ref{as:EX},
the threshold functions $\pi_{d}(z)$ are identified by the distribution of the observed treatment conditional on the instrument, 
$\pi_{d}(z)=F_{D \mid Z}(d \mid z)$ for $d\in\mathcal{D}$.
For the identification analysis, consider
\begin{multline}
\Pr[Y \leq y, D = d \mid Z=z]  = \Pr[Y_{d}\le y,\pi_{d-1}(z)<V_{z}\le\pi_{d}(z) \mid Z=z] \\
  = C(F_{Y_{d}}(y),\pi_{d}(z);\rho_{Y_{d},V_{z}}(y,\pi_{d}(z))) \\ 
-C(F_{Y_{d}}(y),\pi_{d-1}(z);\rho_{Y_{d},V_{z}}(y,\pi_{d-1}(z))), ~ (y,d,z) \in \mathcal{Y}\times \mathcal{D}\times \{0,1\}, \label{eq:LGR_D=00003Dk}  
\end{multline}
where the first equality follows from \eqref{eq:sel_ordered} and the second equality from \ref{as:EX} and Lemma \ref{lem:LGR}. For each $d \in \mathcal{D}$ and $y \in \mathcal{Y}$, \eqref{eq:LGR_D=00003Dk} is a system of two equations in five unknowns: $F_{Y_{d}}(y)$, $\rho_{Y_{d},V_{0}}(y,\pi_{d-1}(0))$, $\rho_{Y_{d},V_{0}}(y,\pi_{d}(0))$, $\rho_{Y_{d},V_{1}}(y,\pi_{d-1}(1))$, and $\rho_{Y_{d},V_{1}}(y,\pi_{d}(1))$. \ref{as:REL}(iii) guarantees that the two equations of the system are not redundant.\footnote{A weaker version of \ref{as:REL}(iii) (i.e., $F_{D \mid Z}(d \mid 1) \neq F_{D \mid Z}(d \mid 0)$) is sufficient for nonredundancy, but \ref{as:REL}(iii) is used for establishing uniqueness of the solution below.}

For $d \in \{0,K\}$, one of the terms on the right hand side drops out because either $\pi_{d-1}(z) = 0$ or $\pi_{d}(z) =1$, yielding a system of two equations in three unknowns. Consequently, the distribution of the potential outcome and local dependence parameter can be identified by combining \ref{as:REL}(iii) and \eqref{eq:CLD} or \eqref{eq:CLD(ii)-1}. For $d\in \{1,\ldots,K-1\}$, \eqref{eq:CLD} 
reduces the number of unknowns only to four. We impose additionally \eqref{eq:CLD(ii)-2} 
to further reduce the number of unknowns to two: $F_{Y_{d}}(y)$ and $\rho_{Y_d}(y)$. The Jacobian of the resulting system of equations, however, does not satisfy the conditions to apply the global univalence results of \citet{gale1965jacobian} even under \ref{as:REL}(iii). We show uniqueness of the solution using an alternative proof strategy described below.

We summarize the identification result in the following theorem:
\begin{theorem}[Identification for Ordered Treatment]\label{thm:ID_ordered} Suppose $D_z$, $z \in \{0,1\}$,
satisfies \eqref{eq:sel_ordered}. Under  \ref{as:EX},
\ref{as:REL}, and \ref{as:CLD}(ii), the functions $y \mapsto F_{Y_d}(y)$ and
$y \mapsto \rho_{Y_{d}}(y)$ are identified on $y \in \bar \Y_d$, for $d \in \mathcal{D}$.
\end{theorem}

In the proof of Theorem \ref{thm:ID_ordered}, contained in SA \ref{app:proof_thm32}, we proceed as follows to show that \eqref{eq:LGR_D=00003Dk}  has a unique solution. First, we use the monotonicity in the first coordinate $F_{Y_d}(y)$ to profile out this coordinate using the first equation of the system. Second, we show that the profiled second equation is strictly monotonic in the remaining coordinate $\rho_{Y_d}(y)$. Here, we use the fact that the Jacobian of the  system has full rank everywhere in the parameter space (by \ref{as:REL}).\footnote{We rely on the stochastic dominance between $\pi_d(0)=F_{D \mid Z}(d \mid 0)$ and $\pi_d(1)=F_{D \mid Z}(d \mid 1)$ in \ref{as:REL}(iii) to establish full rank of the Jacobian. In contrast, a weaker relevance condition $\pi_d(0)\neq \pi_d(1)$ is not sufficient in general.} Third, by combining the monotonicity in the first and second steps, we conclude that the solution of the system is unique everywhere in the parameter space.

\subsection{Continuous Treatment\label{subsec:Models-with-Continuous}}

Suppose $D\in\mathcal{D} \subseteq\mathbb{R}$, where $\mathcal{D}$ is uncountable and $v\mapsto F^{-1}_{D\mid Z}(v\mid z)$ is strictly increasing on $(0,1)$ for $z\in\{0,1\}$. Then, by 
\ref{as:EX} and $V_z \sim U(0,1)$, the treatment selection equation is
\begin{equation}\label{eq:sel_continuous}
    D_z = h(z,V_z) = F_{D \mid Z}^{-1}(V_z \mid z), \ \ z \in \{0,1\}.
\end{equation} 

For the identification analysis, consider 
\begin{equation}
F_{Y\mid D,Z}(y\mid d,z)=F_{Y_{d}\mid D_{z},Z}(y\mid d,z)=F_{Y_{d}\mid V_{z},Z}(y\mid \pi_d(z),z),\label{eq:ID_cont1}
\end{equation}
where the second equality follows from equation \eqref{eq:sel_continuous}
and a change of variables. By the properties of the conditional distribution,
Lemma \ref{lem:LGR}, and \ref{as:EX},
\begin{multline}\label{eq:ID_cont2}
  F_{Y_{d}\mid V_{z},Z}(y\mid v,z)=\frac{(\partial/\partial v)F_{Y_{d},V_{z}\mid Z}(y,v\mid z)}{(\partial/\partial v)F_{V_{z}\mid Z}(v\mid z)} =C_{2}(F_{Y_{d}}(y),v;\rho_{Y_{d},V_{z}}(y,v)) \\ \quad+C_{\rho}(F_{Y_{d}}(y),v;\rho_{Y_{d},V_{z}}(y,v))(\partial/\partial v)\rho_{Y_{d},V_{z}}(y,v),
\end{multline}
where $C_{2}$ and $C_{\rho}$ are the derivatives of $C(\cdot,v;\rho)$ with respect
to $v$ and $\rho$, respectively, and the differentiability follows from \ref{as:CLD}(iii). Then, by the properties of Gaussian copula and \ref{as:CLD}(iii), combining
\eqref{eq:ID_cont1} and \eqref{eq:ID_cont2} yields
\begin{equation}
\Phi^{-1}\left(F_{Y\mid D,Z}(y\mid d,z)\right)=a_{d,y}+b_{d,y}\Phi^{-1}(\pi_d(z)),\quad z\in\{0,1\},\label{eq:ID_cont}
\end{equation}
where $a_{d,y}:=\Phi^{-1}(F_{Y_{d}}(y))/\sqrt{1-\rho_{Y_{d}}(y)^{2}}$
and $b_{d,y}:= -\rho_{Y_{d}}(y)/\sqrt{1-\rho_{Y_{d}}(y)^{2}}$. Equation \eqref{eq:ID_cont} yields a linear system of two equations in two unknowns, $a_{d,y}$ and $b_{d,y}$, which has solution
\begin{align}
    a_{d,y} &=  \frac{\Phi^{-1}(F_{Y \mid D,Z}(y \mid d,0))\Phi^{-1}(\pi_d(1)) - \Phi^{-1}(F_{Y \mid D,Z}(y \mid d,1))\Phi^{-1}(\pi_d(0))}{\Phi^{-1}(\pi_d(1)) - \Phi^{-1}(\pi_d(0))},\nonumber \\
   b_{d,y} &= \frac{\Phi^{-1}(F_{Y \mid D,Z}(y \mid d,1)) - \Phi^{-1}(F_{Y \mid D,Z}(y \mid d,0))}{\Phi^{-1}(\pi_d(1)) - \Phi^{-1}(\pi_d(0))}, \label{eq:a_b}
\end{align}
under \ref{as:REL}.\footnote{In SA \ref{app:local_rep_copula}, we provide another representation of the joint distribution of $(Y_d,V_z)$ under which identification is achieved with the same set of assumptions.}

This discussion implies the following identification result. 
\begin{theorem}[Identification for Continuous Treatment]\label{thm:ID_continuous} Suppose $D_z$, $z \in \{0,1\}$,
satisfies \eqref{eq:sel_continuous}. Under  \ref{as:EX}, \ref{as:REL},
and \ref{as:CLD}(iii), the functions $y \mapsto F_{Y_{d}}(y)$ and $y \mapsto \rho_{Y_{d}}(y)$ are identified on $y\in \bar \Y_d$, 
for $d\in\iD$, by
$$
F_{Y_d}(y) = \Phi\left( \frac{a_{d,y}}{\sqrt{1+b_{d,y}^2}} \right), \quad \rho_{Y_d}(y) = \frac{-b_{d,y}}{\sqrt{1+b_{d,y}^2}},
$$
where $a_{d,y}$ and $b_{d,y}$ are defined in \eqref{eq:a_b}.
\end{theorem}

It is interesting to compare this identification result to \citet{imbens2009identification} and \citet{torgovitsky2010identification,torgovitsky2015identification}, who also provide identification results for nonseparable models with continuous $D$. Unlike \citet{imbens2009identification}, our approach does not require an instrument with large support or SRI, but instead imposes \ref{as:CLD}.
Unlike \citet{torgovitsky2010identification}, our approach does not require RI nor SRI, but again imposes \ref{as:CLD}.

\section{Discussions on Constant Local Dependence\label{sec:Discussions-on-Copula}}

To further understand \ref{as:CLD}, we provide
 an equivalent  condition and interpret its implications for treatment selection (Sections \ref{sssec:SI} and \ref{subsec:selection}), showcase its flexibility in the context of Roy models (Section \ref{ssec:Roy}), compare it with RS (Section \ref{ssec:RS}), and explain how it extrapolates to achieve identification for the overall population (Section \ref{ssec:extrapolation}).

\subsection{\ref{as:CLD} is Equivalent to a Local Single Index Condition}\label{sssec:SI}
We start with a representation of conditional probabilities that follows from the LGR in Lemma \ref{lem:LGR}. We shall use this representation to provide a local single index condition equivalent to \ref{as:CLD} and to interpret  \ref{as:CLD} in terms of treatment selection. Although the representation holds generally, we specialize it to our case with $V \sim U(0,1)$ for concreteness.

\begin{lemma}[LGR for Conditional Probabilities]\label{lem:lgr2} If $V \sim U(0,1)$, $\rho_{Y,V}(y,v)$ is the local dependence parameter of the  LGR  $F_{Y,V}(y,v) \equiv C(F_{Y}(y),v;\rho_{Y,V}(y,v))$ if and only if
    \begin{equation*}
    \Pr[Y \leq y \mid V \leq v]  = \frac{1}{v} \int_0^{v} \Phi\left(a_{Y,V}(y,v) + b_{Y,V}(y,v) \Phi^{-1}(\tilde v)\right) \mathrm{d} \tilde v,
\end{equation*}
where 
$$
a_{Y,V}(y,v) := \frac{\Phi^{-1}(F_{Y}(y))}{\sqrt{1-\rho_{Y,V}(y,v)^2}}, \quad b_{Y,V}(y,v) := - \frac{\rho_{Y,V}(y,v)}{\sqrt{1-\rho_{Y,V}(y,v)^2}}.
%\footnote{If $V \not\sim U(0,1)$, the term $1/v$ becomes $1/F_V(v)$, where $F_V$ is the distribution function of $V$.}
$$
\end{lemma}

Lemma \ref{lem:lgr2} shows that any conditional probability has an aggregated local single index representation with coefficients that depend on the values of $y$ and $v$. The following assumption on the coefficients of this representation for $Y_d$ and $V_z$ is equivalent to \ref{as:CLD}, as shown in Theorem \ref{thm:equivalence} below.

\begin{myas}{SI}[Local Single Index]\label{as:SI} Let $a_{Y_d,V_0}(y,\pi_d(0))  = a_{Y_d,V_1}(y,\pi_d(1)) =: a_{d,y}$ and $b_{Y_d,V_0}(y,\pi_d(0)) = b_{Y_d,V_1}(y,\pi_d(1)) =: b_{d,y}$, so that
\begin{equation}\label{eq:si1}
       \Pr[Y_d \leq y \mid V_z \leq \pi_d(z)]  
    = \frac{1}{ \pi_d(z)} \int_0^{\pi_d(z)} \Phi\left(a_{d,y} + b_{d,y} \Phi^{-1}(v)\right) \mathrm{d} v, \qquad z\in\{0,1\}.  
\end{equation} 
For $y \in \Y$,  (i) if  $\D = \{0,1\}$, \eqref{eq:si1} holds for $d=0$, $a_{Y_1,V_0}(y,\pi_0(0))  = a_{Y_1,V_1}(y,\pi_0(1)) $ and $b_{Y_1,V_0}(y,\pi_0(0)) = b_{Y_1,V_1}(y,\pi_0(1))$; (ii) if $\D = \{0,\ldots,K\}$, $K \geq 2$, \eqref{eq:si1} holds for all $d \in \{0,\ldots,K-1\}$, and $a_{Y_K,V_0}(y,\pi_{K-1}(0))  = a_{Y_K,V_1}(y,\pi_{K-1}(1)) $, $b_{Y_K,V_0}(y,\pi_{K-1}(0)) = b_{Y_K,V_1}(y,\pi_{K-1}(1))$, $a_{Y_d,V_z}(y,\pi_{d-1}(z))  = a_{d,y}$ and $b_{Y_d,V_z}(y,\pi_{d-1}(z)) = b_{d,y}$ for $z\in\{0,1\}$ and $d \in   \{1,\ldots, K-1\}$, so that, for $z \in \{0,1\}$,
\begin{multline*}
    \Pr[Y_d \leq y \mid \pi_{d-1}(z) < V_z \leq \pi_d(z)] 
    = \frac{1}{ \pi_d(z) - \pi_{d-1}(z)} \int_{\pi_{d-1}(z)}^{\pi_d(z)} \Phi\left(a_{d,y} + b_{d,y} \Phi^{-1}(v)\right) \mathrm{d} v;
\end{multline*}
or (iii) if $\D$ is uncountable, \eqref{eq:si1} holds, $v \mapsto a_{Y_d,V_z}(y,v)$ and $v \mapsto b_{Y_d,V_z}(y,v)$ are differentiable and $(\partial/\partial v) a_{Y_d,V_z}(y,v) = (\partial/\partial v) b_{Y_d,V_z}(y,v) = 0$ at $v = \pi_d(z)$ for all $z \in \{0,1\}$ and $d \in \iD$, so that
$$
F_{Y_d \mid V_z}(y \mid \pi_d(z)) =  \Phi\left(a_{d,y} + b_{d,y} \Phi^{-1}(\pi_d(z))\right), \quad z\in\{0,1\}, \quad d \in \iD.
$$
\end{myas}

The first condition of \ref{as:SI} in \eqref{eq:si1} imposes that the index coefficients of the representation of the conditional probabilities are the same at the two $d$-thresholds induced by the instrument. \ref{as:SI}(ii) additionally imposes that the conditional probability in between thresholds of consecutive treatment levels has an aggregated single index form with the same coefficients across instrument levels, whereas \ref{as:SI}(iii) additionally imposes that the distribution of $Y_d$ conditional on $V_z$ has a single index form locally at the $d$-thresholds. 
\begin{theorem}[\ref{as:CLD} $\iff$ \ref{as:SI} ]\label{thm:equivalence} \ref{as:SI} is a necessary and sufficient condition for \ref{as:CLD}. 
\end{theorem}

We focus on \ref{as:SI}(iii) as a way to understand \ref{as:CLD}(iii). \ref{as:CLD}(i) is discussed in the subsequent sections and \ref{as:CLD}(ii) can be viewed as an intermediate case. We show that \ref{as:SI}(iii) is weaker than Gaussianity of the copula of $Y_d$ and $V_z$. We make the comparison under SRI, $V_1=V_0=V$ a.s., because this assumption is usually maintained together with Gaussianity. Under this assumption,  \ref{as:SI}(iii) becomes
\begin{equation}\label{eq:si}
    F_{Y_d \mid V}(y \mid \pi_d(z)) =  \Phi\left(a_{d,y} + b_{d,y} \Phi^{-1}(\pi_d(z))\right), \quad z\in\{0,1\}.
\end{equation}
Gaussianity imposes that
\begin{equation}\label{eq:gauss}
    F_{Y_d \mid V}(y \mid v) =  \Phi\left(a_{d,y} + b_{d} \Phi^{-1}(v)\right), \quad v \in (0,1).
\end{equation}
Compared to \eqref{eq:si}, \eqref{eq:gauss} requires that the single index restriction be \textit{global} in $v\in(0,1)$ with the same index coefficient $b_d$ for every value of $y$; we show in Section \ref{subsec:selection} that this restriction implies monotone treatment selection. Based on this comparison and Lemma \ref{lem:lgr2}, we can interpret \ref{as:SI}(iii) as imposing  Gaussian behavior only locally at the $d$-thresholds induced by the instrument, with the added flexibility that this local behavior can exhibit heterogeneous treatment selection patterns at different levels of $y$. Moreover, \ref{as:SI}(iii) does not require SRI, which we have imposed here only for simplicity.  Finally, we note that given equation \eqref{eq:ID_cont}, $b_{d,y}$ in \ref{as:SI}(iii) can be viewed as a control function. While we have $b_{d,y}=0$ without endogeneity (i.e., $\rho_{Y_d}(y)=0$), $b_{d,y}$ corrects for endogeneity as $\rho_{Y_d}(y)$ departs from zero. We provide additional examples and counterexamples of \ref{as:CLD}/\ref{as:SI} in SA \ref{app:counterex}.

\subsection{Treatment Selection Patterns Under \ref{as:CLD}}\label{subsec:selection}

In this section, we discuss the implications of \ref{as:CLD} for treatment selection. Assume SRI for simplicity and $\pi_d(0) > \pi_d(1)$ without loss of generality. We can assess the sign of the treatment selection by comparing  the distribution of $Y_d$ conditional on $V=v$  at different values of $v$. In particular, the function $v \mapsto \Pr[Y_d \leq y \mid V = v]$ measures the proportion of individuals with $Y_d \leq y$ at each treatment propensity $v$. It is constant under random treatment assignment, increasing under negative treatment selection, that is, when $Y_d$ and $V$ are negatively related, and decreasing under positive treatment selection. Moreover, the sign of the derivative of  $v \mapsto \Pr[Y_d \leq y \mid V = v]$ determines minus the sign of the selection locally at each value of $v$, provided that the function is differentiable.
%\footnote{Note that $v \mapsto \Pr[Y_d \leq y \mid V = v]$ is differentiable by the argument given in footnote \ref{fn:diff}.} 

When the copula of $Y_d$ and $V$ is Gaussian with correlation $\rho$,   
$v \mapsto \Pr[Y_d \leq y \mid V = v] =\Phi\left(a_{d,y} + b_{d} \Phi^{-1}(v)\right)$ is globally constant, increasing or decreasing for all the values of $v$ and $y$ depending on the sign of $b_d$ (i.e., $-\rho$). Gaussianity therefore imposes monotone treatment selection, that is, a first-order stochastic dominance relationship in the distribution of $Y_d \mid V=v$ at different values of $v$. \ref{as:CLD}(iii) imposes restrictions on treatment selection \emph{locally} at the two $d$-treatment thresholds induced by $Z$; see \eqref{eq:si}. Thus, the sign of $\partial \Pr[Y_d \leq y \mid V = v]/\partial v$ is the same at $\pi_d(z)$, $z \in \{0,1\}$ for each $y$. However,  this sign does not need to be the same for all the values of $y$ because the parameter $\rho_{Y_d}(y)$ can change with $y$, allowing for non-monotone treatment selection. For example, if $\rho_{Y_d}(y)>0$, there is positive selection at $y$, but there might be adverse selection at $y'$ if $\rho_{Y_d}(y')<0$. In other words, while switching $Z$ from $0$ to $1$ changes the treatment propensity from $\pi_d(0)$ to $\pi_d(1)$, it does not change the sorting mechanism locally, but the mechanism and extent of the sorting can vary at different parts of the distribution of $Y_d$. Hence, \ref{as:CLD} does not impose any stochastic order relationship between the distribution of $Y_d$ conditional on $V = \pi_d(0)$ and on $V=\pi_d(1)$.\footnote{Note that the sign of $\Pr[Y_d \leq y \mid V = \pi_d(0)] - \Pr[Y_d \leq y \mid V = \pi_d(1)]$ is determined by the sign of $b_{d,y}$ (i.e., $- \rho_{Y_d}(y)$).}

\subsection{Roy Models and Control Functions Under \ref{as:CLD}}\label{ssec:Roy}

Figure \ref{fig:ci} in Section \ref{subsec:Assumptions} demonstrates the flexibility of \ref{as:CLD} in modeling the joint distribution of $(Y_d,V_z)$. Our analyses show how point identification can be achieved under substantially weaker modeling restrictions than Gaussianity, which can be appealing to empirical researchers. To illustrate this further, consider a generalized Roy model for two sectors indexed by binary indicator $D$. Let the potential log wage in sector $d$ be $Y_d=\Ep[Y_d]+\epsilon_d$ for $d\in\{0,1\}$, and suppose that individuals choose sector $D=1$ if the benefit $Y_1 -Y_0$ exceeds the cost $C_z$ where $Z=z$ is a cost shifter. This implies that $D_z=1\{V_z> \pi_0(z)\}$,
where $V_z := F_{Y_1-Y_0-C_z}(Y_1-Y_0-C_z)$ and $\pi_0(z) := F_{Y_1-Y_0-C_z}(0)$.\footnote{This model is more flexible than the standard generalized Roy model as the cost $C_z$ can include a $z$-specific unobservable, which our framework accommodates.} Suppose $Z \indep (Y_0,Y_1,C_z)$. Then, $\Ep[Y\mid D=0, Z=z]=\Ep[Y_0]+\Ep[\epsilon_0\mid V_z \le \pi_0(z)]$, where $\Ep[\epsilon_0\mid V_z \le \pi_0(z)]$ is the \emph{control function} that captures endogenous selection into sector $d=0$; we can obtain a similar expression for $d=1$. For ease of illustration, focus on sector 0, and let $\Ep[Y_0]=0$, and impose SRI, so that the control function simply becomes $\Ep[\epsilon_0\mid V_z \le \pi_0(z)]=\Ep[Y_0\mid V \le \pi_0(z)]$.

\begin{figure}[h!]
\begin{center}
\caption{Examples of Control Functions That Satisfy \ref{as:CLD}}\label{fig:rs-illustration}

\includegraphics[width=0.49\textwidth]{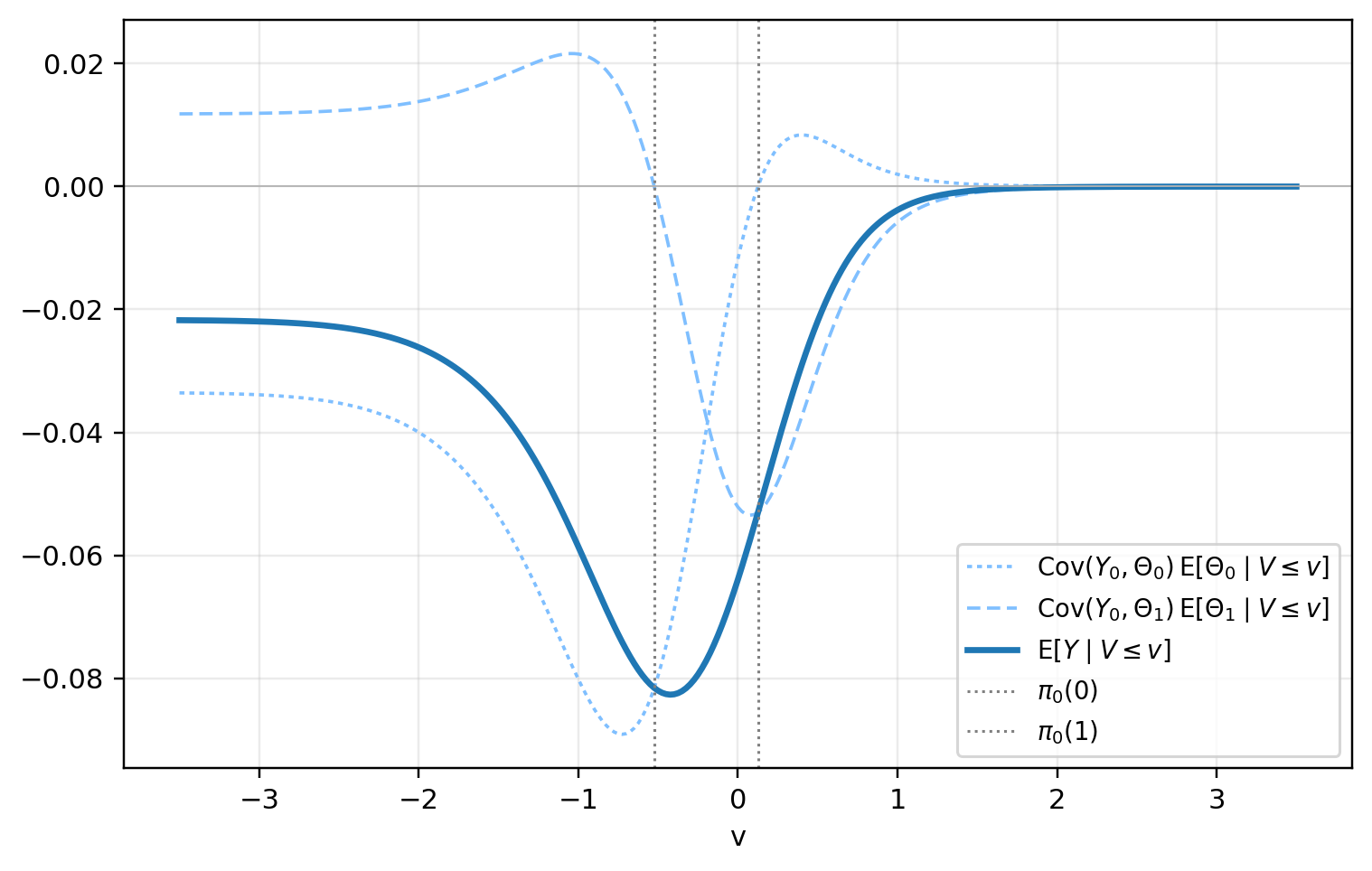}\hfill\includegraphics[width=0.49\textwidth]{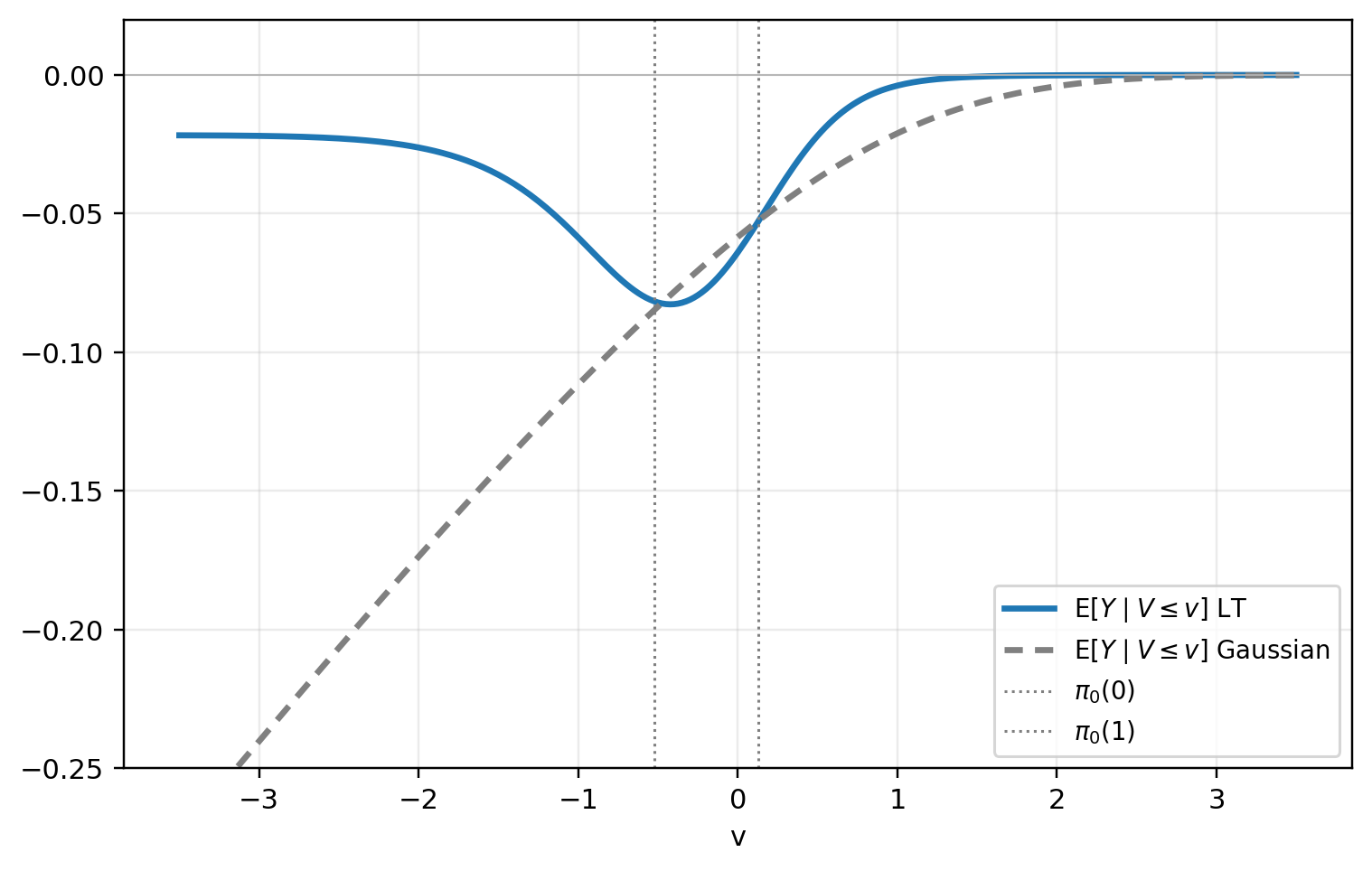}

\vspace{0.1cm}
(a)\hspace{6.5cm}(b)
\end{center}

\vspace{0.1cm}
\footnotesize{\textit{Notes:} Figure \ref{fig:ci}(b) in Section \ref{subsec:Assumptions} depicts the isoquants of the joint density of $(Y_0,V)$ that satisfies \ref{as:LT} and \eqref{eq:as-SR-w1} with $\pi_0(0)$ and $\pi_0(1)$ marked by vertical dotted lines (top) and the correlation function $\rho_{Y_0,V}(y,v)$ with the highlighted slices at $\pi_0(0)$ and $\pi_0(1)$ (bottom). The current figure is based on the same DGP. Panel (a): the control function $\Ep[Y_0\mid V\le v]$ and its decomposition \eqref{eq:cf2}. Panel (b): the control function under Gaussianity (i.e., inverse-Mills ratio) versus \ref{as:LT} (i.e., \ref{as:CLD}). All the figures are drawn in terms of $\Phi^{-1}$-transformed $V$ for readability.}
\end{figure}

When $(Y_0,V)$ follows a Gaussian copula with Gaussian marginal for $Y_0$, $\Ep[Y_0\mid V \le \pi_0(z)]$ is proportional to the inverse Mills ratio \citep{Heckman1979SampleSelection}. This function has a specific shape: it is strictly increasing (decreasing) with respect to the threshold $\pi_0(z)$ when the correlation coefficient is positive (negative), as depicted in Figure \ref{fig:rs-illustration}(b) below.\footnote{The control function $\Ep[Y_0\mid V \le \pi_0(z)]$ remains strictly monotone even when $(Y_0,V)$ follows a more general log-concave distribution \citep{heckman1990empirical}.} By contrast, \ref{as:CLD} allows for general non-monotone selection patterns in the control function, analogous to the non-monotonicity of the conditional distribution in Section \ref{subsec:selection}. 

To illustrate, we introduce a particular class of DGPs that is consistent with \ref{as:CLD}(i) and still yields an economically interpretable closed-form expression for the control function with potentially opposite selection patterns at the two threshold values. Recall that $\pi_0(z)$ ($z\in \{0,1\}$) are the two threshold values for which \ref{as:CLD}(i) holds with $d=0$. Let $U_0:=F_{Y_0}(Y_0)$ denote the sector-0 wage \emph{rank} assuming a continuous $Y_0$. We consider a model in which latent traits shift the distribution of wage rank $U_0$. For $z\in\{0,1\}$, let $C\bigl(u,\pi_0(z);r(u)\bigr)$ be the LGR of $(U_0,V)$ under \ref{as:CLD}(i). Define its deviation from independence at $\pi_0(z)$ as $\Delta_z(u):=C\bigl(u,\pi_0(z);r(u)\bigr)-u\,\pi_0(z) \in (-1/4,1/4)$.

\begin{myas}{LT}[Model of Latent Traits]\label{as:LT}
Let $\Theta=(\Theta_0,\Theta_1)$ be latent traits with
\begin{equation}
\Ep[\Theta_z]=0,\quad\Ep[\Theta_z^2]=1,\quad\Ep[\Theta_0\Theta_1]=0,\quad\textrm{and } U_0 \indep V \mid \Theta.
\label{eq:rs-traits0}
\end{equation} 
Suppose the traits tilt the conditional rank and propensity distributions according to
\begin{equation}
    F_{U_0\mid\Theta}(u\mid\theta)
    =
    u+\theta_0\delta_0(u)+\theta_1\delta_1(u),
    \qquad
    F_{V\mid\Theta}(v\mid\theta)
    =
    v+\theta_0\ell_0(v)+\theta_1\ell_1(v),
    \label{eq:rs-traits}
\end{equation}
where $\ell_z(v):=\kappa_z L_z(v)$ and
$\delta_z(u):=\Delta_z(u)/\kappa_z$ with nonzero constants $(\kappa_0,\kappa_1)$, and $L_0,L_1:[0,1]\to\mathbb{R}$ are differentiable functions with $L_z(0)=L_z(1)=0$,\footnote{\label{fn:copula} For the expressions in \eqref{eq:rs-traits} to be valid (nondecreasing) CDFs for every realization of $\Theta$, the support of $\Theta$ should be bounded, with $1+\theta_0\delta_0'(u)+\theta_1\delta_1'(u)\ge 0$ and $1+\theta_0\ell_0'(v)+\theta_1\ell_1'(v)\ge 0$ over that support. Subject to this restriction, $\Delta_z(u)$ and $L_z(v)$ can take both positive and negative values.}
\begin{equation}\label{eq:rs-cardinal}
    L_0(\pi_0(0))=1,\quad L_0(\pi_0(1))=0,\quad L_1(\pi_0(0))=0,\quad L_1(\pi_0(1))=1.
\end{equation} 
\end{myas}

In this assumption, \eqref{eq:rs-traits0} is motivated by the matching on unobservable factors in \citet[p. 5173]{abbring2007econometric}; see also \citet[pp. 5039--5040]{heckman2007chapter71}. Given $U_0\indep V \mid \Theta$, the latent traits $\Theta$ are the source of endogenous selection under \ref{as:LT}.

\ref{as:LT} implies a particular distribution for $(U_0,V)$:
\begin{equation}
    F_{U_0,V}(u,v)
    =
    \Ep\!\left[
        F_{U_0\mid\Theta}(u\mid\Theta)
        F_{V\mid\Theta}(v\mid\Theta)
    \right]
    =
    uv+L_0(v)\Delta_0(u)+L_1(v)\Delta_1(u),
    \label{eq:rs-traits-copula}
\end{equation}
where the second equality is by \eqref{eq:rs-traits0}. Note that \eqref{eq:rs-traits-copula} is a valid copula capturing an endogenous deviation from an independence benchmark $uv$, where the deviation is decomposed into two selection channels.\footnote{By \ref{as:LT}, the conditions in footnote \ref{fn:copula} imply the 2-increasingness $1+L_0'(v)\Delta_0'(u)+L_1'(v)\Delta_1'(u)\ge 0$ for the copula. This copula can be viewed as a \cite{sarmanov1966generalized} family, which is a generalization of the Farlie-Gumbel-Morgenstern family.} Each term $L_z(v)\Delta_z(u)$ is the selection channel generated by latent trait $\Theta_z$, which affects both the sector wage rank, through $\Delta_z$, and the sector-choice propensity, through $L_z$. Given \eqref{eq:rs-cardinal}, the value of $Z$  determines which trait is responsible for selection at each of the thresholds. We can show that \eqref{eq:rs-traits-copula} and thus \ref{as:LT} implies \eqref{eq:CLD} in \ref{as:CLD}:
\begin{theorem}[\ref{as:LT} $\Longrightarrow$ \ref{as:CLD} \eqref{eq:CLD}]\label{thm:rs-slices}
Under \ref{as:LT}, the LGR of $(Y_0,V)$ satisfies
\[
    \rho_{Y_0,V}(y,\pi_0(0))\;=\;\rho_{Y_0,V}(y,\pi_0(1))\;=\;r\bigl(F_{Y_0}(y)\bigr)
    \quad\text{for all }y\in\Y.
\]
\end{theorem}
An assumption analogous to \ref{as:LT} but with $d=1$ implies \eqref{eq:CLD(i)} in \ref{as:CLD}.

To interpret \ref{as:LT}, let sector $0$ be the technology sector, sector $1$
be the non-technology (e.g., retail) sector, $\Theta_0$ be technical
productivity (e.g., coding skills), $\Theta_1$ be organizational capital
(e.g., communication skills, match quality), and $Z$ be an indicator for low tech-sector entry costs so that $\pi_0(0) < \pi_0(1)$. The anchoring conditions
\eqref{eq:rs-cardinal} then imply that coding skills are the sole source of
selection into technology when entry is restricted ($\pi_0(0)$) and communication skills when entry is easier ($\pi_0(1)$). In model \eqref{eq:rs-traits}, if coding skills shift mass toward the upper tail of
the tech-sector wage distribution, then $\delta_0(u)<0$ in the upper tail, so
a worker with $\theta_0>0$ is shifted upward there. Similarly, if
communication skills are valuable for team-lead or client-facing tech
jobs, then $\delta_1(u)<0$ over the relevant upper-middle ranks, and workers
with high $\theta_1$ are shifted upward in that region. In terms of sector choice, the tech-sector workforce is positively
selected on coding skills where $\ell_0(v)>0$, and it is negatively
selected on communication skills where $\ell_1(v)<0$.\footnote{When viewed in terms of quantiles, the model \eqref{eq:rs-traits} relates to \cite{chesher2003identification} and \cite{matzkin2003nonparametric}, except that it involves latent explanatory variables.}

In addition, under the latent traits model \ref{as:LT}, the control function admits a closed form with a transparent economic structure. Since $Y_0=Q_{Y_0}(U_0)$ and $U_0\indep V\mid\Theta$, we have $Y_0\indep V\mid\Theta$, and \eqref{eq:rs-traits0} and $F_{V\mid \Theta}$ in \eqref{eq:rs-traits} yield
\begin{equation}
    \lambda_0(v)
    :=
    \Ep[Y_0\mid V\le v]
    =
    \mathrm{Cov}(Y_0,\Theta_0)\Ep[\Theta_0\mid V\le v]
    +
    \mathrm{Cov}(Y_0,\Theta_1)\Ep[\Theta_1\mid V\le v].
    \label{eq:cf2}
\end{equation}
The derivations of \eqref{eq:cf2} and \eqref{eq:slope-decomp} below are contained in SA \ref{subsec:roy_proof}. Each term in \eqref{eq:cf2} is the product of two interpretable factors. The selection factor $\Ep[\Theta_z\mid V\le v]=\ell_z(v)/v$ is the average prevalence of trait $z$ among tech-sector participants with $V\le v$, i.e., trait-specific selection. The payoff factor $\mathrm{Cov}(Y_0,\Theta_z)=\Ep[Y_0\Theta_z]$ is the covariance between the tech-sector wage and trait $z$, interpreted as the mean wage ``payoff'' to the trait. Hence \eqref{eq:cf2} describes the contributions of coding skills and communication skills to the mean tech-sector wage among selected workers, each contribution being the product of what the trait pays and how prevalent it is among participants.

As shown below, the two-channel sorting in \eqref{eq:cf2} generates a more general shape of selection patterns than a single sorting under Gaussianity. In particular, the sign of $\lambda_0'(v)$ can flip between $\pi_0(0)$ and $\pi_0(1)$. Based on \eqref{eq:cf2}, we show in SA \ref{subsec:roy_proof} that
\begin{align}
    \pi_0(z)\,\lambda_0'(\pi_0(z))
    &=
    \underbrace{\mathrm{Cov}(Y_0,\Theta_z)\bigl(\Ep[\Theta_z\mid V=\pi_0(z)]-\Ep[\Theta_z\mid V\le\pi_0(z)]\bigr)}_{=:\,c_{z,\mathrm{own}}}\nonumber \\
    &+
    \underbrace{\mathrm{Cov}(Y_0,\Theta_{1-z})\,\Ep[\Theta_{1-z}\mid V=\pi_0(z)]}_{=:\,c_{z,\mathrm{cross}}}.\label{eq:slope-decomp}
\end{align}

The own-channel term $c_{z,\mathrm{own}}$ is the payoff-weighted change in the prevalence of trait $z$ between marginal entrants and the incumbent workers. The cross-channel term $c_{z,\mathrm{cross}}$ is the payoff-weighted prevalence of the other trait among marginal entrants; it has no incumbent counterpart because the anchoring \eqref{eq:rs-cardinal} implies $\Ep[\Theta_{1-z}\mid V\le\pi_0(z)]=0$. The anchoring thus ties each threshold to a different selection channel in the example: at $\pi_0(0)$, incumbents are selected on coding skills alone, so the slope
combines the difference in coding-skill prevalence between marginal entrants
and incumbents ($c_{0,\mathrm{own}}$) with the communication-skill prevalence
of entrants ($c_{0,\mathrm{cross}}$); at $\pi_0(1)$, the roles of the two traits are reversed. Because the two thresholds are governed by different own/cross configurations, the slope of $\lambda_0$ can take opposite signs:
 
\begin{lemma}[Non-monotonicity of the control function]\label{lem:rs-non-monotone}
Under \ref{as:LT}, suppose
\begin{align}
c_{0,\mathrm{own}}+c_{0,\mathrm{cross}}<0,    \qquad c_{1,\mathrm{own}}+c_{1,\mathrm{cross}}>0.
    \label{eq:as-SR-w1}
\end{align}
Then $\lambda_0'(\pi_0(0))<0$ and $\lambda_0'(\pi_0(1))>0$. In particular, $v\mapsto\Ep[Y_0\mid V\le v]$ is non-monotone.\footnote{Let $m_0(v):=\Ep[Y_0\mid V= v]$. \eqref{eq:as-SR-w1} implies that $v\mapsto m_0(v)$ is also non-monotone: since $\lambda_0(v)$ is the average of $m_0$ over $[0,v]$, a monotone $m_0$ would give $\lambda_0'$ a constant sign.}
\end{lemma}

Figure \ref{fig:rs-illustration} is generated from a simulated DGP that satisfies \ref{as:LT} and \eqref{eq:as-SR-w1}; the same DGP is used for Figure \ref{fig:ci}(b) in Section \ref{subsec:Assumptions}. The control function in panel (a) is non-monotone, with $\lambda_0'(\pi_0(0))<0$ and $\lambda_0'(\pi_0(1))>0$ (Lemma \ref{lem:rs-non-monotone}). Panel (a) also plots its decomposition \eqref{eq:cf2} into the payoff-weighted channel contributions $\mathrm{Cov}(Y_0,\Theta_z)\Ep[\Theta_z\mid V\le v]$: the coding-skills contribution vanishes and switches sign at $\pi_0(1)$, and the communication-skills contribution vanishes and switches sign at $\pi_0(0)$. As entry costs fall from the high-cost margin $\pi_0(0)$ to the low-cost margin $\pi_0(1)$, the trait mixture of the tech-sector workforce shifts from the coding channel toward the communication channel, and the payoff-weighted composition of marginal entrants moves from less favorable than the incumbents' (the first condition in \eqref{eq:as-SR-w1}) to more favorable (the second condition), producing the slope reversal.

\ref{as:CLD} only constrains $\rho_{Y_0,V}(y,v)$ at the two propensity-score thresholds, leaving various selection channels flexible to redistribute mass across the propensity-score support. The shape of the control function is then driven by the varying composition of different selection-channel forces. In contrast, joint Gaussianity for $(Y_0,V)$ (panel (b)) allows for only a single channel of selection, resulting in a monotone shape of the control function. Note that, compared to Section \ref{subsec:selection}, the flexibility exhibited here survives aggregation over $y$: although $\rho_{Y_0}(y)$ is integrated out in computing the conditional mean, substantial flexibility remains in $\lambda_0$. This flexibility is what allows \ref{as:CLD} to rationalize a Roy model and sector choices in which adverse selection at one margin and positive selection at another margin can coexist. The \ref{as:CLD} Roy economy does not necessarily imply the conclusions on dispersion and skewness of sector wages in \citet[][Theorems 2 and 3]{heckman1990empirical} either. In this sense, economic implications under \ref{as:CLD} are more nuanced than under Gaussianity. Moreover, the analysis in Section \ref{subsec:Models-with-Binary} implies that the parameters in the sector wage equations (as well as $\rho_{Y_0}(y)$) can be identified even with a binary cost shifter.\footnote{Without any distributional assumptions or assumptions on the dependence structure, one would require IVs with large support to identify the wage parameters  \citep{eisenhauer2015generalized}.}

\begin{remark}[Counterexample: Gaussian Mixture]\label{rem:gaussian_mixture} The primitive model \ref{as:LT} consists of two latent components, which leads to a control function generated by a flexible mixture of latent skills in our running example. Although reminiscent of this mixture DGP, a DGP described by a bivariate Gaussian mixture of subpopulations with different skill sets (i.e., $\omega C(u,v;\rho_1) +(1-\omega)C(u,v;\rho_2)$ with $0< \omega<1$) does not satisfy \ref{as:CLD}. The formal statement and its proof are contained in SA \ref{subsec:gaussian_mixture}. \qed
\end{remark}

\subsection{Comparison to IVQR and RS}\label{ssec:RS}

When the outcome is continuous, the IV quantile regression (IVQR) model \citep{chernozhukov2005iv}
provides an alternative set of conditions under which $QSF_{\tau}$
is point identified. Here, we discuss how IVQR is related to our approach when $D$ is binary. 

The
IVQR model is based on the Skorohod representation, $Y_{d}=Q_{Y_{d}}(U_{d})$, $U_{d}\sim U(0,1)$, $d\in\{0,1\}$.
\citet{chernozhukov2005iv} consider a general selection mechanism,
$D=\delta(Z,V)$, where $V$ can be vector-valued and $Z$
satisfies $U_{d}\indep Z$ (which is implied by \ref{as:EX}).
The key assumption of the IVQR model is RS, $U_{1}\overset{d}{=}U_{0}\mid Z,V$.
RS weakens the classical RI assumption, which requires
$U_{1}=U_{0}$ a.s. 

The IVQR model yields the following conditional moment restriction \citep[Theorem 1]{chernozhukov2005iv},
\begin{align}
\tau=\Pr[Y_{1}\le Q_{Y_{1}}(\tau),D_{z}=1\mid Z=z]+\Pr[Y_{0}\le Q_{Y_{0}}(\tau),D_{z}=0\mid Z=z],~z\in\{0,1\}.\label{eq:testable_implication_CH05}
\end{align}
Under the  selection model \eqref{eq:gen_sel_binary} and  \ref{as:EX}, \eqref{eq:testable_implication_CH05} can be rewritten as
\begin{align*}
\Pr[Y_{1}\le Q_{Y_{1}}(\tau),V_{z}\le\pi_0(z) ]=\Pr[Y_{0}\le Q_{Y_{0}}(\tau),V_{z}\le\pi_0(z)],\quad z\in\{0,1\}.
\end{align*}
Using Lemma \ref{lem:LGR}, we can further rewrite it as 
\begin{align*}
C(\tau,\pi_0(z);\rho_{Y_{1},V_{z}}(Q_{Y_{1}}(\tau),\pi_0(z)))=C(\tau,\pi_0(z);\rho_{Y_{0},V_{z}}(Q_{Y_{0}}(\tau),\pi_0(z))),\quad z\in\{0,1\}.
\end{align*}
This shows that, interestingly, IVQR also relies on a version of constant local dependence: 
\begin{align}
\rho_{Y_{1},V_{z}}(Q_{Y_{1}}(\tau),\pi_0(z))=\rho_{Y_{0},V_{z}}(Q_{Y_{0}}(\tau),\pi_0(z)),\quad z\in\{0,1\}.\label{eq:CI_IVQR}
\end{align}
The IVQR model imposes restrictions across potential outcomes for each instrument level, whereas \ref{as:CLD}(i) imposes restrictions across instrument levels for each potential outcome.

This discussion shows that the IVQR model and our approach rely on non-nested and complementary constant local dependence assumptions to identify causal effects for the overall population using instruments. Relative to the IVQR model, the proposed approach has three main advantages. First, it does not rely on continuity of the outcome to achieve point identification, and it naturally accommodates discrete and mixed discrete-continuous outcomes. Second, it imposes fewer restrictions on the relationship across potential outcomes and thus effect heterogeneity and selection on gains, compared to RS (see SA \ref{app:Rank-Similarity-Restricts}). Finally, it does not require additional full rank and completeness conditions for identification.\footnote{The moment restriction \eqref{eq:testable_implication_CH05} is not sufficient for point identification. For point identification with a binary $D$, \citet{chernozhukov2005iv}  require the Jacobian of \eqref{eq:testable_implication_CH05} to be of full rank and continuous. \citet{vuong2017counterfactual} provide weaker conditions, requiring piecewise strict monotonicity of $y\mapsto (-1)^{d}(\Pr[Y\le y,D=d\mid Z=0]-\Pr[Y\le y,D=d\mid Z=1])$
for some $d$ (in addition to \ref{as:REL}). Both  conditions restrict how the distribution of $(Y,D) \mid Z=z$ changes with $z$. By contrast, under \ref{as:CLD}, we only require \ref{as:REL}.}

\subsection{Gaussian Relaxation and Extrapolation}\label{ssec:extrapolation}

Here, we provide an interpretation of \ref{as:CLD} as an approach to extrapolation. We discuss two related aspects. First,  we show how our approach can be viewed as a relaxation of Gaussianity  by comparing \ref{as:CLD} to assuming a Gaussian copula. Second, we shed light on how \ref{as:CLD} extrapolates from the compliers to the overall population under the LATE assumptions using a first-order Taylor approximation of the LGR. 

\vspace{8pt}
\noindent\textbf{Gaussian Relaxation.} To understand how our approach extrapolates from the observable potential outcome distributions $F_{Y_d|D_z}(y\mid d)=F_{Y|D,Z}(y\mid d,z)$ (under \ref{as:EX}) to $F_{Y_{d}}(y)$, we begin by illustrating how extrapolation works with a Gaussian copula and then show how \ref{as:CLD} relaxes Gaussianity while preserving point identification and extrapolation. We focus on the binary $D$ case and \ref{as:CLD}(i) for the purpose of illustration, and on the distribution of $Y_0$; the analysis for the distribution of $Y_1$ is symmetric.

If $(Y_d,V_z)$ have a Gaussian copula with dependence parameter $\rho_{Y_d,V_z}$, which can be different for different values of $(d,z)$, then\footnote{The notion of Gaussian copula here is weaker than in the treatment effects literature because we do not impose SRI, which restricts selection heterogeneity, and work with separate copula representations for $(Y_0,V_1)$ and $(Y_0,V_0)$.}
\begin{align*}
\Pr[Y\le y\mid D=0,Z=z]
= \frac{1}{\pi_0(z)} \int_0^{\pi_0(z)} \Phi\left(\frac{\Phi^{-1}(F_{Y_0}(y)) - \rho_{Y_0,V_z} \Phi^{-1}(\tilde v)}{\sqrt{1-\rho_{Y_0,V_z}^2}} \right)\mathrm{d} \tilde v, ~ z \in \{0,1\}.
\end{align*}
If $Y_0$ takes on $L\ge 2$ values and $\rho_{Y_0,V_z} = \rho_{Y_0}$, then this yields a system of $2\times (L-1)$ equations with $L$ unknowns ($L-1$ values of $F_{Y_0}(y)$ and $\rho_{Y_0}$). The system is exactly determined if $Y_0$ is binary ($L=2$) and overdetermined if $L>2$.

By contrast, under \ref{as:CLD}(i), we use a local version of the previous system,
$$
\Pr[Y\le y\mid D=0,Z=z]  = \frac{1}{\pi_0(z)} \int_0^{\pi_0(z)} \Phi\left(\frac{\Phi^{-1}(F_{Y_0}(y)) - \rho_{Y_0}(y) \Phi^{-1}(\tilde v)}{\sqrt{1-\rho_{Y_0}(y)^2}} \right)\mathrm{d} \tilde v, \quad z \in \{0,1\},
$$
which also has $2\times (L-1)$ equations, but has $2\times (L-1)$ unknowns and therefore it is always exactly determined.\footnote{Note that, without \ref{as:CLD}(i), there are $3\times (L-1)$  unknowns  ($L-1$ values of each $F_{Y_0}(y)$, $\rho_{Y_0,V_0}(y,\pi_0(0))$ and $\rho_{Y_0,V_1}(y,\pi_0(1))$) by Lemma \ref{lem:lgr2}, and therefore the system is underdetermined.} \ref{as:CLD} is local in that the Gaussian copula does not need to be the same at every value of $y$, and the system only needs to hold at the two treatment propensities induced by $Z$. In this sense, \ref{as:CLD} can be seen as a relaxation of Gaussianity. It shows how far we can deviate from Gaussianity while preserving point identification. 

\vspace{8pt}
\noindent\textbf{Extrapolation from Compliers to Overall Population.} \ref{as:CLD} identifies treatment effects for the overall population with a binary IV. This raises the question of how it extrapolates from the compliers to the overall population in the LATE framework of \citet{imbens1994identification} under SRI \citep{vytlacil2002independence}. There are several different approaches to extrapolation in the LATE/MTE framework. For example, \citet{brinch2017beyond} and \citet{kowalski2023reconciling} impose a linear model for $\Ep[Y_d \mid V=v]$ to extrapolate from the compliers to the overall population.\footnote{Other approaches are based on structural models \citep[e.g.,][]{heckman2003simple}, covariates \citep[e.g.,][]{angrist2013extrapolate}, RS \citep[e.g.,][]{wuthrich2020comparison}, and the smoothness of the marginal treatment effect function \citep[e.g.,][]{mogstad2018using,han2024sharp}.}  By contrast, as shown in Section \ref{sssec:SI}, \ref{as:CLD} imposes a flexible model for the entire conditional distribution, $F_{Y_{d}\mid V}(y\mid v)$, locally at $v \in \{\pi_d(0),\pi_d(1)\}$. 

To better understand extrapolation under \ref{as:CLD}, we relate $F_{Y_0}$ to the potential outcome distribution for the compliers, 
$F_{Y_{0}|D_1>D_0}$, using a first-order Taylor approximation of the Gaussian copula around $\rho=0$ (i.e., around exogeneity).\footnote{This analysis is inspired by \citet{chiburis2011practical,chiburis2012practical}, who use a Taylor expansion to relate the LATE to the ATE obtained from a bivariate Probit model.} The relationship between $F_{Y_1}$ and $F_{Y_{1}|D_1>D_0}$ can be characterized similarly.\footnote{A similar argument to below can be used to characterize the relationship between $F_{Y_0}(y)$ and $F_{Y_0\mid V_z}(y\mid \pi_0(z))$ under \ref{as:CLD}(iii).} Under \ref{as:CLD}(i) and using Lemma \ref{lem:LGR}, $F_{Y_{0}\mid D_1>D_0}(y)$ can be written as
\begin{align*}
F_{Y_{0}\mid D_1>D_0}(y)=\frac{C(F_{Y_0}(y),\pi_0(0);\rho_{Y_0}(y))-C(F_{Y_0}(y),\pi_0(1);\rho_{Y_0}(y))}{\pi_0(0)-\pi_0(1)}.
\end{align*}
Then, using a first-order Taylor approximation of the Gaussian copula around $\rho=0$, $C(u,v;\rho)\approx uv + \rho \phi(\Phi^{-1}(u))\phi(\Phi^{-1}(v))$, and rearranging, we get
\begin{equation}
\underbrace{F_{Y_0}(y)}_{\text{population CDF}}\approx \underbrace{F_{Y_{0}\mid D_1>D_0}(y)}_{\text{complier CDF}}+\underbrace{\rho_{Y_0}(y)\phi\left(\Phi^{-1}(F_{Y_0}(y)) \right)\frac{\phi(\Phi^{-1}(\pi_0(1)))-\phi(\Phi^{-1}(\pi_0(0)))}{\pi_0(0)-\pi_0(1)}}_{\text{correction term}}.\label{eq:approximation}
\end{equation}

Equation \eqref{eq:approximation} sheds light on how our approach extrapolates from the compliers to the overall population. Specifically, it starts with the complier distribution and adds a correction term, which has three components. The first component, $\rho_{Y_0}(y)$, captures the degree of endogeneity, which under \ref{as:CLD} is specific to the  point $y$. If $\rho_{Y_0}(y)=0$, then the correction term is zero, and the complier CDF coincides with the population CDF at $y$. The second component, $\phi(\Phi^{-1}(F_{Y_0}(y)))$, captures the impact of the point $y$ at which we are extrapolating. It is large around the median where $F_{Y_{0}}(y)=0.5$ and smaller in the tails. This reflects the fact that there is more room for variation in the CDF around the median than in the tails given that it is bounded between zero and one. The third component is equal to $\Ep[\Phi^{-1}(V)\mid D_1>D_0]-\Ep[\Phi^{-1}(V)]$ and thus captures the average position of the compliers in the distribution of $V$ relative to the overall population. It is equal to zero under perfect compliance ($\pi_0(0)=1$ and $\pi_0(1)=0$), consistent with the fact that $F_{Y_0}$ is point identified via $F_{Y_0 \mid D_1>D_0}$ in this case. The third component is also equal to zero if the first stage is symmetric, $\pi_0(0)=1-\pi_0(1)$. This provides an interesting connection to \citet{angrist2004treatment} who shows that the LATE equals the ATE under symmetry of the first stage (combined with symmetry of the latent error distribution); see also \citet[][Section 7]{heckman2000local}.

\section{Estimation and Inference\label{sec:Estimation-and-inference} }

The identification results in Section \ref{identification-analyses} are constructive and lead to tractable estimators. Here, we propose semiparametric estimators of $F_{Y_d}$ for the three types of treatments based on distribution regression (DR) and show how to construct estimators
of treatment effect parameters using the plug-in rule. We focus on target parameters that yield $\sqrt{n}$-consistent and asymptotically normal estimators. We do not consider parameters like  $\partial QSF_{\tau}(d)/\partial d$, because estimating such parameters would involve nonparametric methods, which require choosing tuning parameters and exhibit slower convergence rates.

In this section, we make the role of the covariates $X$ explicit. We assume we have access
to a random sample of size $n$ from $(Y,D,Z,X)$, $\left\{ (Y_{i},D_{i},Z_{i},X_{i})\right\} _{i=1}^{n}$. Let $B(X_{i})$, $B(Z_i,X_{i})$, and $B(D_{i},Z_{i},X_{i})$ denote
vectors of transformations of $X_{i}$, $(Z_i,X_{i})$, and $(D_{i},Z_{i},X_{i})$,
respectively. Define the indicators $I_{i}(y):=1\{Y_{i}\le y\}$ and $J_i(d) := 1\{D_i \leq d\}$. Let $\D_h$ and $\Y_h$ be two finite grids covering $\D$ and $\Y$ with spacing $h$.\footnote{We can set $\Y_h = \Y$ when $\Y$ is finite and $ \D_h=\D$ for binary and ordered $D$.}

\subsection{Binary Treatments}
We consider separate DR models for the conditional potential outcome
distributions, 
\begin{equation}
F_{Y_{d}|X}(y|x)=\Phi(B(x)'\beta_{d}(y)),\quad d\in\{0,1\},\label{eq:dr_model_binary}
\end{equation}
where $y \mapsto \beta_d(y)$ is a function-valued unknown parameter, and a Probit model for the propensity score, 
\begin{equation}
1-\pi_0(z,x) := \Pr[D=1 \mid Z=z,X=x] = \Phi(B(z,x)'\pi).\label{eq:probit_model_binary}
\end{equation}
We model the local dependence parameter at the $d$-treatment thresholds induced by $Z$  as 
\begin{equation}
\rho_{Y_{d}; X}(y;x)=\rho(B(x)'\gamma_{d}(y)),\quad d\in\{0,1\},\label{eq:correlation_model_binary}
\end{equation}
where $\rho(u)=\tanh(u)\in(-1,1)$, the inverse Fisher transformation, and $y \mapsto \gamma_d(y)$ is a function-valued unknown parameter.\footnote{To simplify the notation, we use the same vector of transformations in
models \eqref{eq:dr_model_binary} and \eqref{eq:correlation_model_binary}.
This is not essential, and one can use different specifications in
both models.} Together, \eqref{eq:dr_model_binary}, \eqref{eq:probit_model_binary},
and \eqref{eq:correlation_model_binary} imply the bivariate DR model
\begin{align*}
\Pr[Y \leq y, D = 1 \mid Z=z, X=x] & =\Phi_{2}(B(x)'\beta_{1}(y),B(z,x)'\pi;-\rho(B(x)'\gamma_{1}(y))),\\
\Pr[Y \leq y, D = 0 \mid Z=z, X=x] & =\Phi_{2}(B(x)'\beta_{0}(y),-B(z,x)'\pi;\rho(B(x)'\gamma_{0}(y))),
\end{align*}
where we have used the symmetry properties of the bivariate Gaussian
distribution to simplify the previous expressions.

We propose a computationally tractable two-step maximum likelihood estimator,
building on \citet{chernozhukov2025distribution}.

\begin{algorithm}[Estimation with Binary $D$]\label{algo:binary} We compute the estimator in two stages:
\begin{enumerate}[noitemsep]
    \item Treatment equation: estimate $\pi$ using a Probit regression 
\begin{eqnarray*}
\widehat{\pi}\in\arg\max_{c}\sum_{i=1}^{n}\left[D_{i}\log\Phi(B(Z_{i},X_{i})'c)+(1-D_{i})\log(1-\Phi(B(Z_i,X_{i})'c))\right].
\end{eqnarray*}
\item Outcome equation: for $y\in \Y_h$ and $d\in\{0,1\}$, $\widehat{F}_{Y_{d}|X}(y|x)=\Phi(B(x)'\widehat{\beta}_{d}(y))$ and $\widehat\rho_{Y_{d}; X}(y;x)=\rho(B(x)'\widehat \gamma_{d}(y))$, where
\begin{align*}
(\widehat \beta_1(y), \widehat \gamma_1(y)) \in & \arg\max_{b,g}\sum_{i=1}^{n}D_{i}\big[I_{i}(y)\log\Phi_{2}(B(X_{i})'b,B(Z_{i},X_{i})'\widehat{\pi};-\rho(B(X_{i})'g))\\
   & +(1-I_{i}(y))\log\Phi_{2}(-B(X_{i})'b,B(Z_{i},X_{i})'\widehat{\pi};\rho(B(X_{i})'g))\big],\\
(\widehat \beta_0(y), \widehat \gamma_0(y)) \in & \arg\max_{b,g}\sum_{i=1}^{n}(1-D_{i})\big[I_{i}(y)\log\Phi_{2}(B(X_{i})'b,-B(Z_i,X_{i})'\widehat{\pi};\rho(B(X_{i})'g))\\
   & +(1-I_{i}(y))\log\Phi_{2}(-B(X_{i})'b,-B(Z_i,X_{i})'\widehat{\pi};-\rho(B(X_{i})'g))\big].
\end{align*}
Rearrange the estimates $y \mapsto \widehat F_{Y_d \mid X}(y \mid x)$ on $\Y_h$ if needed. 
\end{enumerate}
\end{algorithm}
\begin{remark}[Computation] The first stage of Algorithm \ref{algo:binary} is a conventional Probit regression, and estimation can proceed using existing software. The second stage is computationally more expensive since it involves a nonlinear smooth optimization problem. This optimization problem can be solved using standard algorithms such as Newton-Raphson. To improve efficiency, a one-step correction can be added to Algorithm \ref{algo:binary}  \citep[see, e.g., Section 3.4 in][]{newey1994handbook}. \qed
\end{remark}
 
\subsection{Ordered Discrete Treatments}
As for binary treatments, we model all the components using flexible generalized linear and DR models: $F_{Y_d \mid X}(y \mid x) = \Phi(B(x)'\beta_d(y))$, $\pi_d(z,x) = F_{D \mid Z,X}(d \mid z,x) = \Phi(B(z,x)'\pi(d))$, and  $\rho_{Y_d ; X}(y ; x) = \rho(B(x)'\gamma_d(y))$ with $\rho(u) = \tanh(u)$.

\begin{algorithm}[Estimation with Ordered $D$] \label{algo:ordered}We compute the estimator in two stages:
\begin{enumerate}[noitemsep]
    \item Treatment equation: set $\widehat \pi_{-1}(z,x) = 0$ and $\widehat \pi_K(z,x) = 1$ for all $(z,x)$. For $d \in \{0,\ldots,K-1\}$, $\widehat \pi_d(z,x) = \Phi(B(z,x)'\widehat \pi(d))$, where
    $$
    \widehat \pi(d) \in \arg \max_{p} \sum_{i=1}^n \left[J_i(d) \log \Phi(B(Z_i,X_i)'p) + (1-J_i(d)) \log \Phi(-B(Z_i,X_i)'p)\right].
    $$
    Rearrange the estimates $d \mapsto \widehat \pi_d(z,x)$ on $\mathcal{D}$ and trim and renormalize if needed to ensure $\widehat\pi_d(z,x)>\widehat\pi_{d-1}(z,x)$. This rearrangement is important to avoid having logarithms of nonpositive numbers in the second stage.\footnote{In the second stage, $g_{d,i}(b,g) > 0$ and $\bar g_{d,i}(b,g) > 0$ a.s. if $\widehat \pi_d(Z_i,X_i) > \widehat \pi_{d-1}(Z_i,X_i)$ a.s.} 
    \item Outcome equation: for $y \in \Y_h$ and $d \in \D$, let $\widehat F_{Y_d \mid X}(y \mid x) = \Phi(B(x)'\widehat \beta_d(y))$ and $\widehat\rho_{Y_{d}; X}(y;x)=\rho(B(x)'\widehat \gamma_{d}(y))$, where
    \begin{align*}
    (\widehat \beta_d(y), \widehat \gamma_d(y)) &\in \arg \max_{b,g} \sum_{i=1}^n 1\{ D_i =d \} \left[ I_i(y) \log g_{d,i}(b,g) + (1-I_i(y)) \log \bar g_{d,i}(b,g) \right],
    \end{align*}
with $g_{0,i}(b,g) :=\Phi_2(B(X_i)'b, \Phi^{-1}\left(\widehat \pi_0(Z_i,X_i)\right); \rho(B(X_i)'g))$, for $d\in \mathcal{D}\setminus \{0,K\}$, $g_{d,i}(b,g) := \Phi_2(B(X_i)'b, \Phi^{-1}\left(\widehat \pi_d(Z_i,X_i)\right); \rho(B(X_i)'g)) - \Phi_2(B(X_i)'b,\Phi^{-1}\left(\widehat \pi_{d-1}(Z_i,X_i)\right); \rho(B(X_i)'g))$, $g_{K,i}(b,g):= \Phi(B(X_i)'b) - \Phi_2(B(X_i)'b,\Phi^{-1}\left(\widehat \pi_{K-1}(Z_i,X_i)\right); \rho(B(X_i)'g))$ and $\bar g_{d,i}(b,g) := \widehat \pi_d(Z_i,X_i) - \widehat \pi_{d-1}(Z_i,X_i) - g_{d,i}(b,g)$.
    Rearrange the estimates $y \mapsto \widehat F_{Y_d \mid X}(y \mid x)$ on $\Y_h$ if needed. 
\end{enumerate}
    
\end{algorithm}

\begin{remark}[Computation] The first stage is a sequence of Probit regressions that can be solved using standard software. The second stage is a nonlinear smooth optimization problem that can be solved using standard algorithms such as Newton-Raphson. As in Algorithm \ref{algo:binary}, a one-step correction can be added to Algorithm \ref{algo:ordered} to improve efficiency. \qed
\end{remark}

\subsection{Continuous Treatments}
\label{subsec:Estimation-Continuous-Treatment}
We construct plug-in estimators based on the closed-form solutions
in Section \ref{subsec:Models-with-Continuous}. We consider DR
models for $F_{Y \mid D,Z,X}$ and $F_{D \mid Z,X}$, $F_{Y \mid D,Z,X}(y \mid d,z,x)  = \Phi(B(d,z,x)'\beta(y))$ and   $F_{D \mid Z,X}(d \mid z,x)  = \Phi(B(z,x)'\pi(d))$, respectively, where $y \mapsto \beta(y)$ and $d \mapsto \pi(d)$ are unknown function-valued parameters.

\begin{algorithm}[Estimation with Continuous $D$]\label{algo:continuous} We compute the estimator in two stages:
\begin{enumerate}[noitemsep]
    \item Observable conditional distributions: for $y\in \Y_h$ and $d\in\D_h$, $
\widehat{F}_{Y|D,Z,X}(y|d,z,x)  =  \Phi(B(d,z,x)'\widehat{\beta}(y))$ and $
\widehat{F}_{D|Z,X}(d|z,x) =  \Phi(B(z,x)'\widehat{\pi}(d))$,
where 
\begin{align*}
\widehat{\beta}(y) & \in \arg\max_{b}\sum_{i=1}^{n}\left[I_{i}(y)\log\Phi(B(D_{i},Z_{i},X_{i})'b)+(1-I_{i}(y))\log(1-\Phi(B(D_{i},Z_{i},X_{i})'b))\right],\\
\widehat{\pi}(d) & \in \arg\max_{p}\sum_{i=1}^{n}\left[J_{i}(d)\log\Phi(B(Z_{i},X_{i})'p)+(1-J_{i}(d))\log(1-\Phi(B(Z_{i},X_{i})'p))\right].
\end{align*}
\item Potential outcome distributions: for $y\in \Y_h$ and $d\in\D_h$, $\widehat{F}_{Y_{d}|X}(y|x)=\Phi\Big(\widehat{a}_{d,y;x}/\sqrt{1+\widehat{b}_{d,y;x}^{2}}\Big)$ and $\widehat\rho_{Y_{d}; X}(y;x)=-\widehat{b}_{d,y;x}/\sqrt{1+\widehat{b}_{d,y;x}^{2}}$,
where
\begin{align*}
\widehat{a}_{d,y;x} & = \frac{(B(d,0,x)'\widehat{\beta}(y))(B(1,x)'\widehat{\pi}(d))-(B(d,1,x)'\widehat{\beta}(y))(B(0,x)'\widehat{\pi}(d))}{B(1,x)'\widehat{\pi}(d)-B(0,x)'\widehat{\pi}(d)},\\
\widehat{b}_{d,y;x} & = \frac{B(d,1,x)'\widehat{\beta}(y)-B(d,0,x)'\widehat{\beta}(y)}{B(1,x)'\widehat{\pi}(d)-B(0,x)'\widehat{\pi}(d)}.
\end{align*}
Rearrange the estimates $y \mapsto \widehat F_{Y_d \mid X}(y \mid x)$ on $\Y_h$ if needed.
\end{enumerate}
\end{algorithm}

\subsection{Marginal Distributions and Functionals}
\label{subsec:Estimation-Functionals}
With covariates, the marginal distributions of the potential outcomes are identified as
$$
F_{Y_d}(y) = \int F_{Y_d \mid X}(y \mid x) \mathrm{d} F_X(x),  \quad d \in \D,
$$
where $F_X$ is the distribution of $X$. We can use this expression to construct estimators of $F_{Y_d}$ and functionals of interest, such as the QSF and QTE, by plugging in the estimators obtained above. To provide a unified approach to all the treatment cases, we set $ \D_h = \D$ when $D$ is binary or discrete ordered. 

\begin{algorithm}[Estimation of $F_{Y_d}$, QSF and QTE]\label{algo:target}
 Estimation proceeds in two steps.
    \begin{enumerate}[noitemsep]
        \item Unconditional distribution: for $y\in \Y_h$ and $d\in\D_h$, 
        $
        \widehat{F}_{Y_d}(y)=\frac{1}{n}\sum_{i=1}^n\widehat{F}_{Y_{d} \mid X}(y \mid X_i).
        $
        Values $y$ between grid points can be obtained by step or linear interpolation, with extension outside the grid satisfying the CDF boundary constraints.
        
        %For $y\in \Y \setminus \Y_h$ and $d\in\D_h$,
        %$
        %\widehat{F}_{Y_d}(y) = \max \{\widehat{F}_{Y_d}(\bar y) : \bar y < y, \bar y \in  \Y_h\}. \footnote{In practice, one can also use linear interpolation when $Y$ is continuous.}
        %$
        \item Unconditional QSF and QTE: $\widehat{QSF}_{\tau}(d) =\mathcal{Q}_{\tau}(\widehat{F}_{Y_{d}})$ and  $\widehat{QTE}_{\tau}(d,d') = \{\widehat{QSF}_{\tau}(d)-\widehat{QSF}_{\tau}(d')\}/(d-d')$.

    \end{enumerate}
\end{algorithm}

\begin{remark}[Estimating ASF and ATE]
Based on Step 1 of Algorithm \ref{algo:target}, the ASF and ATE can be estimated using the following plug-in estimators,
$\widehat{ASF}(d)=\int y\mathrm{d}\widehat{F}_{Y_d}(y),$
and $\widehat{ATE}(d,d') = \{\widehat{ASF}(d)-\widehat{ASF}(d')\}/(d-d')$. Implementing these estimators is straightforward when the outcomes and treatments are discrete. With continuous outcomes or treatments, additional computational challenges arise because we approximate $\Y$ and $\D$ using grids $ \Y_h$ and $\D_h$. The resulting estimators of the ASF and ATE can be sensitive to the choice of grid points in the tails, especially when the outcome and treatment are unbounded and the sample size is small or moderate.\footnote{One can also construct plug-in estimators based on other representations of the ASF, such as $ASF(d)=\int_0^1 QSF_\tau(d)\mathrm{d}\tau$ 
or $ASF(d)=\int_0^\infty (1-F_{Y_d}(y))\mathrm{d}y-\int_{-\infty}^0 F_{Y_d}(y)\mathrm{d}y$.}
In such settings, we recommend focusing on estimands that admit more robust estimators, such as the QSF and QTE for $\tau\in\mathcal{T}\subset (c,1-c)$, for $c>0$, as illustrated in our empirical application. \qed
\end{remark}

\subsection{Asymptotic Theory and Inference} 
The target parameters in Section \ref{subsec:Estimation-Functionals} are function-valued. Inference on these parameters can be performed using resampling methods. To provide a unified framework, we denote the functional parameters by $u\mapsto \delta_u, u\in \mathcal{U}$, where $\mathcal{U}\subset \tilde{\mathcal{Y}}\times \tilde{\mathcal{D}}\times \mathcal{T}$, with  $\mathcal{T}\subset (c,1-c)$ for $c>0$, $\tilde{\mathcal{Y}}$  is a compact subset of $\mathcal{Y}$ when $\mathcal{Y}$ is uncountable and equal to  $\mathcal{Y}$ otherwise, and  $\tilde{\mathcal{D}}$ is defined analogously. For example, if we are interested in $\tau \mapsto QSF_{\tau}(d)$ on $[.05,.95]$, then $u=\tau$, $\delta_u = QSF_{u}(d)$ and $\mathcal{U} = [.05,.95]$ for fixed $d$. In practice, we approximate $\mathcal{U}$ using a fine grid $\mathcal{U}_h$. We denote the estimator of $\delta_u$ obtained from Algorithms \ref{algo:binary}--\ref{algo:target} as $\widehat\delta_u$.

We show in SA \ref{app:asymptotic_theory} that $\sqrt{n}(\widehat\delta_u-\delta_u)$ converges in distribution to a mean-zero Gaussian process $Z_\delta$ and the bootstrap is valid for a wide range of parameters of interest, including $F_{Y_d}$ and its (Hadamard differentiable) functionals. These results imply that the bootstrap algorithm below is theoretically valid. When the outcomes are discrete or mixed discrete-continuous, the estimators of the QSF or QTE are not asymptotically Gaussian in general. In this case, one can use the inference methods proposed by \citet{chernozhukov2020generic}.

We focus on constructing pointwise confidence intervals (CIs) and uniform confidence bands (CBs), $CI_{(1-\alpha)}(\delta_{\bar u})$ and $CB_{(1-\alpha)}(\delta_u)$ respectively, satisfying $\lim_{n\rightarrow \infty }\Pr[\delta_{\bar u}\in CI_{(1-\alpha)}(\delta_{\bar u})] = 1-\alpha$ and $\lim_{n\rightarrow \infty }\Pr[\delta_u\in CB_{(1-\alpha)}(\delta_u) \text{ for all }u\in \mathcal{U}] = 1-\alpha$.
Uniform CBs can be used to test a variety of hypotheses of interest, such as the hypotheses of no effect or constant effects when applied to QTE, or stochastic dominance.  

The following algorithm provides a generic bootstrap construction of $CB_{(1-\alpha)}(\delta_u)$. A similar algorithm can be used for $CI_{(1-\alpha)}(\delta_{\bar u})$, which we omit.
\begin{algorithm}[Uniform Confidence Bands for Functional Parameters\footnote{See, for example, \citet{chernozhukov2013inference,chernozhukov2020generic,chernozhukov2025distribution} for similar algorithms.}]\label{algo:bootstrap}\text{ }
    \begin{enumerate}[noitemsep]
        \item For $u\in \mathcal{U}_h$, obtain $B$ bootstrap draws of the estimator $\widehat\delta_u$, $\{\widehat\delta^{(b)}_u:1\le b\le B\}$. \label{step:bootstrap_draws}
        \item For $u\in \mathcal{U}_h$, compute the robust standard error,
        $
        SE(\widehat\delta_u)=(\widehat{Q}_{0.75}(\widehat\delta_u)-\widehat{Q}_{0.25}(\widehat\delta_u))/(\Phi^{-1}(0.75)-\Phi^{-1}(0.25)),
        $
        where $\widehat{Q}_{\tau}(\widehat\delta_u)$ is the $\tau$-quantile of $\{\widehat\delta^{(b)}_u:1\le b\le B\}$.
        \item Compute $cv(1-\alpha)$ as the $(1-\alpha)$-quantile of 
$\left\{\max_{u\in \mathcal{U}_h} SE(\widehat\delta_u)^{-1}
|\widehat\delta^{(b)}_u-\widehat\delta_u|:1\le b\le B\right\}$.
        \item Compute the $(1-\alpha)$ uniform CB as $CB_{(1-\alpha)}(\delta_u)=[\widehat\delta_u\pm cv(1-\alpha)SE(\widehat\delta_u)]$, $u\in \mathcal{U}_h$.

    \end{enumerate}
\end{algorithm}

The algorithms for binary and ordered treatments (Algorithms \ref{algo:binary} and \ref{algo:ordered}) involve nonlinear optimization problems in the second step. Therefore, we recommend using the multiplier bootstrap in Step \ref{step:bootstrap_draws} of Algorithm \ref{algo:bootstrap}, which avoids re-estimating parameters in each of the $B$ bootstrap iterations. The algorithm for continuous treatments (Algorithm \ref{algo:continuous}) does not involve a nonlinear optimization problem in the second step and is computationally less expensive than Algorithms \ref{algo:binary} and \ref{algo:ordered}. Therefore, the standard empirical bootstrap is a natural alternative to the multiplier bootstrap, provided that the sample size is not too large, as for example in Section \ref{sec:Applications}.

\begin{remark}[Additional theoretical complications when estimating the ASF and ATE] Additional theoretical complications arise when estimating the ASF and ATE with unbounded $\Y$ and $\D$. Standard functional central limit theorems for DR estimators \citep[e.g.,][]{chernozhukov2013inference} guarantee convergence only over a compact subset of the support. One approach to establish functional central limit theorems over the entire support is to impose restrictions on the tails of the distribution \citep[e.g.,][]{chernozhukov2025crrr}. \qed
\end{remark}

\begin{remark}[Combining different identification approaches for efficiency] The proposed method can be combined with alternative identification approaches to obtain more efficient estimators. If the alternative approaches identify the same estimands and yield nonredundant moment conditions, then such moment conditions can be combined with the first-order conditions from our estimators in a GMM framework. The choice of whether to combine different approaches is subject to a trade-off between robustness (i.e., relying on fewer assumptions) and obtaining more efficient estimators. We recommend using overidentification tests whenever multiple approaches are combined. \qed
\end{remark}

\section{Empirical Illustration\label{sec:Applications}}
We illustrate our methods by estimating the distributional effects of sleep on well-being.\footnote{The empirical results were obtained using \texttt{Stata} \citep{statacorp2025} and \texttt{R} \citep{R26}.} We reanalyze the data from \citet{bessone2021economic}, who studied the effects of randomized interventions to increase sleep time of low-income adults in India.\footnote{We downloaded the data from the Harvard Dataverse replication package \citep{bessone2021replication}.} According to an expert survey by \citet{bessone2021economic}, increasing sleep could have large benefits in this empirical context, which is characterized by low baseline levels of sleep per night and sleep efficiency.

\citet{bessone2021economic} considered two main treatments (see their Section III for details): (i) \emph{devices $+$ encouragement} (information, encouragement and sleep trackers, various devices for improving sleep environment) and (ii) \emph{devices $+$ incentives} (same as (i) and payments for each minute of sleep increase). In addition, they cross-randomized a \emph{nap treatment} (the opportunity to nap at work).

The outcome of interest ($Y$) is an overall index of individual well-being. The  treatment ($D$) is sleep time per night (in hours).  We use an indicator for whether an individual received either of the main treatments as an instrument ($Z$) for sleep. The vector of covariates ($X$) includes controls for gender, three age indicators, and the baseline well-being index, as in \citet[][Table A.XVII]{bessone2021economic}. Following \citet{dong2023nonparametric}, we restrict the sample to individuals who did not receive the nap treatment, so that the total sample size is $n=226$.

The treatment  $D$ takes on many values in this application and is treated as continuous. We therefore use the estimators for continuous treatments described in Algorithm \ref{algo:continuous}. 
We choose fully saturated (in $Z_i$) specifications for $B(d,z,x)$ and $B(z,x)$. The resulting DR models are
$F_{Y \mid D,Z,X}(y \mid d,z,x)  = \Phi((d,x')\beta_z(y))$ and $F_{D \mid Z,X}(d \mid z,x)  = \Phi(x'\pi_z(d))$ for $z\in \{0,1\}.$
Our flexible estimators allow us to analyze the full distributional impact of sleep on well-being. Our results thus complement the analyses of the average effect of sleep on well-being in \citet{bessone2021economic} and \citet{dong2023nonparametric}.

\begin{figure}[h!]
\caption{Quantile Treatment Effects and Local Dependence}
\label{fig:fs_qte_rho}
\begin{center}

	\begin{center}

\subfigure[Distributional first stage]{%
  \includegraphics[width=0.32\textwidth,trim=0cm 0.6cm 0cm 3.7cm]{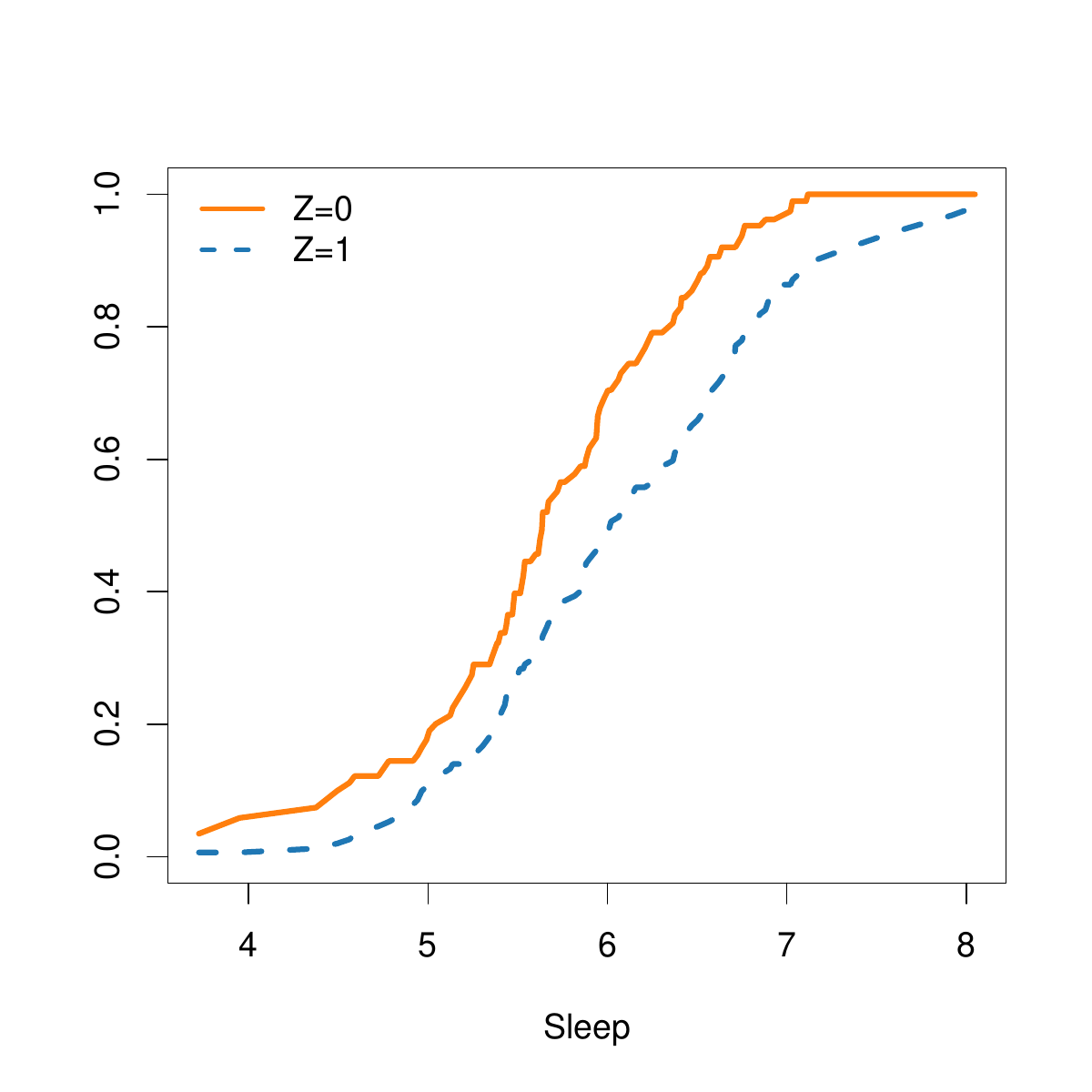}
}
\subfigure[Quantile treatment effect]{%
  \includegraphics[width=0.32\textwidth,trim=0cm 0.6cm 0cm 3.7cm]{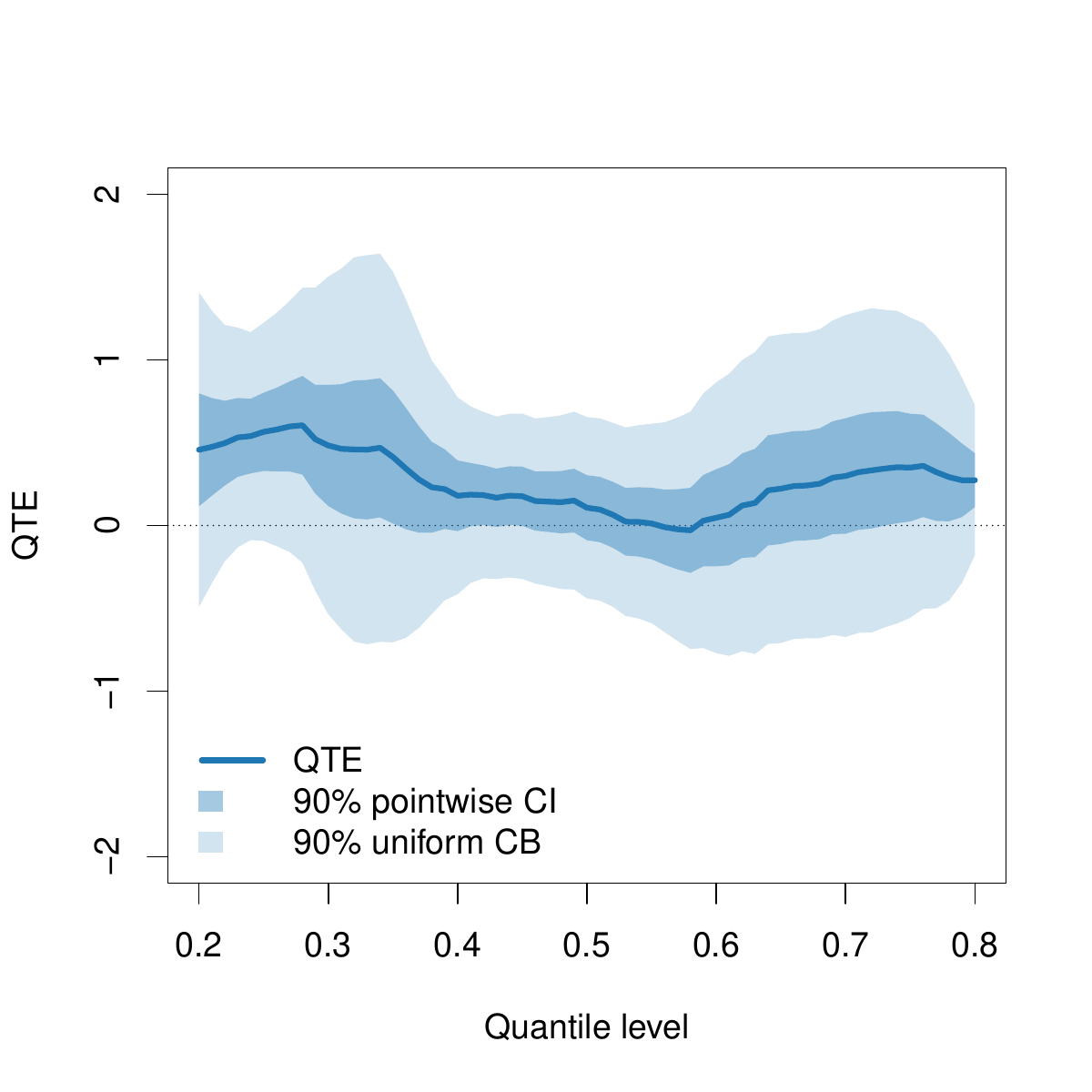}
}\hfill
\subfigure[Local dependence]{%
  \includegraphics[width=0.32\textwidth,trim=0cm 0.6cm 0cm 3.7cm]{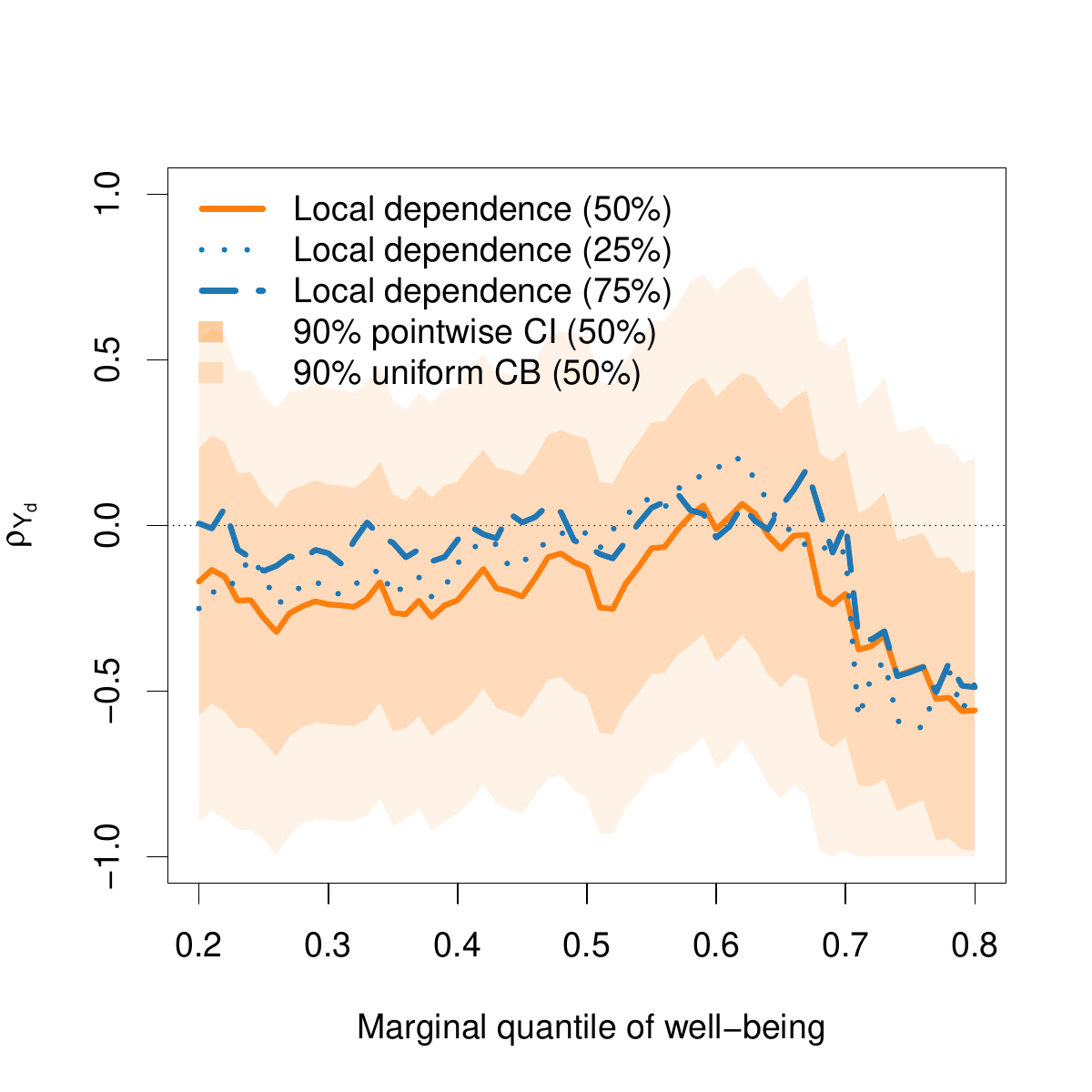}
}
\end{center}

\end{center}
\footnotesize{\textit{Notes:} Pointwise CIs and uniform CB for the QTE and local dependence are computed using the empirical bootstrap with 5,000 repetitions. All specifications control for gender, three age indicators, and the baseline well-being index.}
\end{figure}

We begin by analyzing the (distributional) first-stage relationship between $D$ and $Z$ to shed light on the plausibility of \ref{as:REL}. Figure \ref{fig:fs_qte_rho}(a) plots $\widehat{F}_{D \mid Z}(\cdot \mid 1) = n^{-1}\sum_{i=1}^n\widehat{F}_{D\mid Z,X}(\cdot \mid 1,X_i)$ and  $\widehat{F}_{D \mid Z}(\cdot \mid 0) =  n^{-1}\sum_{i=1}^n\widehat{F}_{D\mid Z,X}(\cdot \mid 0,X_i)$. It shows that the instrument shifts the distribution of $D$: sleep under the experimental treatments first-order stochastically dominates sleep without treatments.

In addition to \ref{as:REL}, our method relies on \ref{as:EX} and \ref{as:CLD}. The experimental random assignment renders the independence assumptions in \ref{as:EX} plausible.\footnote{Note that random assignment does not automatically imply the (implicit) exclusion restriction, which requires that the instrument has no direct effect on well-being.} \ref{as:CLD} allows the local dependence between potential well-being and unobserved sleep propensity (i.e., the selection sorting) to depend on the level of well-being (e.g., underlying health), while restricting it to be constant in sleep propensity (e.g., biological sleep factors) at and locally around the $d$-treatment thresholds. A sufficient condition for \ref{as:CLD} to hold is that selection sorting depends on underlying health but does not change with biological sleep factors.  \ref{as:CLD} would be violated when selection sorting changes with sleep propensity level in an unrestricted fashion. For example, sleep propensity may be negatively related to underlying health for those with high propensity, while positively related for those with low propensity.

Figure \ref{fig:fs_qte_rho}(b) plots estimates of the QTE, $\{\widehat{QTE}_{\tau}(d,d'): \tau \in [.2,.8]\}$, including 90\% pointwise CIs and uniform CBs computed using the empirical bootstrap. We set $d=\widehat{Q}_{D}(0.75)$ and $d'=\widehat{Q}_{D}(0.25)$, where $\widehat{Q}_{D}(\tau)$ is the $\tau$-quantile of the empirical distribution of sleep. Figure \ref{fig:fs_qte_rho}(b) suggests interesting effect heterogeneity across the distribution of well-being. The QTEs are the largest and pointwise significant in the tails and smaller and insignificant around the median.
While the uniform CB includes zero at all quantiles, it allows us to rule out large negative effects at the lower tail.

An interesting feature of our method is that it allows estimation of the local dependence parameter $\rho_{Y_d;X}(y;x)$, which can be interpreted as a local measure of endogeneity or self-selection. Figure \ref{fig:fs_qte_rho}(c) plots the average local dependence parameter, 
$
\tau \mapsto \frac{1}{n}\sum_{i=1}^n\widehat\rho_{Y_d;X}(\widehat{Q}_{Y}(\tau);X_i),$ for $(d,\tau)\in \left\{\widehat{Q}_{D}(0.25),\widehat{Q}_{D}(0.50),\widehat{Q}_{D}(0.75)\right\}\times [.2,.8],
$
where $\widehat{Q}_{Y}(\tau)$ is the $\tau$-quantile of the empirical distribution of well-being. The average local dependence is mostly negative, and there is interesting heterogeneity both across well-being and sleep levels. These predominantly negative point estimates suggest negative local dependence between the unobserved sleep propensity and potential well-being. This pattern is consistent with individuals with relatively poor underlying health conditions sleeping more. Figure \ref{fig:fs_qte_rho}(c) also shows that, at $d=\widehat{Q}_{D}(0.50)$, average local dependence is pointwise significant at the upper tail, but the uniform CB includes zero at all quantiles.

\begin{figure}[h!]
\caption{Comparison to Other Methods}
	\label{fig:comparison}
	\begin{center}
\subfigure[QTE under exogeneity]{%
  \includegraphics[width=0.32\textwidth,trim=0cm 0.6cm 0cm 3.7cm]{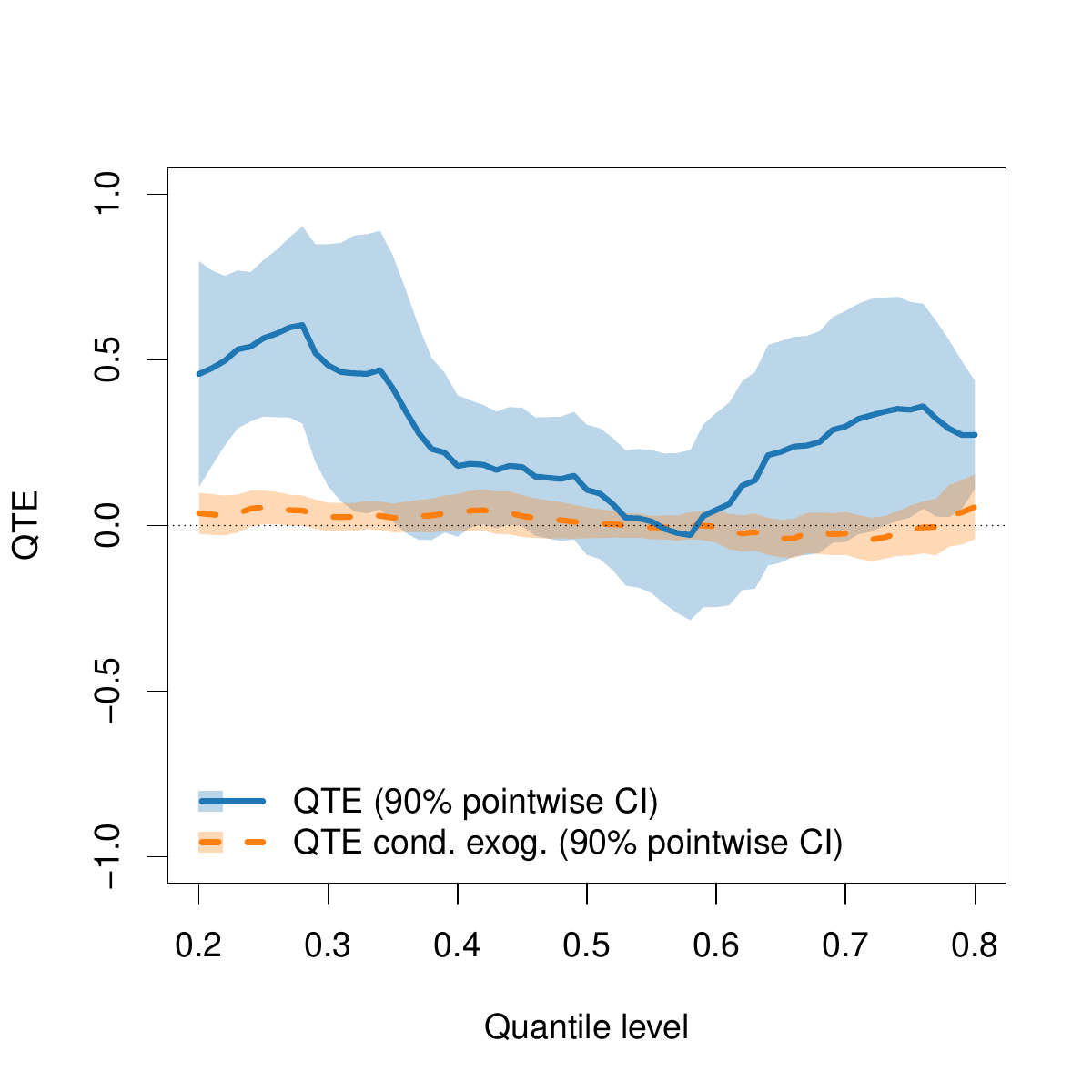}
}
\subfigure[2SLS]{%
  \includegraphics[width=0.32\textwidth,trim=0cm 0.6cm 0cm 3.7cm]{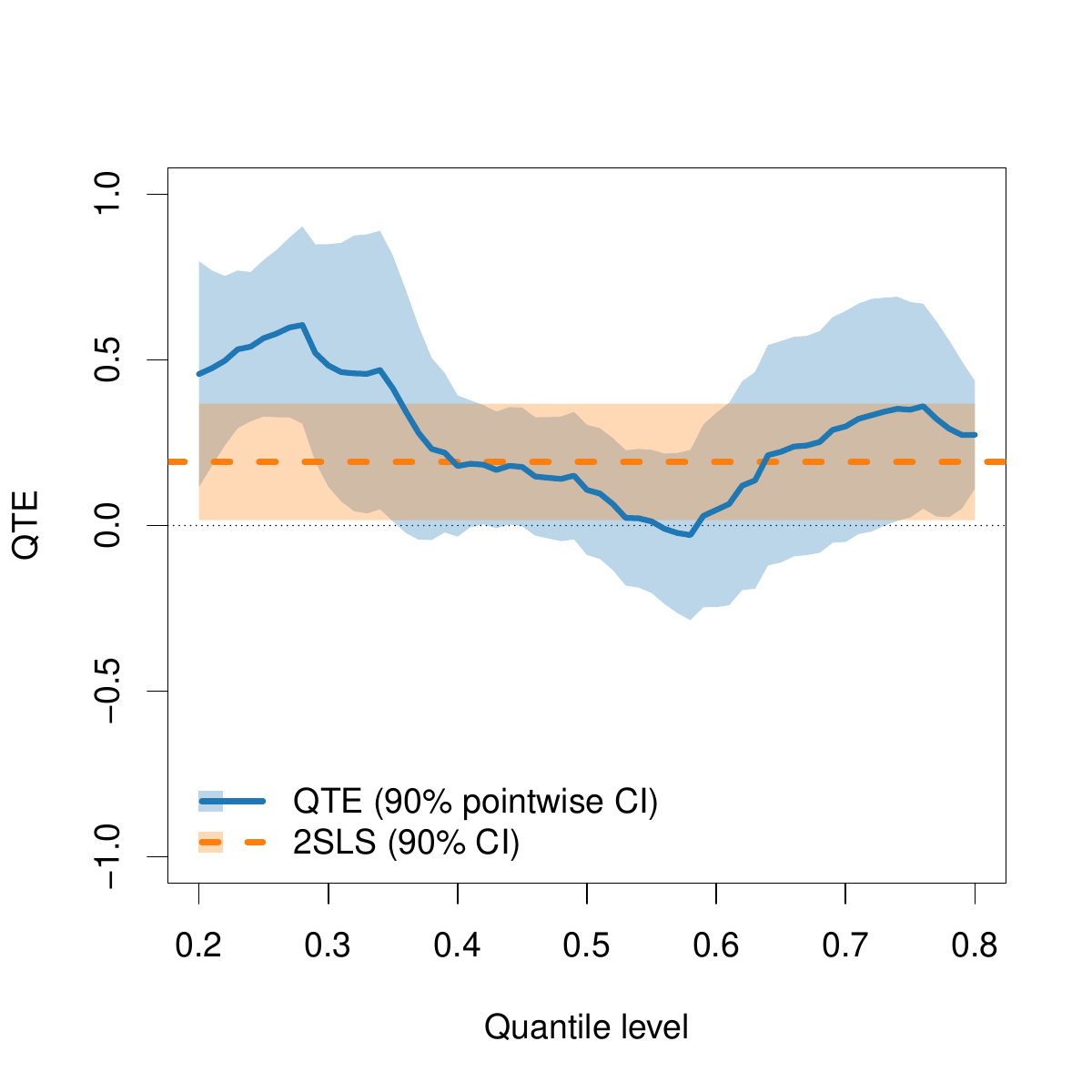}
}
\subfigure[IVQR]{%
  \includegraphics[width=0.32\textwidth,trim=0cm 0.6cm 0cm 3.7cm]{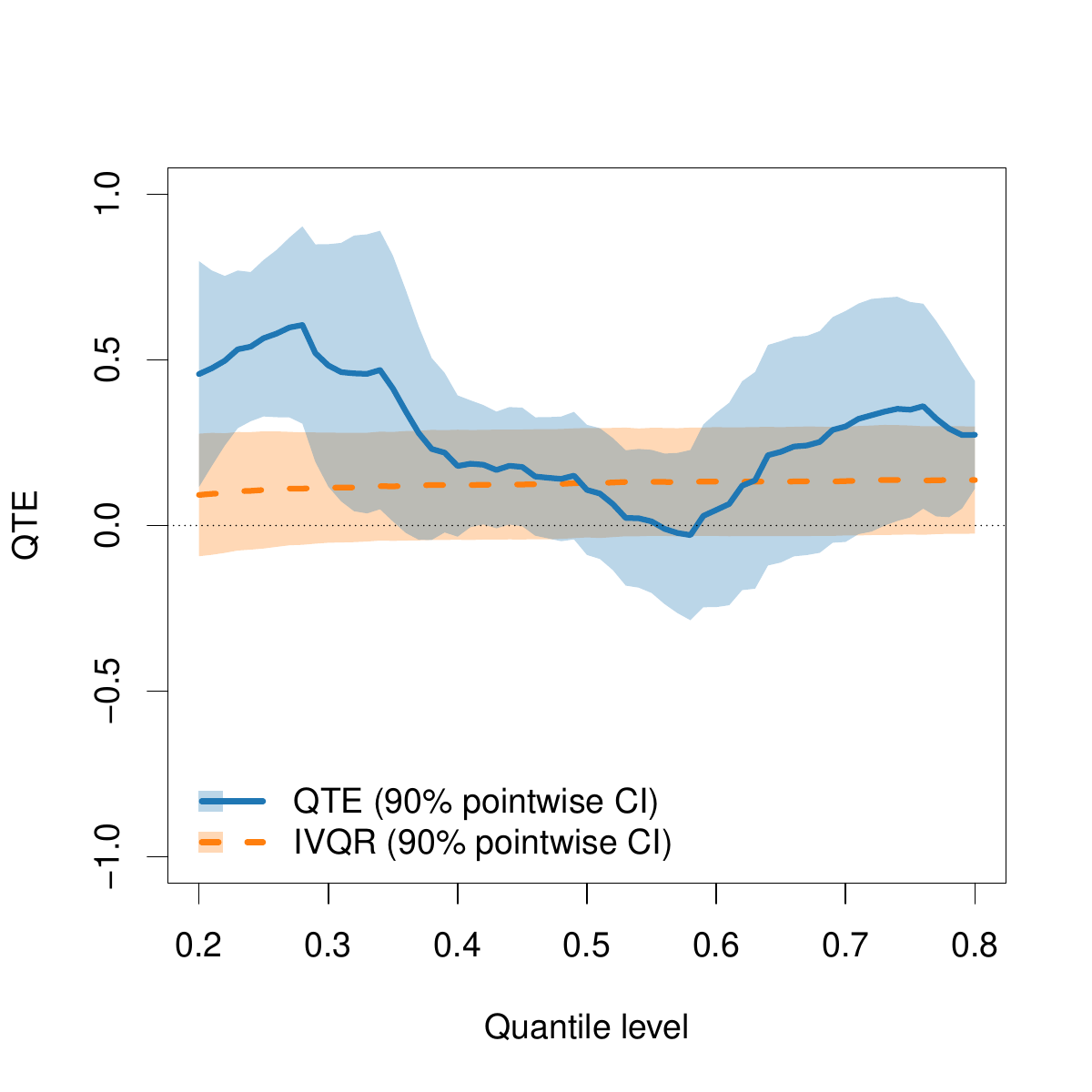}
}
\end{center}

\footnotesize{\textit{Notes:} Pointwise CIs for the QTE are computed using the empirical bootstrap with 5,000 repetitions.  All specifications control for gender, three age indicators, and the baseline well-being index.}	
\end{figure}

Negative local dependence suggests that estimators ignoring the endogeneity of sleep underestimate the effect of sleep on well-being. Figure \ref{fig:comparison}(a) illustrates this by comparing our (IV) QTE estimates with those under conditional exogeneity, $Y_d\indep D\mid X $ for $d\in \mathcal{D}$.\footnote{Under conditional exogeneity, $F_{Y_d}$ is identified as $F_{Y_d}(y)=\int  F_{Y|D,X}(y|d,x)dF_{X}(x).$
For estimation, we consider the DR model $F_{Y|D,X}(y|d,x)=\Phi(d\gamma(y)+x'\beta(y)).$
The DR estimator is $\widehat{F}_{Y|D,X}(y|d,x)=\Phi(d\widehat\gamma(y)+x'\widehat\beta(y)),$ where $\widehat\gamma(y)$ and $\widehat\beta(y)$ are the coefficients obtained from a Probit regression of $1\{Y_i\le y\}$ on $D_i$ and $X_i$. The final estimator of $F_{Y_d}(y)$ under conditional exogeneity is $\widehat{F}_{Y_d}(y)=\frac{1}{n}\sum_{i=1}^n\Phi(d\widehat\gamma(y)+X_i'\widehat\beta(y)).$
} The estimates under conditional exogeneity are small and insignificant at most quantiles, missing the positive well-being effects of sleep in the tails of the well-being distribution. This highlights the importance of using IV methods in this empirical context. 

Figure \ref{fig:comparison}(b) further showcases the value added that our method can bring to standard empirical analyses focusing on average effects. It compares the QTE estimates obtained from our method to standard 2SLS estimates using the same set of covariates. While the 2SLS analysis suggests that sleep has moderate average effects on well-being, our method uncovers interesting patterns of heterogeneity in the effect of sleep on well-being.

Finally, we compare our results with estimates from the IVQR model \citep{chernozhukov2005iv}. As discussed in Section \ref{ssec:RS}, in addition to IV validity, the key identifying assumption underlying IVQR is RS. RS requires the individuals' ranks in the potential well-being distributions to be the same across different levels of sleep, up to ``unsystematic deviations,'' in the language of \citet{chernozhukov2005iv}, thereby restricting the heterogeneity in the effect of sleep on well-being. For example, RS requires an individual's well-being ranking  when sleeping few hours to remain similar when sleeping many hours, up to unsystematic deviations. This discussion shows how \ref{as:CLD} and RS are complementary assumptions that impose non-nested restrictions. \ref{as:CLD} restricts how the nature of endogeneity or selection can depend on unobservable determinants of sleep, while RS restricts the effect of sleep on well-being.

Figure \ref{fig:comparison}(c) compares the QTE estimates obtained from our method to unconditional QTE estimates based on the IVQR model.\footnote{To obtain unconditional QTE estimates based on the IVQR model, we proceed as follows. We estimate linear IVQR models using the inverse quantile regression estimator of \citet{chernozhukov2006instrumental} and obtain the unconditional QTE estimates as described in \citet[][Section 9.3]{chernozhukov2017handbook} using the plug-in approach proposed in \citet{chernozhukov2013inference}.} The IVQR estimates suggest that there is much less heterogeneity across quantiles than our method, and the pointwise CIs include zero at all quantiles.

\section{Concluding Remarks}\label{sec:Conclusions}

When identifying the causal effect of endogenous treatments, researchers inevitably face modeling trade-offs. This paper proposes a new direction to deal with these trade-offs by imposing assumptions on the local dependence between potential outcomes and unobservables determining treatment assignment. In doing so, we make minimal assumptions on the equations determining potential outcomes and treatments, thereby allowing for rich heterogeneity in these equations. The proposed framework applies to  binary, discrete ordered, and continuous treatments, where point identification is achieved even with instruments that have limited variation (e.g., a binary instrument). In all three cases, the identification analysis leads to practical estimation and inference procedures that might appeal to practitioners.

\newpage

\begin{center}
\Large{Supplemental Appendix to ``Estimating Causal Effects of Discrete and Continuous Treatments with Binary Instruments''}
\end{center}
\begin{abstract}
    In this supplemental appendix we provide omitted discussions, results, and proofs by section, in the same order they are referred to in the paper.
\end{abstract}

\begin{appendix}

%\setcounter{page}{1} 

%\startcontents[sections]
%\printcontents[sections]{l}{1}{\setcounter{tocdepth}{2}}

\section{Supplements to Section \ref{sec:Models-and-Identification}}

\subsection{Discussion on Stochastic Dominance in \ref{as:REL}(iii)}\label{app:REL(iii)}

It is interesting to see what type of compliance behavior with respect to $D_0$ and $D_1$
is ruled out by the stochastic dominance between $\pi_d(0)=F_{D \mid Z}(d \mid 0)$ and $\pi_d(1)=F_{D \mid Z}(d \mid 1)$ in \ref{as:REL}(iii). To explore this, define the compliers and defiers of order $j\in\mathcal{D}\backslash\{0\}$ as $C_{j}  :=\bigcup_{d=0}^{K-j}\{D_{0}=d,D_{1}=d+j\}$ and $B_{j}  :=\bigcup_{d=0}^{K-j}\{D_{1}=d,D_{0}=d+j\}$.

\begin{lemma}[Compliance Shares]\label{lem:compliance}Suppose $V_{0}$ and $V_{1}$
are exchangeable, i.e., $F_{V_0,V_1}(v_0,v_1) = F_{V_0,V_1}(v_1,v_0)$ for all $v_0,v_1 \in (0,1)$. Then, $\pi_d(0) >\pi_d(1)$ (resp. $<$) for all $d\in\D\setminus \{K\}$
implies that the share of all complier groups is larger (resp. smaller)
than the share of all defier groups, that is, $\Pr[\bigcup_{j=1}^{K}C_{j}]\geq\Pr[\bigcup_{j=1}^{K}B_{j}]$
(resp. $\leq$). Moreover, the weak inequality becomes strict either when the joint distribution of $(V_0,V_1)$ has
full support on $(0,1)^2$ or under SRI.\end{lemma}

The exchangeability, which trivially holds if $V_0 = V_1$ a.s., states that the distribution of $(V_{0},V_{1})$
is symmetric; most known copulas are symmetric. The condition about the share of compliers and defiers is reminiscent
of a similar assumption used in \citet{de2017tolerating} in the case
of a binary treatment. Another simple interpretation of Lemma \ref{lem:compliance}
can be made under the restriction $V_{0}=V_{1}$ a.s.:  there are no defier groups (i.e., $\Pr[D_{1}<D_{0}]=0$)
if and only if $F_{D \mid Z}(d \mid 0)>F_{D \mid Z}(d \mid 1)$ for all $d\in\D\setminus \{K\}$.
In general, when $V_{z}$ is not restricted, the exchangeability of $(V_{0},V_{1})$
alone does not eliminate compliers or defiers.

\subsubsection{Proof of Lemma \ref{lem:compliance}}

Before presenting a formal proof, it is helpful to consider an illustrative
example with $K=2$. In this case we have two complier groups and
two defier groups: 
\begin{align*}
C_{1} & :=\{D_{0}=0,D_{1}=1\}\cup\{D_{0}=1,D_{1}=2\},\\
C_{2} & :=\{D_{0}=0,D_{1}=2\},\\
B_{1} & :=\{D_{1}=0,D_{0}=1\}\cup\{D_{1}=1,D_{0}=2\},\\
B_{2} & :=\{D_{1}=0,D_{0}=2\}.
\end{align*}
Note that the union of $C_1$ and $C_2$ is identical to the multiple-counting union of 
\begin{align*}
C_{1} & :=\{D_{0}=0,D_{1}=1\}\cup\{D_{0}=1,D_{1}=2\},\\
C_{2} & :=\{D_{0}=0,D_{1}=2\},\\
C_{2} & :=\hphantom{\{D_{0}=0,D_{1}=1\}\cup{}}\{D_{0}=0,D_{1}=2\}.
\end{align*}
Suppose $\pi_d(0)>\pi_d(1)$ for $d=0,1$. An individual belongs
to a complier group if and only if her treatment crosses at least one
of the two treatment thresholds from below. Thus,
\begin{align*}
\Pr[C_1 \cup C_2] & =\Pr[\{0<V_{0}\le\pi_{0}(0),\pi_{0}(1)<V_{1}\le1\}\cup\{0<V_{0}\le\pi_{1}(0),\pi_{1}(1)<V_{1}\le1\}]\\
 & =\Pr[\{0<V_{1}\le\pi_{0}(0),\pi_{0}(1)<V_{0}\le1\}\cup\{0<V_{1}\le\pi_{1}(0),\pi_{1}(1)<V_{0}\le1\}]\\
 & \geq\Pr[\{0<V_{1}\le\pi_{0}(1),\pi_{0}(0)<V_{0}\le1\}\cup\{0<V_{1}\le\pi_{1}(1),\pi_{1}(0)<V_{0}\le1\}]\\
 & =\Pr[B_1 \cup B_2],
\end{align*}
where the second equality follows from exchangeability and the weak inequality
is by $\pi_{d}(0)>\pi_{d}(1)$ for all $d=0,1$.\\
The following is the formal proof of the lemma. With $\pi_{-1}(z)=0$
and $\pi_{K}(z)=1$ for all $z$, 
\begin{align*}
 &\Pr\left[\bigcup_{j=1}^{K}C_{j}\right] =\Pr\left[\bigcup_{j=1}^{K}\bigcup_{s=0}^{K-j}\{\pi_{s-1}(0)<V_{0}\le\pi_{s}(0),\pi_{s+j-1}(1)<V_{1}\le\pi_{s+j}(1)\}\right]\\
 & =\Pr\left[\bigcup_{j=0}^{K-1}\{0<V_{0}\le\pi_{j}(0),\pi_{j}(1)<V_{1}\le1\}\right]=\Pr\left[\bigcup_{j=0}^{K-1}\{0<V_{1}\le\pi_{j}(0),\pi_{j}(1)<V_{0}\le1\}\right]\\
 & \geq\Pr\left[\bigcup_{j=0}^{K-1}\{0<V_{1}\le\pi_{j}(1),\pi_{j}(0)<V_{0}\le1\}\right]=\Pr\left[\bigcup_{j=1}^{K}B_{j}\right].
\end{align*}
The second equality follows from the same derivation as in the case
of $K=2$, and the third equality is by exchangeability. The
weak inequality follows from $\pi_{d}(0)> \pi_{d}(1)$ for all $d\in\{0,\ldots,K-1 \}$. Finally, the weak inequality becomes strict either when the joint distribution of $(V_0,V_1)$ has full support or under SRI (which implies $\Pr[B_j]=0$ $\forall j$). The proof of the opposite direction of inequality is symmetric.

\subsection{CLD and LATE Monotonicity}\label{app:LATE_Monotonicity}

The LATE framework of \citet{imbens1994identification} provides conditions
for identifying causal effects for the subpopulation of compliers.
The compliers are the individuals who react to the instrument so that
$D_{1} > D_{0}$. Compared to the assumptions in Section \ref{subsec:Models-with-Binary},
the LATE framework restricts the selection model by setting $V_{1}=V_{0}=V$ a.s.\ \citep{vytlacil2002independence}, but does not impose any constant local dependence. 

Note that the specification
of $D_{z}$ in \eqref{eq:gen_sel} allows for rich compliance patterns due to the inclusion of two unobservables: $V_{1}$ and $V_{0}$. For example with binary $D$, \eqref{eq:gen_sel_binary} is weaker than LATE monotonicity, and Remark \ref{rem:restriction_v0v1} shows that \ref{as:CLD} does not imply $V_0=V_1$ a.s. Consequently, we can
generate any compliance patterns from the joint distribution of $(V_{1},V_{0})$:
\begin{align*}
\Pr[D_{1}=1,D_{0}=1] & =\Pr[V_{1}>\pi_0(1),V_{0}>\pi_0(0)]\\
\Pr[D_{1}=0,D_{0}=1] & =\Pr[V_{1}\le \pi_0(1),V_{0}>\pi_0(0)]\\
\Pr[D_{1}=1,D_{0}=0] & =\Pr[V_{1}>\pi_0(1),V_{0}\le\pi_0(0)]\\
\Pr[D_{1}=0,D_{0}=0] & =\Pr[V_{1}\le\pi_0(1),V_{0}\le\pi_0(0)]
\end{align*}

In addition to allowing for richer selection patterns, \ref{as:CLD} also allows us to identify treatment effects for the entire population rather than the subpopulation of compliers. See Section \ref{ssec:extrapolation} for a discussion of how \ref{as:CLD} extrapolates from the compliers to the overall population.

\begin{remark}[\ref{as:CLD} and SRI] \label{rem:restriction_v0v1}
One might wonder if \ref{as:CLD} implies SRI, $V_0=V_1$, a.s. The following example shows that this is not the case.  Let $\tilde V_0:=\Phi^{-1}(V_0)$ and $\tilde V_1:=\Phi^{-1}(V_1)$ be the normal scores corresponding to $V_0$ and $V_1$,\footnote{Note that it is equivalent to work with the normal scores because they are a one-to-one transformation of $V_0$ and $V_1$.} and
\[
\left(\begin{array}{c}
Y_d\\
\tilde V_0\\
\tilde V_1
\end{array}\right)\sim\mathcal{N}_{3}\left(\left[\begin{array}{c}
0\\
0\\
0
\end{array}\right],\left[\begin{array}{ccc}
1 & \rho_{Y_d,V_0} & \rho_{Y_d,V_1}\\
\rho_{Y_d,V_0} & 1 & \rho_{V_0,V_1}\\
\rho_{Y_d,V_1} & \rho_{V_0,V_1} & 1
\end{array}\right]\right).
\]
Under \ref{as:CLD}, $\rho_{Y_d,V_0}=\rho_{Y_d,V_1}$, so that $(Y_d,\tilde V_0)$ and $(Y_d,\tilde V_1)$ have the same distribution.
Moreover, the covariance matrix is positive definite if $\rho_{V_0,V_1} > 2 \rho_{Y_d,V_0}^2 - 1$. In other words,  $\rho_{Y_d,V_0}=\rho_{Y_d,V_1}$ does not  imply that $\rho_{V_0,V_1} = 1$ and therefore $ V_1= V_0$, if $\rho_{Y_d,V_0}^2 < 1$. 
\qed
\end{remark}

\subsection{CLD and (In)dependence}\label{app:cld&indep}

Consider applying \ref{as:CLD} to the LGR. \ref{as:CLD} allows for dependence between $Y_d$ and $V_z$ (i.e., endogeneity) when the DGP is a Gaussian copula and $\rho_{Y_d,V_z}(y,v)=\rho_{Y_d}$. Here, we formally show that \ref{as:CLD} is compatible with dependence for many different non-Gaussian DGPs when the outcome is binary. We also provide an example of a restricted class of DGPs under which \ref{as:CLD} implies independence.

We consider the case in which $D$ and $Y$ are binary and focus on the dependence between $Y_0$ and $V_z$; the argument for the dependence between $Y_1$ and $V_z$ is analogous. In this case, the object of interest is $F_{Y_0}(0)\in (0,1)$, and the relevant restriction in \ref{as:CLD} is
\begin{equation}
\rho_{Y_0,V_0}(0,\pi_0(0))=\rho_{Y_0,V_1}(0,\pi_0(1)).\label{eq:CLD_relevant}
\end{equation}

If $Y_0$ and $V_z$ are independent, then $F_{Y_0,V_z}(0,v)=F_{Y_0}(0)v$. Suppose that the dependence between $Y_0$ and $V_z$ can be modeled using a parametric copula, $\tilde{C}_z(\cdot,\cdot;\theta_z)$, $
F_{Y_0,V_z}(0,v;\theta_z)=\tilde{C}_z(F_{Y_0}(0),v;\theta_z),
$
and define $\theta_z^0$ to be the parameter at which independence is achieved, i.e., $\tilde{C}_z(u,v;\theta_z^0)=uv$. Define the following map
$$
m_z(\theta_z):=\text{unique $\rho$ s.t.\ } \tilde{C}_z(F_{Y_0}(0),\pi_0(z);\theta_z)=C(F_{Y_0}(0),\pi_0(z);\rho).
$$

Using this definition, \eqref{eq:CLD_relevant} requires that $m_0(\theta_0)=m_1(\theta_1)$. A sufficient condition for \eqref{eq:CLD_relevant} to be compatible with dependence is that there exists an $m^\ast\ne 0$ in 
$
\operatorname{Range}(m_0(\cdot))\cap \operatorname{Range}(m_1(\cdot))
$. 
We refer to this condition as the \emph{range-overlap condition}. If such $m^\ast$ exists, then one can choose $\theta_1^\ast$ and $\theta_0^\ast$ such that $m_0(\theta_0^\ast)=m_1(\theta_1^\ast)=m^\ast\ne 0$ and \eqref{eq:CLD_relevant} holds. But since $m^\ast\ne 0$ and $C(F_{Y_0}(0),\pi_0(z);m^\ast)\ne F_{Y_0}(0)\pi_0(z)$, we have that 
$
\tilde{C}_z(F_{Y_0}(0),\pi_0(z);\theta_z^\ast)=F_{Y_0,V_z}(0,\pi_0(z);\theta^\ast_z)\ne F_{Y_0}(0)\pi_0(z),
$ i.e., $Y_0$ and $V_z$ are not independent. 

The range-overlap condition holds for many parametric copula families. A simple sufficient condition is the existence of an admissible path from independence to positive (or negative) dependence, referred to as \emph{positive (or negative) dependence path}. Here we focus on positive dependence paths; an analogous argument applies to negative dependence paths. Suppose that, for $z\in \{0,1\}$, there exists a path $\theta_z(t)$, $t\in [0,\epsilon]$, starting at $\theta_z^0$, such that
$
\tilde{C}_z(F_{Y_0}(0),\pi_0(z);\theta_z(t))>F_{Y_0}(0)\pi_0(z)$ for $ t>0.$
Because $\rho\mapsto C(F_{Y_0}(0),\pi_0(z);\rho)$ is strictly increasing by the properties of the Gaussian copula, $m_z(\theta_z(t))>0$ for $t>0$. If $t\mapsto m_z(\theta_z(t))$ is continuous, the interval $[0,m_z(\theta_z(\bar t_z))]$ is contained in $\operatorname{Range}(m_z(\cdot))$ for some $\bar t_z>0$.\footnote{This is implied by continuity of $t\mapsto \tilde{C}_z(F_{Y_0}(0),\pi_0(z);\theta_z(t))$, which holds for many parametric copulas, since $\rho\mapsto C(F_{Y_0}(0),\pi_0(z);\rho)$ is continuous and strictly increasing.} Consequently, the ranges of $m_1$ and $m_0$ overlap away from zero, and the range-overlap condition is satisfied.

Positive (and negative) dependence paths exist for many copula families, including FGM and Fr\'echet-Mardia copulas, as we show in the following examples.
\begin{example}[FGM copula: $\tilde{C}_z(u,v;\theta_z)=uv+\theta_z uv(1-u)(1-v)$] Take $\theta_z(t)=t$ with $t\in [0,\epsilon]$ and $\epsilon>0$ sufficiently small. For $t>0$, we have that 
$$
\tilde{C}_z(F_{Y_0}(0),\pi_0(z);t)-F_{Y_0}(0)\pi_0(z)=t F_{Y_0}(0)(1-F_{Y_0}(0))\pi_0(z)(1-\pi_0(z))>0.
$$ 
Moreover, the map $t\mapsto \tilde{C}_z(F_{Y_0}(0),\pi_0(z);t)$ is continuous. Thus, a positive dependence path exists and the range-overlap condition holds.
    
\end{example}

\begin{example}[Fr\'echet-Mardia copula: $\tilde{C}_z(u,v;a_z,b_z)= a_zM(u,v) + (1-a_z-b_z)uv + b_zW(u,v)$, where $M(u,v)=\min\{u,v\}$ and $W(u,v)=\max\{u+v-1,0\}$]
Take $(a_z(t),b_z(t))=(t,0)$ with $t\in [0,\epsilon]$ and $\epsilon>0$ sufficiently small. For $t>0$, we have that
$$
\tilde{C}_z(F_{Y_0}(0),\pi_0(z);t,0)-F_{Y_0}(0)\pi_0(z)=t(M(F_{Y_0}(0),\pi_0(z))-F_{Y_0}(0)\pi_0(z))>0.
$$
Moreover, the map $t\mapsto \tilde{C}_z(F_{Y_0}(0),\pi_0(z);t,0)$ is continuous. Thus, a positive dependence path exists and the range-overlap condition holds.
\end{example}

The Clayton and Frank copula families also satisfy the range-overlap condition, once they are defined via their continuous extensions at independence.\footnote{For a single-parameter copula $\tilde C$, the independence copula is the unique continuous extension at $\theta=0$ if and only if $\lim_{\theta\rightarrow0}\tilde C(u_1,u_2;\theta)=u_1 u_2$ for every $(u_1,u_2)\in[0,1]^2$. The latter holds for both Clayton and Frank copulas, whose expressions are shown in SA \ref{app:local_rep_copula}, although they are mentioned in that section as examples for local representation rather than DGPs.}

This discussion demonstrates that \ref{as:CLD} is compatible with dependence and endogeneity for many parametric copula families.  At the same time, it is possible to construct examples of classes of DGPs for which \ref{as:CLD} implies independence. Suppose that $\tilde{C}_0$ is an FGM family that only allows for positive dependence,
$
\tilde{C}_0(u,v;\theta_0)=uv+\theta_0uv(1-u)(1-v),$ $\theta_0\in [0,1]$,
 and $\tilde{C}_1$ is an FGM family that only allows for negative dependence,
$
\tilde{C}_1(u,v;\theta_1)=uv+\theta_1uv(1-u)(1-v), $ $ \theta_1\in [-1,0]$.
Note that since $\tilde{C}_0(F_{Y_0}(0),\pi_0(0);\theta_0)-F_{Y_0}(0)\pi_0(0)\ge 0$ and $\tilde{C}_0(F_{Y_0}(0),\pi_0(0);\theta_0)-F_{Y_0}(0)\pi_0(0)= 0$ if and only if $\theta_0=0$, it follows that $m_0(\theta_0)\ge 0$ with $m_0(\theta_0)=0$ if and only if $\theta_0=0$ because the Gaussian copula is strictly increasing in $\rho$. Similarly, we have that $m_1(\theta_1)\le 0$ with $m_1(\theta_1)=0$ if and only if $\theta_1=0$. Therefore, $m_0(\theta_0)=m_1(\theta_1)$ can only occur when $\theta_1=\theta_0=0$ in which case $Y_0\indep V_0$ and $Y_0\indep V_1$. Thus, \ref{as:CLD} implies independence in this sign-separated FGM class.

The formal argument above is specific to binary outcomes. With multi-valued discrete or continuous outcomes, \ref{as:CLD} needs to hold at more values of $y$ and the same pair $(\theta_0,\theta_1)$ must satisfy all of these restrictions, which can be more restrictive. For example, one can show that \ref{as:CLD} requires independence with FGM copulas as DGPs when the outcome is continuous. Non-Gaussian DGPs satisfying \ref{as:CLD} with multi-valued discrete or continuous outcomes, such as the latent-traits model of Section \ref{ssec:Roy}, therefore generally rely on richer dependence structures than those allowed by one-parameter copula families.

\section{Supplements to Section \ref{identification-analyses}}

\subsection{Proof of Theorem \ref{thm:ID_binary}}
\label{app:proof_ID_binary}

Note that $\pi_0(z)$ is identified
as a reduced-form parameter. Let $F_{d,y} := F_{Y_{d}}(y)$ and $\rho_{d,y} := \rho_{Y_{d}}(y)$ be the structural parameters of interest. Consider the following mapping between
the structural and reduced-form parameters: 
\begin{align}
F_{Y\mid D,Z}(y\mid 0,0)\pi_0(0) & =C(F_{0,y},\pi_0(0);\rho_{0,y}),\label{eq:sys_Y1_1}\\
F_{Y\mid D,Z}(y\mid 0,1)\pi_0(1) & =C(F_{0,y},\pi_0(1);\rho_{0,y}),\label{eq:sys_Y1_2}
\end{align}
which we can express as $\pi_{y} =G(\theta_{y})$, where $\theta_{y}:= (F_{0,y},\rho_{0,y})'$, $\pi_{y}:=(F_{Y|D,Z}(y|0,0)\pi_0(0),$ $F_{Y|D,Z}(y|0,1)\pi_0(1))'$, and $G:(0,1)\times (-1,1)\rightarrow(0,1)^2$. 
Let $C_{1}$, $C_{2}$ and $C_{\rho}$ denote the derivatives of copula
$C(u_{1},u_{2};\rho)$ with respect to $u_{1}$, $u_{2}$ and $\rho$,
respectively. Consider the Jacobian of the system of nonlinear equations
\eqref{eq:sys_Y1_1}--\eqref{eq:sys_Y1_2}: 
\begin{align*}
J=\frac{\partial G}{\partial\theta_{y}} & =\begin{bmatrix}C_{1}(F_{0,y},\pi_0(0);\rho_{0,y}) & C_{\rho}(F_{0,y},\pi_0(0);\rho_{0,y})\\
C_{1}(F_{0,y},\pi_0(1);\rho_{0,y}) & C_{\rho}(F_{0,y},\pi_0(1);\rho_{0,y})
\end{bmatrix}.
\end{align*}
The matrix has full rank if and only if 
\begin{align}
\frac{C_{\rho}(F_{0,y},\pi_0(1);\rho_{0,y})}{C_{1}(F_{0,y},\pi_0(1);\rho_{0,y})} & \neq\frac{C_{\rho}(F_{0,y},\pi_0(0);\rho_{0,y})}{C_{1}(F_{0,y},\pi_0(0);\rho_{0,y})},\label{eq:full_rank}
\end{align}
which is true by Assumption \ref{as:REL} and Lemma 4.1 in \citet{han2017identification}
as the Gaussian copula satisfies the stochastically increasing ordering
condition (Assumption 5 in \citet{han2017identification}). Moreover, the Jacobian is a P-matrix for all $(F_{0,y},\rho_{0,y})\in(0,1)\times(-1,1)$ by the proof of Theorem 2.1 in \cite{chernozhukov2025distribution}. Moreover, the domain of the mapping $G$ (i.e., $(0,1)\times (-1,1)$) is open and rectangular and $G$ is differentiable as $C(u_1,u_2;\rho)$ is differentiable with respect to $(u_1,\rho)$. Therefore, one can apply \cite[Theorem 4w]{gale1965jacobian}'s global
univalence theorem, which identifies $\theta_{y}$.\footnote{An alternative identification strategy would be to apply a similar argument as in the proof for the case of ordered $D$ (SA \ref{app:proof_thm32}). In particular, if the Jacobian has full rank everywhere in the parameter space that is open, Steps 1--3 of that proof can be applied to the system of equations \eqref{eq:sys_Y1_1}--\eqref{eq:sys_Y1_2} here.} Analogously, we have 
\begin{align*}
F_{Y\mid D,Z}(y \mid 1,z)(1-\pi_0(z)) & =\Pr[Y_{1}\le y \mid Z=z]-\Pr[Y_{1}\le y,V_{z}\le\pi_0(z) \mid Z=z]\\
 & =\Pr[Y_{1}\le y \mid Z=z]-C(F_{Y_{1} \mid Z}(y|z),\pi_0(z);\rho_{Y_{1},V_{z}}(y,\pi_0(z)))\\
 & =\Pr[Y_{1}\le y]-C(F_{Y_{1}}(y),\pi_0(z);\rho_{Y_{1}}(y))
\end{align*}
and 
\begin{align}
F_{Y\mid D,Z}(y \mid 1,0)(1-\pi_0(0)) & =F_{Y_{1}}(y)-C(F_{Y_{1}}(y),\pi_0(0);\rho_{1,y}),\label{eq:sys_Y0_1}\\
F_{Y\mid D,Z}(y \mid 1,1)(1-\pi_0(1)) & =F_{Y_{1}}(y)-C(F_{Y_{1}}(y),\pi_0(1);\rho_{1,y}),\label{eq:sys_Y0_2}
\end{align}
and the mapping has a unique solution for $\tilde{\theta}_{y}:=(F_{Y_{1}}(y),\rho_{1,y})'$
by a similar argument as above.

\subsection{Proof of Theorem \ref{thm:ID_ordered}}
\label{app:proof_thm32}

Consider $d\in\{1,\ldots,K-1\}$. Write $F_{d,y}:=F_{Y_d}(y)$. Recall from the text that we additionally
impose constant local dependence between a pair of levels:
\begin{align*}
\rho_{Y_{d},V_{z}}(y,\pi_{d}(z)) & =\rho_{Y_{d},V_{z}}(y,\pi_{d-1}(z))=:\rho_{Y_{d}}(y)=:\rho_{d,y}.
\end{align*}
For fixed $(d,y)$ and for $z\in\{0,1\}$, define
$
G_z(F,\rho) :=C(F,\pi_d(z);\rho)-C(F,\pi_{d-1}(z);\rho)$ and $\quad 
q_z  := \Pr[Y\leq y,D=d\mid Z=z]$.
Then \eqref{eq:LGR_D=00003Dk} can be written as a system of two
equations in $(F,\rho)=(F_{d,y},\rho_{d,y})$:
\begin{equation}\label{app:sys_eq}
    q_z
=G_z(F,\rho),\qquad z\in\{0,1\}.
\end{equation}
\emph{Step 1}. For every $\rho\in(-1,1)$, $F\mapsto G_0(F,\rho)$ is strictly
increasing because
\[
\frac{\partial G_0(F,\rho)}{\partial F}
=
C_1(F,\pi_d(0);\rho)
-C_1(F,\pi_{d-1}(0);\rho)>0.
\]
Moreover, $\lim_{F\downarrow0}G_0(F,\rho)=0$ and $\lim_{F\uparrow1}G_0(F,\rho)
=\pi_d(0)-\pi_{d-1}(0)$. Hence, for every $q_0\in
(0,\pi_d(0)-\pi_{d-1}(0))$ and every $\rho\in(-1,1)$, there exists \emph{globally} a
unique $F_{q_0}(\rho)\in (0,1)$ such that
$
    G_0(F_{q_0}(\rho),\rho)=q_0,
$
and the global map
$\rho\mapsto F_{q_{0}}(\rho)$ is continuously differentiable on $(-1,1)$ by the
implicit function theorem, with $F_{q_{0}}'(\rho)=-G_{0,\rho}/G_{0,F}$.

\noindent\emph{Step 2}. It follows that
\begin{align*}
\frac{\mathrm{d}}{\mathrm{d}\rho}G_1(F_{q_0}(\rho),\rho)
&=
G_{1,\rho}
-G_{1,F}\frac{G_{0,\rho}}{G_{0,F}} =
\frac{
G_{0,F}G_{1,\rho}-G_{0,\rho}G_{1,F}
}{
G_{0,F}
}
=
\frac{\det J_d}{G_{0,F}}
\end{align*}
with all derivatives and $J_d$ in the display being evaluated at
$(F_{q_0}(\rho),\rho)$ and $J_d$ being the Jacobian for \eqref{app:sys_eq} displayed below, where we show \ref{as:REL}(iii) implies that $\det J_d\neq0$ throughout $(F,\rho)\in(0,1)\times(-1,1)$. Since $\det J_d$ is continuous and
its domain is connected, it has the same sign everywhere.  Because
$G_{0,F}>0$, the function
$
\rho\mapsto G_1(F_{q_0}(\rho),\rho)
$
is therefore strictly monotone.

\noindent\emph{Step 3}. Suppose $(F,\rho)$ and $(\widetilde{F},\widetilde{\rho})$ solve \eqref{app:sys_eq}. The first equation with $z=0$ implies that
$F=F_{q_{0}}(\rho)$ and $\widetilde{F}=F_{q_{0}}(\widetilde{\rho})$ for their common value $q_{0}$. The second equation with $z=1$ and strict monotonicity of $\rho\mapsto G_1(F_{q_0}(\rho),\rho)$ implies
$\rho=\widetilde{\rho}$, and therefore $F=\widetilde{F}$.  A solution exists because,
under \ref{as:EX}, \ref{as:REL}, and \ref{as:CLD}(ii), the data are generated by the
model at $(F_{Y_{d}}(y),\rho_{Y_{d}}(y))$, so that $q_{0}\in(0,\pi_d(0)-\pi_{d-1}(0))$ whenever
$F_{Y_{d}}(y)\in(0,1)$. Therefore $F_{Y_{d}}(y)$ and $\rho_{Y_{d}}(y)$ are
identified for every $y$ with $F_{Y_{d}}(y)\in(0,1)$.

The cases $d=0$ and $d=K$ reduce to the binary-treatment system and follow from the argument in SA \ref{app:proof_ID_binary}. 
%The cases $d=0$ and $d=K$ follow from the arguments given for the
%extreme treatment levels.  
This completes the proof.

Now, it remains to prove the full rank of the Jacobian of \eqref{app:sys_eq} over $(F,\rho)\in(0,1)\times(-1,1)$:
\begin{align*}
J_{d} & =\begin{bmatrix}C_{1}(F,\pi_{d}(0);\rho)-C_{1}(F,\pi_{d-1}(0);\rho) & C_{\rho}(F,\pi_{d}(0);\rho)-C_{\rho}(F,\pi_{d-1}(0);\rho)\\
C_{1}(F,\pi_{d}(1);\rho)-C_{1}(F,\pi_{d-1}(1);\rho) & C_{\rho}(F,\pi_{d}(1);\rho)-C_{\rho}(F,\pi_{d-1}(1);\rho)
\end{bmatrix}.
\end{align*}
For any $(F,\rho)\in(0,1)\times(-1,1)$, this Jacobian has full rank if and only if
\begin{align}
\frac{C_{\rho}(F,\pi_{d}(1);\rho)-C_{\rho}(F,\pi_{d-1}(1);\rho)}{C_{1}(F,\pi_{d}(1);\rho)-C_{1}(F,\pi_{d-1}(1);\rho)} & \neq\frac{C_{\rho}(F,\pi_{d}(0);\rho)-C_{\rho}(F,\pi_{d-1}(0);\rho)}{C_{1}(F,\pi_{d}(0);\rho)-C_{1}(F,\pi_{d-1}(0);\rho)}.\label{eq:full_rank2}
\end{align}
Showing this is more involved than showing \eqref{eq:full_rank} with
the binary treatment, because the equality can arise due to two points
on the indifference curve. Nonetheless, the full-rank condition \eqref{eq:full_rank2}
can be expressed as $\lambda(0)\neq\lambda(1)$, where 
\[
\lambda(z):=\frac{\phi(r_{d}(z))-\phi(r_{d-1}(z))}{\Phi(r_{d}(z))-\Phi(r_{d-1}(z))},\quad r_{\ell}(z):=\frac{\Phi^{-1}(\pi_{\ell}(z))-\rho\Phi^{-1}(F)}{\sqrt{1-\rho^{2}}}.
\]
 To interpret this condition, we note that it
can be related to the mean of a truncated Gaussian random variable:
for $A\sim N(\mu,\sigma^{2})$,
\[
\Ep[A\mid l<A<u]=\mu-\left(\frac{\phi\left(\frac{u-\mu}{\sigma}\right)-\phi\left(\frac{l-\mu}{\sigma}\right)}{\Phi\left(\frac{u-\mu}{\sigma}\right)-\Phi\left(\frac{l-\mu}{\sigma}\right)}\right)\sigma.
\]
Therefore, the full-rank condition $\lambda(0)\neq\lambda(1)$
can be equivalently expressed as
\begin{align*}
\Ep[A\mid\pi_{d-1}(0)<\Phi(A)<\pi_{d}(0)] & \neq \Ep[A\mid\pi_{d-1}(1)<\Phi(A)<\pi_{d}(1)]
\end{align*}
with $\mu=\rho\Phi^{-1}(F)$ and $\sigma^{2}=1-\rho^{2}$.
For example, this holds when threshold functions are such that $\pi_{d-1}(0)>\pi_{d-1}(1)$
and $\pi_{d}(0)>\pi_{d}(1)$, which is guaranteed by \ref{as:REL}(iii).

\subsection{Further Details on Ordered Choice Models}\label{subsec:HV07}

It is interesting to discuss \citet{heckman2007chapter71}'s model (i.e., $D_z = k$ if $\pi_{k-1}<\mu(z)+V\le \pi_{k}$ for $k\in\{0,...,K\}$ with $\pi_{-1}=-\infty$, $\pi_{K}=\infty$ and $V \mid Z\sim N(0,1)$) in comparison to our ordered choice models \eqref{eq:sel_ordered} in Section \ref{subsec:Models-with-Ordered}  \emph{with the normalization} $V_{z}\mid Z \sim N(0,1)$. The cutoffs of the two models are related as $\pi_{\ell}(z)=\pi_{\ell}-\mu(z)$.
Under this model, we have 
$r_{\ell}(z)=\frac{\pi_{\ell}-\mu(z)-\rho_{d,y}\Phi^{-1}(F_{d,y})}{\sqrt{1-\rho_{d,y}^{2}}},$
which is particularly easy to interpret because $r_{d}(z)-r_{d-1}(z)=\frac{\pi_{d}-\pi_{d-1}}{\sqrt{1-\rho_{d,y}^{2}}}$. On the other hand, in the general model,
we have $
r_{d}(z)-r_{d-1}(z)=\frac{\pi_{d}(z)-\pi_{d-1}(z)}{\sqrt{1-\rho_{d,y}^{2}}}.
$
Note that in the simplified model, the full-rank condition holds by
construction. Specifically, since $r_{d}(z)-r_{d-1}(z)$ does not
depend on $z$, we can write $r_{d}(z)=r_{d-1}(z)+c$ for $c>0$.
Therefore, we can rewrite $\lambda(z)$ as $
\frac{\phi(r_{d-1}(z)+c)-\phi(r_{d-1}(z))}{\Phi(r_{d-1}(z)+c)-\Phi(r_{d-1}(z))}.$
This function is monotonically decreasing in $r_{d-1}(z)$. Thus,
as long as $\mu(1)\ne\mu(0)$ so that $r_{d-1}(1)\ne r_{d-1}(0)$,
the full rank condition holds.

\subsection{Identification under Other Local Representations}\label{app:local_rep_copula}
 In the main text, we work with a local Gaussian representation. Here, we ask what other
single-parameter copulas can be used for local representation and whether identification can be similarly achieved.

First, consider the case of binary $D$. Let $C(u_1,u_2;\rho)$ denote the copula which the local representation analogous to Lemma \ref{lem:LGR} is based on. For such a representation to work, the copula needs to be a \emph{comprehensive copula}; see below for examples. We want to show that, for a given copula, the Jacobian of the system \eqref{eq:sys_Y1} has full rank. The Jacobian having full rank is a necessary condition for it to be a P-matrix in applying \cite{gale1965jacobian}. Moreover, one can show that, if the Jacobian has full rank everywhere in the parameter space that is open, an identification strategy analogous to that in SA \ref{app:proof_thm32} (i.e., Steps 1--3) can be alternatively applied. In showing full rank, the following quantity typically arises once the Jacobian
is transformed using elementary operations: fixing $y\in \Y$,
\begin{align*}
\frac{C_{\rho}(F_{Y_{0}}(y),\pi_0(0);\rho_{Y_{0}}(y))}{C_{1}(F_{Y_{0}}(y),\pi_0(0);\rho_{Y_{0}}(y))}-\frac{C_{\rho}(F_{Y_{0}}(y),\pi_0(1);\rho_{Y_{0}}(y))}{C_{1}(F_{Y_{0}}(y),\pi_0(1);\rho_{Y_{0}}(y))},
\end{align*}
where $C_{\rho}$ and $C_{1}$ are the derivatives with respect to $\rho$
and the first argument, respectively. Therefore, any copula that satisfies
\begin{align}
\frac{C_{\rho}(F_{Y_{0}}(y),\pi_0(0);\rho_{Y_{0}}(y))}{C_{1}(F_{Y_{0}}(y),\pi_0(0);\rho_{Y_{0}}(y))} & \neq\frac{C_{\rho}(F_{Y_{0}}(y),\pi_0(1);\rho_{Y_{0}}(y))}{C_{1}(F_{Y_{0}}(y),\pi_0(1);\rho_{Y_{0}}(y))}\label{eq:condi_Jac}
\end{align}
for $\pi_0(0)\neq\pi_0(1)$ (\ref{as:REL}(iii)) will yield the full rank Jacobian. \citet{han2017identification}
show that any single-parameter copula that follows the ordering of
stochastic increasingness with respect to $\rho$ satisfies \eqref{eq:condi_Jac}.
Let $C_L$, $C_U$ and $C_I$ denote the lower and upper Fr\'echet-Hoeffding copula bounds and independence copula. The following Archimedean copulas are comprehensive
copulas:

\begin{example}[Clayton copula] \label{ex:clayton}
\begin{align*}
C(u_{1},u_{2};\rho) & =\max\{u_{1}^{-\rho}+u_{2}^{-\rho}-1,0\}^{-1/\rho},\quad\rho\in[-1,\infty)\backslash\{0\}
\end{align*}
and $C\rightarrow C_{L}$ when $\rho\rightarrow-1$, $C\rightarrow C_{U}$
when $\rho\rightarrow\infty$, and $C\rightarrow C_{I}$ when $\rho\rightarrow0$.\end{example}

\begin{example}[Frank copula] \label{ex:frank}
\begin{align*}
C(u_{1},u_{2};\rho) & =-\frac{1}{\rho}\ln\left(1+\frac{(e^{-\rho u_{1}}-1)(e^{-\rho u_{2}}-1)}{e^{-\rho}-1}\right),\quad\rho\in(-\infty,\infty)\backslash\{0\}
\end{align*}
and $C\rightarrow C_{L}$ when $\rho\rightarrow-\infty$, $C\rightarrow C_{U}$
when $\rho\rightarrow\infty$, and $C\rightarrow C_{I}$ when $\rho\rightarrow0$.\end{example}

Next, consider the case of continuous $D$. Following Proposition 1 of \cite{anjos2005representation}, consider the following
local representation for an arbitrary copula $\tilde{C}(u_1,u_2)$
\begin{align}
\tilde{C}(u_1,u_2) & = u_1 u_2+\rho(u_1,u_2)\sqrt{u_1 u_2(1-u_1)(1-u_2)}.\label{eq:LCR}
\end{align}
 Then, the local dependence function satisfies
\begin{align*}
\rho(u_1,u_2) & =\frac{\tilde{C}(u_1,u_2)-u_1 u_2}{\sqrt{u_1 u_2(1-u_1)(1-u_2)}},
\end{align*}
which captures the normalized deviation from the independent copula.
In fact, $\rho(u_1,u_2)$ can be interpreted as a local Spearman correlation coefficient that satisfies
many interesting properties \citep[Sections 3.1 and 3.2]{anjos2005representation}. For the identification analysis, by the properties of the conditional
distribution, \eqref{eq:LCR}, \ref{as:EX}, and \ref{as:CLD} applied to the representation in \eqref{eq:LCR}, that is $\rho_{Y_d,V_0}(y,\pi_d(0)) = \rho_{Y_d,V_1}(y,\pi_d(1)) = \rho_{Y_d}(y)$ and $\partial \rho_{Y_d,V_z}(y,v)/\partial v|_{v = \pi_d(z)} = 0$, we have for $z\in \{0,1\}$
\begin{align}\label{eq:anjos}
F_{Y\mid D,Z}(y\mid d,z) & =F_{Y_{d}}(y)+\frac{\rho_{Y_{d}}(y)}{2} w(d,z)\sqrt{F_{Y_{d}}(y)(1-F_{Y_{d}}(y))},
\end{align}
where $w(d,z) := (1-2\pi_d(z))/\sqrt{\pi_d(z)(1-\pi_d(z))}$, which is assumed to be nonzero. Note that
\begin{align*}
\frac{F_{Y\mid D,Z}(y\mid d,1)-F_{Y_{d}}(y)}{w(d,1)} & =\frac{F_{Y\mid D,Z}(y\mid d,0)-F_{Y_{d}}(y)}{w(d,0)},
\end{align*}
which implies that
\begin{align*}
F_{Y_{d}}(y) & =\frac{F_{Y\mid D,Z}(y\mid d,1)w(d,0)-F_{Y\mid D,Z}(y\mid d,0)w(d,1)}{w(d,0) - w(d,1)}.
\end{align*}
Also, combining this result with \eqref{eq:anjos} gives
\begin{align*}
\rho_{Y_{d}}(y) & =2\frac{F_{Y\mid D,Z}(y\mid d,z)-F_{Y_{d}}(y)}{w(d,z)\sqrt{F_{Y_{d}}(y)(1-F_{Y_{d}}(y))}}.
\end{align*}

\subsection{Alternative Identification Strategies}
For the case of binary $D$, we show there can be other identification
strategies using alternative versions of constant local dependence. The analysis can
be extended to the ordered treatment case. Here we assume \ref{as:EX}
and \ref{as:REL} and the treatment selection equation $D_{z}=1\{V_{z} > \pi_0(z)\}$
where $V_{z} \sim U(0,1)$, and focus on local identification. It might be possible to establish global identification under additional assumptions using global univalence results.

\subsubsection{Restrictions Within Treatment Levels}

We focus here on the treatment level  $d=0$. A similar analysis follows for
$d=1$. For $y\in\mathcal{Y}$, consider the LGR of the observed probabilities,
that is 
\[
\Pr[Y_{0}\leq y,D=0\mid Z=z]=C(F_{Y_0}(y),\pi_0(z);\rho_{0}(y,\pi_0(z))),\quad z\in\{0,1\},
\]
where $\rho_{0}(y,\pi_0(z)):=\rho_{Y_{0},V_{z}}(y,\pi_0(z))$.
The identification problem is that we have two probabilities to identify
three parameters: $F_{Y_0}(y)$, $\rho_{0}(y,\pi_0(0))$ and $\rho_{0}(y,\pi_0(1))$.
So far, we have reduced the number of parameters by imposing the condition
$
\rho_{0}(y,\pi_0(0))=\rho_{0}(y,\pi_0(1)).
$
This restriction was imposed separately for each value of $y$. However,
it is also possible to impose restrictions across values of $y$.
Assume that there exists $y'\in\mathcal{Y}$ such that $F_{Y_{0}}(y)\neq F_{Y_{0}}(y')$
and
\begin{equation}
\rho_{0}(y,\pi_0(z))=\rho_{0}(y',\pi_0(z)),\quad z\in\{0,1\}.\label{eq:CI4}
\end{equation}
This condition leads to the following system of four equations with
four unknowns: 
\begin{align*}
\Pr[Y_{0}\leq y,D=0\mid Z=z] & = C(F_{Y_0}(y),\pi_0(z);\rho_{0}(y,\pi_0(z))),\quad z\in\{0,1\},\\
\Pr[Y_{0}\leq y',D=0\mid Z=z] & = C(F_{Y_0}(y'),\pi_0(z);\rho_{0}(y,\pi_0(z))),\quad z\in\{0,1\}.
\end{align*}
Then, it is possible to find conditions under which the solution to
this system exists and is locally unique.  \eqref{eq:CI4} is appealing in that
it does not impose restrictions across levels of $z$.

Let $C_{1}$, $C_{2}$, and $C_{\rho}$ denote the partial derivatives of the Gaussian copula
$C$ with respect to the first and second arguments and $\rho$. The Jacobian
of the system of equations is
\[
J(y,y')=\left(\begin{array}{cccc}
C_{1}(y,1) & 0 & C_{\rho}(y,1) & 0\\
C_{1}(y,0) & 0 & 0 & C_{\rho}(y,0)\\
0 & C_{1}(y',1) & C_{\rho}(y',1) & 0\\
0 & C_{1}(y',0) & 0 & C_{\rho}(y',0)
\end{array}\right),
\]
where $C_{j}(k,z):=C_{j}(F_{Y_0}(k),\pi_0(z);\rho_{0}(k,\pi_0(z)))>0$
for $j\in\{1,\rho\}$, $k\in\{y,y'\}$ and $z\in\{0,1\}$. By the Laplace
expansion, the Jacobian determinant is 
\[
\det(J(y,y'))=C_{1}(y,0)C_{1}(y',1)C_{\rho}(y,1)C_{\rho}(y',0)-C_{1}(y,1)C_{1}(y',0)C_{\rho}(y',1)C_{\rho}(y,0),
\]
which does not vanish if 
\[
\frac{C_{\rho}(y,1)}{C_{1}(y,1)}\frac{C_{\rho}(y',0)}{C_{1}(y',0)}\neq\frac{C_{\rho}(y,0)}{C_{1}(y,0)}\frac{C_{\rho}(y',1)}{C_{1}(y',1)}.
\]
Let 
\[
\lambda(k,z):=\frac{\phi\left(u(k,z)\right)}{\Phi\left(u(k,z)\right)},\quad u(k,z):=\frac{\Phi^{-1}(\pi_0(z))-\rho_{0}(k,\pi_0(z))\Phi^{-1}(F_{Y_0}(k))}{\sqrt{1-\rho_{0}(k,\pi_0(z))^{2}}}.
\]
Then, using that $C_{\rho}(k,z)\sqrt{1-\rho_{0}(k,\pi_0(z))^{2}} =\lambda(k,z) \phi\left( \Phi^{-1}(F_{Y_0}(k)) \right) C_{1}(k,z)$, the previous
condition can be expressed as
\begin{equation}
\frac{\lambda(y,1)}{\lambda(y,0)}\neq\frac{\lambda(y',1)}{\lambda(y',0)}\quad\text{ or }\quad\frac{\lambda(y,1)}{\lambda(y',1)}\neq\frac{\lambda(y,0)}{\lambda(y',0)},\label{eq:ID4}
\end{equation}
that is, the change in the conditional inverse Mills ratio from $z=0$
to $z=1$ is different at $y$ and $y'$, or the change in the conditional
inverse Mills ratio from $y$ to $y'$ is different at $z=0$ and
$z=1$. For example, if the identification condition holds locally
for $y'=y+\mathrm{d}y$, then the condition becomes 
$
(\partial/\partial y) \log\lambda(y,1) \neq (\partial/\partial y) \log\lambda(y,0).
$

\begin{theorem}\label{thm:ID_binary4} Suppose Assumptions \ref{as:EX} and \ref{as:REL}
hold, and $D_{z}=1[V_{z} > \pi_0(z)]$ with $V_{z} \mid Z=z\sim U[0,1]$. Given $y\in\mathcal{Y}$, suppose that there exists $y'\in\mathcal{Y}$
such that \eqref{eq:CI4}, \eqref{eq:ID4} and the analogous conditions for $d=1$ hold. Then, $F_{Y_{d}}(y)$
and $\rho_{Y_{d},V_z}(y,\pi_0(z))$ are locally identified for $d\in\{0,1\}$ and $z\in\{0,1\}$.\end{theorem}

The proof of this theorem follows from the arguments
appearing before the theorem. In general, let $d_{y}$ be the number of  values of $Y$ that we use to
construct the system of equations. Then, we have $2d_{y}$ equations
and $3d_{y}$ unknowns. Therefore, we need to reduce the number of
parameters by $d_y$ using some
combination of \eqref{eq:CI4} and
the alternative assumptions.

\subsubsection{Restrictions Between Treatment Levels}\label{ssec:btw_trt_levels}

As an alternative to the previous subsection, we can impose restrictions
involving parameters for different treatment levels. This strategy
is based on the system of equations 
\begin{align*}
\Pr[Y\leq y,D=0\mid Z=z] & = C(F_{Y_0}(y),\pi_0(z);\rho_{0}(y,\pi_0(z))),\quad z\in\{0,1\},\\
\Pr[Y\leq y,D=1\mid Z=z] & = C(F_{Y_1}(y),1-\pi_0(z);-\rho_{1}(y,\pi_0(z))),\quad z\in\{0,1\},
\end{align*}
where $\rho_{d}(y,\pi_0(z)) := \rho_{Y_{d},V_{z}}(y,\pi_0(z))$. Assume that the local dependence function is the same across treatment levels,
\begin{equation}
\rho_{0}(y,\pi_0(z))=\rho_{1}(y,\pi_0(z)),\quad z\in\{0,1\}.\label{eq:CI5}
\end{equation}
This condition is similar to the RS condition in \eqref{eq:CI_IVQR}, but does not require the potential outcomes to be continuous.
It leads to the following system of four equations with
four unknowns: 
\begin{align*}
\Pr[Y\leq y,D=0\mid Z=z] & = C(F_{Y_0}(y),\pi_0(z);\rho_{0}(y,\pi_0(z))),\quad z\in\{0,1\},\\
\Pr[Y\leq y,D=1\mid Z=z] & = C(F_{Y_1}(y),1-\pi_0(z);-\rho_{0}(y,\pi_0(z))),\quad z\in\{0,1\}.
\end{align*}
Then, it is possible to find conditions under which the solution to
this system exists and is locally unique. Like \eqref{eq:CI4}, \eqref{eq:CI5}  is appealing in that
it does not impose restrictions across levels of $z$. Let $C_{1}$, $C_{2}$, and $C_{\rho}$ denote the partial derivatives of the Gaussian copula
$C$ with respect to the first and second arguments and $\rho$. Also let $C_{j}(d,z):= C_{j}(F_{Y_d}(y),\pi_0(z);\rho_{d}(y,\pi_0(z)))>0$ and $\bar C_{j}(d,z):= C_{j}(F_{Y_d}(y),1-\pi_0(z);-\rho_{d}(y,\pi_0(z)))>0$
for $j\in\{1,\rho\}$, $d\in\{0,1\}$ and $z\in\{0,1\}$. The Jacobian
of the system of equations is
\[
J =\left(\begin{array}{cccc}
C_{1}(0,1) & 0 & C_{\rho}(0,1) & 0\\
C_{1}(0,0) & 0 & 0 & C_{\rho}(0,0)\\
0 & \bar C_{1}(1,1) & - \bar C_{\rho}(1,1) & 0\\
0 & \bar C_{1}(1,0) & 0 & - \bar C_{\rho}(1,0)
\end{array}\right).
\]
By the Laplace
expansion, the Jacobian determinant is 
\[
\det(J)=C_{1}(0,1)\bar C_{1}(1,0)C_{\rho}(0,0)\bar C_{\rho}(1,1) - C_{1}(0,0)\bar C_{1}(1,1)C_{\rho}(0,1)\bar C_{\rho}(1,0),
\]
which does not vanish if 
\[
\frac{C_{\rho}(0,1)}{C_{1}(0,1)}\frac{\bar C_{\rho}(1,0)}{\bar C_{1}(1,0)}\neq\frac{C_{\rho}(0,0)}{C_{1}(0,0)}\frac{\bar C_{\rho}(1,1)}{\bar C_{1}(1,1)}.
\]
Let 
\[
\lambda(d,z):=\frac{\phi\left(u(d,z)\right)}{\Phi\left(u(d,z)\right)} \quad \text{ and } \quad 
\bar \lambda(d,z):=\frac{\phi\left(u(d,z)\right)}{\Phi\left(-u(d,z)\right)},
\]
with $u(d,z):=\{\Phi^{-1}(\pi_0(z))-\rho_{0}(y,\pi_0(z))\Phi^{-1}(F_{Y_d}(y))\}/\sqrt{1-\rho_{0}(y,\pi_0(z))^{2}}$. Then, using that $C_{\rho}(0,z) \sqrt{1-\rho_{0}(y,\pi_0(z))^{2}} =\lambda(0,z)\phi\left( \Phi^{-1}(F_{Y_0}(y)) \right) C_{1}(0,z)$ and $\bar C_{\rho}(1,z) \sqrt{1-\rho_{0}(y,\pi_0(z))^{2}} =\bar \lambda(1,z) \phi\left( \Phi^{-1}(F_{Y_1}(y)) \right)\bar C_{1}(1,z)$, the previous
condition can be expressed as
\begin{equation}
\frac{\lambda(0,1)}{\lambda(0,0)}\neq\frac{\bar \lambda(1,1)}{\bar \lambda(1,0)}\quad\text{ or }\quad\frac{\lambda(0,1)}{\bar \lambda(1,1)}\neq\frac{\lambda(0,0)}{\bar \lambda(1,0)},\label{eq:ID5}
\end{equation}
that is, the change in the conditional inverse Mills ratio from $z=0$
to $z=1$ is different at $d=1$ and $d=0$, or the change in the conditional
inverse Mills ratio from $d=1$ to $d=0$ is different at $z=0$ and
$z=1$.

\begin{theorem}\label{thm:ID_binary5} Suppose Assumptions \ref{as:EX} and \ref{as:REL}
hold, and $D_{z}=1[V_{z} > \pi_0(z)]$ with $V_{z} \mid Z=z\sim U[0,1]$. Assume also that \eqref{eq:CI5} and \eqref{eq:ID5} hold for all $y \in \mathcal{Y}$. Then, $y \mapsto F_{Y_{d}}(y)$
and $y \mapsto \rho_{Y_{d},V_z}(y,\pi_0(z))$ are locally identified on $\mathcal{Y}$, for $d\in\{0,1\}$ and $z\in\{0,1\}$.\end{theorem}

The proof of this theorem follows from the arguments
appearing before the theorem.

\section{Supplements to Section \ref{sec:Discussions-on-Copula}}

\subsection{Additional Examples and Counterexamples of \ref{as:CLD}}\label{app:counterex}

Assume first that SRI holds and focus on $d=0$. Let $\tilde V := \Phi^{-1}(V)$, $v_d(z) := \Phi^{-1}(\pi_d(z))$ and $Y_d \mid \tilde V = v \sim N(\mu_d + \beta_d \ \mathrm{sgn}(|v|-c)  v, \sigma_d^2 - \beta_d^2)$, where $\mathrm{sgn}$ is the sign function and $\beta_d \neq 0$, $\sigma_d>0$ and $c\geq 0$ are constants such that $|\beta_d| < \sigma_d$. In this case $Y_d \sim N(\mu_d, \sigma_d^2)$ because $v \mapsto \mathrm{sgn}(|v|-c)v$ preserves normality and $F_{Y_d,\tilde V}(y,v)=\Phi_2\big((y-\mu_d)/\sigma_d,v;\beta_d/\sigma_d\big)$ for $|v|\geq c$, so that $\rho_{Y_d,\tilde V}(y,v)=\beta_d/\sigma_d$.\footnote{Note that $F_{Y_d,\tilde V}(y,v) = \int_{-\infty}^v \Phi\left( \frac{y - \mu_d - \beta_d \ \mathrm{sgn}(|\tau|-c)  \tau}{\sqrt{\sigma_d^2 - \beta_d^2}} \right)\phi(\tau) \mathrm{d}\tau $. If $v \leq -c$,  $\mathrm{sgn}(|\tau|-c)=1$ for $\tau<v$. If $v \geq c$, $\mathrm{sgn}(|\tau|-c)=-1$ for $-c < \tau < c \leq v$, but the integral over $\tau \in (-c,c)$ does not change. } Hence \ref{as:CLD} holds if $|v_d(0)|\geq c$ and $|v_d(1)|\geq c$ (for instance if $c=0$, i.e. Gaussian copula, or $v_d(0)<-c<c<v_d(1)$).\footnote{\ref{as:CLD}(iii) does not hold when either $|v_d(0)| = c$ or $|v_d(1)| = c$ because $v \mapsto F_{Y_d,\tilde V}(y,v)$ is not differentiable at $v \in \{-c,c\}$, unless $c=0$.} For $|v|<c$ the local dependence parameter differs from $\beta_d/\sigma_d$ at every $y$, so \ref{as:CLD} fails if exactly one threshold lies in $(-c,c)$.

The conditions for \ref{as:CLD} change without SRI, which we do not assume in the paper. In this case, the analog of the previous example is $Y_d \mid \tilde V_z = v \sim N(\mu_d + \beta_d \ \mathrm{sgn}(|v|-c_z)  v, \sigma_d^2 - \beta_{d}^2)$. This covers the previous example as a special case with $c_0=c_1=c$. Then, \ref{as:CLD} holds if $|v_d(0)|\geq c_0$ and $|v_d(1)|\geq c_1$.

\subsection{Rank Similarity Restricts Effect Heterogeneity}
\label{app:Rank-Similarity-Restricts}

In Section \ref{ssec:RS}, we compared our approach to \citet{chernozhukov2005iv}'s IVQR approach based on RS. Unlike \ref{as:CLD}, IVQR imposes constant local dependence assumptions across potential outcomes. Here we illustrate that these restrictions across potential outcomes restrict the treatment effect heterogeneity in an example. 

For ease of notation, we consider a slightly different version of the treatment selection equation \eqref{eq:gen_sel_binary}, $D_{z} =1\{\tilde{V}_{z}> q_0(z)\}$, where $\tilde{V}_{z} \sim \mathcal{N}(0,1)$ is an alternative normalization such that $q_0(z):=\Phi^{-1}(\pi_0(z))$.\footnote{This is because of the normalization $V_{z}\mid Z\sim U[0,1]$ and thus $\tilde V_z = \Phi^{-1}(V_{z}) \mid Z \sim \mathcal{N}(0,1)$.}  Suppose that the LATE assumptions hold, $\tilde{V}_{1}=\tilde{V}_{0}=\tilde{V}$ and $(Y_{1},Y_{0},\tilde{V})\indep Z$, and suppose further that $(Y_{1},Y_{0},\tilde{V})$ are jointly normal with zero means and unit variances, i.e.,
\[
\left(\begin{array}{c}
Y_{0}\\
Y_{1}\\
\tilde V
\end{array}\right)\mid Z=z\sim\mathcal{N}_{3}\left(\left(\begin{array}{c}
0\\
0\\
0
\end{array}\right),\left(\begin{array}{ccc}
1 & \rho_{01} & \rho_{0V}\\
\rho_{01} & 1 & \rho_{1V}\\
\rho_{0V} & \rho_{1V} & 1
\end{array}\right)\right).
\]
Here, \ref{as:CLD} holds by construction, and  therefore does not restrict treatment effect heterogeneity since $\rho_{01}$ and hence the relationship between $Y_{1}$ and $Y_{0}$ is unrestricted. By contrast, RS imposes that $\rho_{0V}=\rho_{1V}=\rho_{V}$, which restricts treatment effect heterogeneity.

To see the last point, note that the joint distribution of the treatment effect $Y_1-Y_0$ and treatment propensity $\tilde V$ is
\[
\left(\begin{array}{c}
Y_{1} - Y_0\\
\tilde V
\end{array}\right)\mid Z=z\sim\mathcal{N}_{2}\left(\left(\begin{array}{c}
0\\
0
\end{array}\right),\left(\begin{array}{cc}
2(1 - \rho_{01}) & \rho_{1V} - \rho_{0V}\\
\rho_{1V} - \rho_{0V} & 1
\end{array}\right)\right).
\]
Under RS, $\rho_{1V} - \rho_{0V} =0$ and therefore $Y_{1} - Y_0 \indep \tilde V$, that is, there is no selection on gains.\footnote{If $Y_1$ and $Y_0$ have non-unit variances $\sigma_1^2$ and $\sigma_0^2$, then RS requires $\text{Cov}(Y_1-Y_0,\tilde V) = (\sigma_1 - \sigma_0) \rho_V$, which allows for selection on gains, but the relationship between $Y_1-Y_0$ and $\tilde V$ is restricted. In this case, $\text{Cov}(Y_1-Y_0,\tilde V) = \sigma_1 \rho_{1V} - \sigma_0 \rho_{0V}$ in general.} Indeed, when selection is based on gains, then RS requires that the potential outcomes are perfectly correlated under joint normality. To see this, suppose that $\tilde{V} \propto Y_1 - Y_0$ and
$$
\left(\begin{array}{c}
Y_{0}\\
Y_{1}\\
Y_{1} - Y_{0}
\end{array}\right)\sim\mathcal{N}_{3}\left(\left(\begin{array}{c}
0\\
0\\
0
\end{array}\right),\left(\begin{array}{ccc}
1 & \rho_{01} & \rho_{01}-1\\
\rho_{01} & 1 & 1-\rho_{01}  \\
 \rho_{01}-1 & 1-\rho_{01}  & 2(1-\rho_{01}) 
\end{array}\right)\right).
$$
In this case \ref{as:CLD} holds trivially, whereas RS requires that $\rho_{01}-1 = 1-\rho_{01}$, so that $\rho_{01} = 1$, which corresponds to RI.\footnote{The result still holds when the potential outcomes have non-zero means and non-unit variances.} Given that the means and variances of $Y_1$ and $Y_0$ are the same in this example, $\rho_{01}=1$ further implies that $Y_1=Y_0$ a.s.

\subsection{\ref{as:CLD} and Gaussianity}\label{app:cld_gauss}

Gaussianity of the copula of $(Y_d,V_z)$ with the same dependence parameter for $z\in \{0,1\}$ is sufficient for \ref{as:CLD} but not necessary. Here we show that the sufficient condition for \ref{as:CLD} mentioned in Section \ref{subsec:Assumptions} is equivalent to Gaussianity when imposed on $\mathcal{V}=(0,1)$. To see this, suppose that $\rho_{Y_d,V_z}(y,v)=\rho_{Y_d}(y)$ for all $v\in\mathcal{V}=(0,1)$. By Lemma \ref{lem:lgr2}, $\Pr[Y_d\le y,V_z\le v]=\int_0^v\Phi(a_{d,y}+b_{d,y}\Phi^{-1}(\tilde v))\,\mathrm{d}\tilde v$, where $a_{d,y}=a_{Y_d,V_z}(y,v)$ and $b_{d,y}=b_{Y_d,V_z}(y,v)$ do not depend on $v$. For $y<y'$ with $0<F_{Y_d}(y)$ and $F_{Y_d}(y')<1$, $v\mapsto\Pr[y<Y_d\le y',V_z\le v] = \int_0^v\left[ \Phi(a_{d,y'}+b_{d,y'}\Phi^{-1}(\tilde v)) - \Phi(a_{d,y}+b_{d,y}\Phi^{-1}(\tilde v))\right] \mathrm{d}\tilde v$ is nondecreasing and its integrand is continuous. It follows that $\Phi(a_{d,y'}+b_{d,y'}\Phi^{-1}(v))\ge\Phi(a_{d,y}+b_{d,y}\Phi^{-1}(v))$ and, because $\Phi$ is strictly increasing, this means that $(a_{d,y'}-a_{d,y})+(b_{d,y'}-b_{d,y})\Phi^{-1}(v)\ge0$ for all $v\in(0,1)$. Since $\Phi^{-1}(v)$ ranges over $\mathbb{R}$, $b_{d,y'}=b_{d,y}$, and therefore $\rho_{Y_d}(y) = \rho_{Y_d}$, i.e. $\rho_{Y_d}(y)$ does not depend on $y$,  because $\rho_{Y_d}(y)$ is a one-to-one function of $b_{d,y}$. Thus, $(Y_d,V_z)$ has a Gaussian copula. The converse is immediate. Similarly, one can show that $F_{Y_d\mid V_z}(y\mid v)=\Phi(a_{d,y}+b_{d,y}\Phi^{-1}(v))$ for all $v\in(0,1)$ and $z\in\{0,1\}$ is equivalent to a Gaussian copula. 

The key assumption underlying the above result is that $\mathcal{V}=(0,1)$. If $\mathcal{V}$ is a proper subinterval of $(0,1)$, the argument does not apply. This is why the sufficient condition in Section \ref{subsec:Assumptions} is stated for $\mathcal{V}\subseteq(0,1)$ and \ref{as:CLD} only restricts the local dependence parameter at and around $d$-treatment thresholds and thus accommodates non-Gaussian copulas (e.g., Figure \ref{fig:ci} and Section \ref{ssec:Roy}).

\subsection{Further Details on Roy Models}\label{subsec:roy_proof}
First, we prove \eqref{eq:cf2} in  Section \ref{ssec:Roy}. By $Y_0\indep V\mid\Theta$, the law of iterated expectations, and \eqref{eq:rs-traits},
\[
    \Ep[Y_01\{V\le v\}]
    =
    \Ep\bigl[\Ep[Y_0\mid\Theta]\,F_{V\mid\Theta}(v\mid\Theta)\bigr]
    =
    v\,\Ep[Y_0]+\ell_0(v)\,\Ep[Y_0\Theta_0]+\ell_1(v)\,\Ep[Y_0\Theta_1].
\]
Similarly,
$\Ep[\Theta_z1\{V\le v\}]
=\Ep[\Theta_zF_{V\mid\Theta}(v\mid\Theta)]
=v\,\Ep[\Theta_z]+\ell_0(v)\,\Ep[\Theta_z\Theta_0]+\ell_1(v)\,\Ep[\Theta_z\Theta_1]
=\ell_z(v)$
by the moment normalizations in \ref{as:LT}. Since $V\sim U(0,1)$, dividing both
displays by $v$ gives $\Ep[\Theta_z\mid V\le v]=\ell_z(v)/v$ and, with
$\Ep[Y_0]=0$, yields expression \eqref{eq:cf2}.

Next, we provide further interpretation of the control function $\lambda_0(v)$ under \ref{as:LT} and prove \eqref{eq:slope-decomp} along the way. Consider \eqref{eq:cf2}. Defining $a_z:=-\!\int_{-\infty}^{\infty}\! \Delta_z\bigl(F_{Y_0}(y)\bigr)\,\mathrm{d}y$, \eqref{eq:cf2} becomes
\begin{equation}
    \lambda_0(v)
    =
    \frac{a_0\,L_0(v)+a_1\,L_1(v)}{v},
    \label{eq:rs-H-uniform}
\end{equation}
which can be derived as follows. Conditioning on $\Theta$ in \eqref{eq:rs-traits} as above gives
$\Ep[\Theta_z1\{Y_0\le y\}]=\delta_z(F_{Y_0}(y))$, and hence
$\Ep[\Theta_z1\{Y_0>y\}]=-\delta_z(F_{Y_0}(y))$ by $\Ep[\Theta_z]=0$.
Multiplying the pointwise identity
$Y_0=\int_0^{\infty}1\{Y_0>y\}\,\mathrm{d}y-\int_{-\infty}^{0}1\{Y_0\le y\}\,\mathrm{d}y$
by $\Theta_z$, taking expectations, and interchanging the order of integration
(valid given $\Ep[|Y_0\Theta_z|]<\infty$) yields
$\Ep[Y_0\Theta_z]=-\int_{-\infty}^{\infty}\delta_z(F_{Y_0}(y))\,\mathrm{d}y$.
With $\ell_z=\kappa_zL_z$ and $\delta_z=\Delta_z/\kappa_z$, this equals
$a_z/\kappa_z$, and the product of the payoff and selection factors is
$(a_z/\kappa_z)(\kappa_zL_z(v)/v)=a_zL_z(v)/v$, delivering \eqref{eq:rs-H-uniform}.

\eqref{eq:rs-H-uniform} shows that, under \ref{as:LT}, two-channel sorting in the control function is determined by the wage-rank loading $a_z$ and the anchor $L_z(v)$. Note that the sign of $\lambda_0'(\pi_0(z))$ is determined by the relative magnitude of the mean wage of inframarginal individuals $\lambda_0(v)$ and the mean wage of marginal individuals $m_0(v):=\Ep[Y_0\mid V=v]$ evaluated at $\pi_0(z)$, because
\begin{equation}
    \lambda_0'(\pi_0(z))
    =
    \frac{m_0(\pi_0(z))-\lambda_0(\pi_0(z))}{\pi_0(z)},
    \label{eq:inframargin-margin}
\end{equation}
where, under \ref{as:LT},
\begin{align}
    \lambda_0(\pi_0(z))&=\mathrm{Cov}(Y_0,\Theta_z)\,\Ep[\Theta_z\mid V\le \pi_0(z)], \label{eq:lambda}\\ 
    m_0(\pi_0(z))&=\mathrm{Cov}(Y_0,\Theta_z)\Ep[\Theta_z\mid V= \pi_0(z)]+\mathrm{Cov}(Y_0,\Theta_{1-z})\Ep[\Theta_{1-z}\mid V= \pi_0(z)].\label{eq:m}
\end{align}
While \eqref{eq:inframargin-margin} holds for any joint distribution of
$(Y_0,V)$, \ref{as:LT} gives economic content to its two components,
\eqref{eq:lambda} and \eqref{eq:m}. By the anchoring \eqref{eq:rs-cardinal},
$\Ep[\Theta_{1-z}\mid V\le\pi_0(z)]=0$, so the inframarginal mean wage
\eqref{eq:lambda} is the payoff-weighted prevalence of trait $z$ alone among
incumbent workers, whereas the marginal mean wage \eqref{eq:m} combines the
payoff-weighted prevalences of both traits among marginal entrants.
Substituting \eqref{eq:lambda} and \eqref{eq:m} into
\eqref{eq:inframargin-margin} then yields \eqref{eq:slope-decomp}, whose
economic interpretation is given in the main text.

\subsection{Details of Remark \ref{rem:gaussian_mixture}}\label{subsec:gaussian_mixture}

\newcommand{\R}{\mathbb{R}}
\newcommand{\PhiTwo}{\Phi_2}
\newcommand{\phitwo}{\phi_2}
\newcommand{\eps}{\varepsilon}

We formally show that a two-component Gaussian copula mixture violates \ref{as:CLD}(i). For simplicity, consider the standard two-component Gaussian copula mixture
\begin{equation}
    C_M(u,v)
    =
    \omega C(u,v;\rho_1)
    +(1-\omega)C(u,v;\rho_2),
    \qquad (u,v)\in(0,1)^2,
    \label{eq:mixture-copula}
\end{equation}
where $0\leq \omega\leq1$ and $\rho_1,\rho_2\in(-1,1)$.
Since a convex combination of copulas is a copula, $C_M$ is a valid copula. For each $(u,v)\in(0,1)^2$, define $\rho_M(u,v)$ as the unique solution to
\begin{equation}
    C(u,v;\rho_M(u,v))=C_M(u,v).
    \label{eq:RM-def}
\end{equation}
The question is whether
\begin{equation}
    \rho_M(u,v_0)=\rho_M(u,v_1)
    \qquad\text{for all }u\in(0,1)
    \label{eq:CLD-copula-level}
\end{equation}
can hold nontrivially when $v_0\ne v_1$ where $v_z:= \pi_d(z)$.

\begin{proposition}[Failure of \ref{as:CLD}(i) for Gaussian mixture]
\label{prop:failure}
Let $C_M$ be defined by \eqref{eq:mixture-copula} with $0<\omega<1$, $\rho_1\ne\rho_2$, and $\rho_1,\rho_2\in(-1,1)$. Let $v_0,v_1\in(0,1)$ with $v_0\ne v_1$. Then there exists $u^\ast\in(0,1)$ such that
$
    \rho_M(u^\ast,v_0)\ne \rho_M(u^\ast,v_1).
$
\end{proposition}

\begin{remark}
\ref{as:CLD}(i) would only hold in degenerate cases: (i) $\omega\in\{0,1\}$; (ii) $\rho_1=\rho_2$; (iii) $v_0=v_1$. \qed
\end{remark}

\subsection{Proof of Lemma \ref{lem:lgr2}}

    The result follows from the following property of the Gaussian copula: 
    \begin{equation}\label{eq:cgc}
C(u,v;\rho) = \int_{0}^v \Phi\left(\frac{\Phi^{-1}(u)-\rho \Phi^{-1}(\tilde v)}{\sqrt{1-\rho^2}}\right) \mathrm{d}\tilde v. 
\end{equation}

Indeed, by \eqref{eq:cgc} and $V \sim U(0,1)$,
    \begin{equation*}
        \Pr[Y \leq y \mid V \leq v] = \frac{F_{Y,V}(y,v)}{F_{V}(v)} = \frac{1}{v} \int_0^{v} \Phi\left(\frac{\Phi^{-1}(F_{Y}(y))-\rho_{Y,V}(y, v) \Phi^{-1}( \tilde v)}{\sqrt{1-\rho_{Y,V}(y, v)^2}}\right) \mathrm{d} \tilde v.
    \end{equation*}
    Conversely, by \eqref{eq:cgc} and uniqueness of the LGR
      \begin{multline*}
        F_{Y,V}(y,v) =  \Pr[Y \leq y \mid V \leq v] F_{V}(v) \ \ = \int_0^{v} \Phi\left(\frac{\Phi^{-1}(F_{Y}(y)) - \rho_{Y,V}(y,v) \Phi^{-1}(\tilde v)}{\sqrt{1-\rho_{Y,V}(y,v)^2}}\right) \mathrm{d} \tilde v \\
        =C(F_Y(y),v;\rho_{Y,V}(y,v)).
    \end{multline*}

\subsection{Proof of Theorem \ref{thm:equivalence}}
\label{app:proof_equivalence}
The result follows from Lemma \ref{lem:lgr2} and  noting that (i) $\rho_{Y_d,V_z}(y,v) = - b_{Y_d,V_z}(y,v)/\sqrt{1 + b_{Y_d,V_z}(y,v)^2}$ and (ii) $a_{Y_d,V_z}(y,v)$ only depends on $v$ through $\rho_{Y_d,V_z}(y,v)$. 

\subsection{Proof of Theorem \ref{thm:rs-slices}}
\label{app:proof_rs-slices}
At $v=\pi_0(z)$, \eqref{eq:rs-traits}, \eqref{eq:rs-cardinal} and $\Delta_z(u):=C\bigl(u,\pi_0(z);r(u)\bigr)-u\,\pi_0(z)$ give
\[
    F_{Y_0,V}(y,\pi_0(z))
    =F_{Y_0}(y)\,\pi_0(z)+\Delta_z\bigl(F_{Y_0}(y)\bigr)
    =C\bigl(F_{Y_0}(y),\pi_0(z);r(F_{Y_0}(y))\bigr).
\]
By LGR (Lemma~\ref{lem:LGR}) and strict monotonicity of $r\mapsto\Phi_{2}(\cdot,\cdot;r)$ on $(-1,1)$ at any interior $(F_{Y_0}(y),\pi_0(z))$, $\rho_{Y_{0},V}(y,\pi_0(z))=r(F_{Y_0}(y))$ for $z\in\{0,1\}$ and $F_{Y_0}(y)\in (0,1)$.

\subsection{Proof of Proposition \ref{prop:failure}}

Let $a=\Phi^{-1}(u)$, $b_0=\Phi^{-1}(v_0)$, and $b_1=\Phi^{-1}(v_1)$. Since $v_0\ne v_1$, we have $b_0\ne b_1$. Fix $u\in(0,1)$ or equivalently $a\in\R$. Define the curve
\[
    \Gamma_a
    =
    \left\{
    \bigl(x_a(\rho),y_a(\rho)\bigr):\rho\in(-1,1)
    \right\},
\]
where $x_a(\rho)=\Phi_2(a,b_0;\rho)$ and $y_a(\rho)=\Phi_2(a,b_1;\rho)$. The point on this curve indexed by $\rho$ records the two Gaussian copula slice values at $v_0$ and $v_1$ for the same local dependence parameter $\rho$. For the mixture copula, we have
\begin{align*}
    C_M(u,v_0)
    &=
    \omega x_a(\rho_1)+(1-\omega)x_a(\rho_2),
    \\
    C_M(u,v_1)
    &=
    \omega y_a(\rho_1)+(1-\omega)y_a(\rho_2).
\end{align*}
Thus the pair $\bigl(C_M(u,v_0),C_M(u,v_1)\bigr)$ is a chord point between the two points $\bigl(x_a(\rho_1),y_a(\rho_1)\bigr)$ and $\bigl(x_a(\rho_2),y_a(\rho_2)\bigr)$ on the curve $\Gamma_a$. If \ref{as:CLD}(i) held at this $u$, then there would exist some $r=r(u)\in(-1,1)$ such that $C(u,v_0;r(u))=C_M(u,v_0)$ and $C(u,v_1;r(u))=C_M(u,v_1)$, or equivalently
\begin{align}
    x_a(r)
    &=
    \omega x_a(\rho_1)+(1-\omega)x_a(\rho_2),\label{eq:chord1}
    \\
    y_a(r)
    &=
    \omega y_a(\rho_1)+(1-\omega)y_a(\rho_2).\label{eq:chord2}
\end{align}
That is, the chord point would have to lie again on $\Gamma_a$.

We now show that this is not possible for a suitable interior value of $u$. Since $\frac{\partial}{\partial \rho}\Phi_2(a,b;\rho)=\phi_2(a,b;\rho)>0$, $x_a(\rho)$ is strictly increasing in $\rho$. Thus, the curve $\Gamma_a$ can be written as the graph $y=q_a(x)$ with $x=x_a(\rho)$, whose slope is
\[
    q_a'\bigl(x_a(\rho)\bigr)
    =
    \frac{y_a'(\rho)}{x_a'(\rho)}
    =
    \frac{\phi_2(a,b_1;\rho)}{\phi_2(a,b_0;\rho)}.
\]
By the bivariate normal density, differentiating $\log q_a'\bigl(x_a(\rho)\bigr)$ with respect to $\rho$ gives
\begin{equation}
    \frac{d}{d\rho}
    \log q_a'\bigl(x_a(\rho)\bigr)
    =
    \frac{(b_1-b_0)\{a(1+\rho^2)-\rho(b_0+b_1)\}}
    {(1-\rho^2)^2}.
    \label{eq:log-slope-derivative}
\end{equation}
Note that
\[
\operatorname{sign}\!\left(q_a''(x_a(\rho))\right)
=
\operatorname{sign}\!\left(
\frac{d}{d\rho}q_a'(x_a(\rho))
\right)
=
\operatorname{sign}\!\left(
\frac{d}{d\rho}\log q_a'(x_a(\rho))
\right),
\]
where the second equality is because \(q_a'(x_a(\rho))>0\) and the first equality is because
\[
q_a''(x_a(\rho))
=
\frac{
\dfrac{d}{d\rho}q_a'(x_a(\rho))
}{
x_a'(\rho)
},
\]
but $x_a'(\rho)>0$.

Now, suppose $\rho_1<\rho_2$ without loss of generality, and let $I\equiv [\rho_1,\rho_2]$. Choose $a$ large enough that $a(1+\rho^2)-\rho(b_0+b_1)>0$ for all $\rho\in I$, which is possible because the function $\rho\mapsto \rho(b_0+b_1)/(1+\rho^2)$ is bounded on compact $I\subset (-1,1)$. Since $b_1-b_0\ne0$, equation \eqref{eq:log-slope-derivative} then has a non-zero sign on $I$. Hence $q_a'$ is strictly monotone (i.e., $q_a$ is strictly convex or strictly concave) on the interval between $x_a(\rho_1)$ and $x_a(\rho_2)$. Since the strict convex/concave graph $q_a$ (i.e., the curve $\Gamma_a$) lies strictly below/above the interior of the chord, \eqref{eq:chord1}--\eqref{eq:chord2} cannot happen.\footnote{Since \(x_a\) is strictly increasing, equation \eqref{eq:chord1}
also implies \(r\in(\rho_1,\rho_2)\).} Therefore \ref{as:CLD}(i) fails.

\section{Supplements to Section \ref{sec:Estimation-and-inference}}\label{app:asymptotic_theory}

Here we discuss the asymptotic properties of the estimators $\widehat{F}_{Y_d|X}$ and $\widehat\rho_{Y_d;X}$ in Algorithms \ref{algo:binary}, \ref{algo:ordered}, and \ref{algo:continuous} and the validity of the bootstrap. The asymptotic properties of the estimators of the target parameters in Algorithm \ref{algo:target} and the validity of the bootstrap then follow because all the functionals involved are Hadamard differentiable under standard regularity conditions \citep[e.g.,][]{van1996weak,chernozhukov2013inference}. Specifically, the functional delta method implies that $\sqrt{n}(\widehat\delta_u-\delta_u)$ converges in distribution to a mean-zero Gaussian process $Z_\delta$ and the functional delta method for the bootstrap implies that the bootstrap consistently estimates the limiting law \citep[][Chapter 3.9]{van1996weak}. Note that Hadamard differentiability of the inverse operator fails for discrete and mixed discrete-continuous outcome variables. In this case, one can perform inference on the QSF and QTE using the method proposed by \citet{chernozhukov2020generic}.

To state the formal results, we introduce some additional notation. Let $\tilde{\mathcal{Y}}$ be a compact subset of $\mathcal{Y}$ when $\mathcal{Y}$ is uncountable and  let $\tilde{\mathcal{Y}}$ be equal to  $\mathcal{Y}$ otherwise. Define $\tilde{\mathcal{D}}$ analogously. Let  $\mathcal{AB}:= \{(a,b):a\in \mathcal{A},b\in \mathcal{B}\}$ for sets $\mathcal{A}$ and $\mathcal{B}$. Denote the set of bounded functions on the set $\mathcal{A}$ by $\ell^{\infty}(\mathcal{A})$. Finally, let $\rightsquigarrow$ denote weak convergence. 

Using arguments similar to those in \citet{chernozhukov2025distribution}, one can show that the estimators $\widehat{F}_{Y_d|X}$ and $\widehat{\rho}_{Y_d;X}$ in Algorithms \ref{algo:binary} and \ref{algo:ordered} satisfy functional central limit theorems (FCLTs),
\begin{eqnarray*}
\sqrt{n}\left(\widehat{F}_{Y_{d}|X}(y|x)-F_{Y_{d}|X}(y|x)\right)&\rightsquigarrow &Z_{F_{Y_{d}|X}(y|x)} \text{ in } \ell^\infty(\tilde{\mathcal{D}}\mathcal{X}\tilde{\mathcal{Y}}),\\
\sqrt{n}\left(\widehat{\rho}_{Y_d;X}(y;x)-\rho_{Y_d;X}(y;x)\right)&\rightsquigarrow& Z_{\rho_{Y_d;X}(y;x)} \text{ in } \ell^\infty(\tilde{\mathcal{D}}\mathcal{X}\tilde{\mathcal{Y}}),
\end{eqnarray*}
where $Z_{F_{Y_{d}|X}(y|x)}$ and $Z_{\rho_{Y_d;X}(y;x)}$ are zero-mean Gaussian processes, and that the multiplier bootstrap is valid. We therefore focus on the estimators $\widehat{F}_{Y_d|X}$ and $\widehat\rho_{Y_d;X}$ in Algorithm \ref{algo:continuous}, which have a different structure.

We start by introducing some additional notation. Let $G(u):= \phi(u)/[\Phi(u)\Phi(-u)]$ and define the expected Hessians
\begin{align*}
H(y)&:= -\Ep[G(B(D,Z,X)'\beta(y))\phi(B(D,Z,X)'\beta(y))B(D,Z,X)B(D,Z,X)'],\\
H(d)&:= -\Ep[G(B(Z,X)'\pi(d))\phi(B(Z,X)'\pi(d))B(Z,X)B(Z,X)'].
\end{align*}
We impose two assumptions. The first assumption is essentially a first-stage assumption ensuring that $a_{d,y;x}$ and $b_{d,y;x}$ are well-defined.
\begin{assumption}\label{ass:FS} $|(B(1,x)-B(0,x))'\pi(d)|\ge c>0$ for $(d,x)\in\mathcal{DX}$. 
\end{assumption}
The second assumption is a version of Condition DR in \citet{chernozhukov2013inference}. It ensures that the first-step coefficient estimators $\widehat\beta(y)$ and $\widehat\pi(d)$ satisfy FCLTs. We present the theoretical results for the case where the outcome is continuously distributed. Results for discrete outcomes can be obtained similarly.

\begin{assumption}\label{ass:DR} (i) $\mathcal{D}$ is a compact interval in $\mathbb{R}$ and $\mathcal{X}$ is compact. (ii) The conditional densities $f_{Y|D,Z,X}(y\mid d,z,x)$ and $f_{D|Z,X}(d\mid z,x)$ exist and are uniformly bounded and uniformly continuous. (iii) $B(d,z,x)$ and $B(z,x)$ are bounded over $\tilde{\mathcal{D}}\times \{0,1\}\times\mathcal{X}$ and continuous, $\Ep\left[\|B(D,Z,X) \|^2\right]< \infty$, and $\Ep\left[\|B(Z,X) \|^2\right]< \infty$. (iv) The minimum eigenvalues of $-H(y)$ and $-H(d)$ are bounded away from zero over $\tilde{\mathcal{Y}}$ and $\tilde{\mathcal{D}}$. 
\end{assumption}

The following theorem establishes FCLTs for the estimators $\widehat{F}_{Y_{d}|X}(y|x)$ and $\widehat\rho_{Y_d;X}(y;x)$ in Algorithm \ref{algo:continuous} and the validity of the exchangeable bootstrap.
\begin{theorem}\label{thm:weak_convergence}
Consider the estimators $\widehat{F}_{Y_{d}|X}(y|x)$ and $\widehat\rho_{Y_d;X}(y;x)$ defined in Algorithm \ref{algo:continuous}. Suppose that Assumptions \ref{ass:FS} and \ref{ass:DR} hold. Then,
\begin{eqnarray*}
\sqrt{n}\left(\widehat{F}_{Y_{d}|X}(y|x)-F_{Y_{d}|X}(y|x)\right)&\rightsquigarrow& Z_{F_{Y_{d}|X}(y|x)} \text{ in } \ell^\infty(\tilde{\mathcal{D}}\mathcal{X}\tilde{\mathcal{Y}}),
\\
\sqrt{n}\left(\widehat{\rho}_{Y_d;X}(y;x)-\rho_{Y_d;X}(y;x)\right)&\rightsquigarrow& Z_{\rho_{Y_d;X}(y;x)} \text{ in } \ell^\infty(\tilde{\mathcal{D}}\mathcal{X}\tilde{\mathcal{Y}}),
\end{eqnarray*}
where $Z_{F_{Y_{d}|X}(y|x)}$ and $Z_{\rho_{Y_d;X}(y;x)}$ are zero-mean Gaussian processes defined in the proof in SA \ref{app:proof_weak_convergence}, and the exchangeable bootstrap is consistent for estimating the limiting laws.
\end{theorem}

\subsection{Proof of Theorem \ref{thm:weak_convergence}}
\label{app:proof_weak_convergence}

%The proof proceeds in two steps.

%\bigskip

\noindent \textbf{Step 1: FCLT for $(\widehat{\beta}(y)',\widehat{\pi}(d)')'$ and bootstrap validity.} In this step, we establish FCLTs  for $\widehat{\beta}(y)$ and $\widehat{\pi}(d)$ and the validity of the bootstrap, building on \citet{chernozhukov2013inference}.

Under Assumption \ref{ass:DR}, Corollary 5.3 in \citet{chernozhukov2013inference} implies that
\footnotesize
\begin{align*}
\sqrt{n}(\widehat{\beta}(y)-\beta(y))&= H^{-1}(y) \frac{1}{\sqrt{n}}\sum_{i=1}^n\frac{ \Phi(B(D_i,Z_i,X_i)'\beta(y))-I_i(y)}{\Phi(B(D_i,Z_i,X_i)'\beta(y))\Phi(-B(D_i,Z_i,X_i)'\beta(y))}\phi(B(D_i,Z_i,X_i)'\beta(y))B(D_i,Z_i,X_i)\\
&+o_P(1),\\
\sqrt{n}(\widehat{\pi}(d)-\pi(d))&=H^{-1}(d) \frac{1}{\sqrt{n}}\sum_{i=1}^n\frac{ \Phi(B(Z_i,X_i)'\pi(d))-J_i(d)}{\Phi(B(Z_i,X_i)'\pi(d))\Phi(-B(Z_i,X_i)'\pi(d))}\phi(B(Z_i,X_i)'\pi(d))B(Z_i,X_i)+o_P(1),
\end{align*}
\normalsize
and
$$
\sqrt{n}\left(\begin{pmatrix}\widehat{\beta}(y)\\\widehat{\pi}(d) \end{pmatrix}-\begin{pmatrix}\beta(y)\\\pi(d) \end{pmatrix}\right) \rightsquigarrow \begin{pmatrix}Z_{\beta}(y)\\Z_{\pi}(d)\end{pmatrix}\quad  \text{in}\quad \ell^\infty (\tilde{\mathcal{D}}\tilde{\mathcal{Y}})^{\dim(B(Z,X))+\dim(B(D,Z,X))},
$$
where $Z_{\beta}(y)$ and $Z_{\pi}(d)$ are mean-zero Gaussian processes, and that the bootstrap is valid.

\bigskip

\noindent \textbf{Step 2: FCLT for $\widehat{\rho}_{Y_d;X}(y;x)$ and $\widehat{F}_{Y_{d}|X}(y|x)$ and bootstrap validity.} In this step, we build on Step 1 to establish a FCLT for $\widehat{\rho}_{Y_d;X}(y;x)$ and $\widehat{F}_{Y_{d}|X}(y|x)$ and the validity of the bootstrap. Because all maps involved are Hadamard differentiable, the result follows from the functional delta method and the functional delta method for the bootstrap.

Under Assumption \ref{ass:FS}, by Step 1 and the functional delta method, 
\begin{eqnarray*}
\sqrt{n}\left( \begin{pmatrix} \widehat{a}_{d,y;x}\\ \widehat{b}_{d,y;x}\end{pmatrix} -\begin{pmatrix} a_{d,y;x}\\ b_{d,y;x}\end{pmatrix}\right)\rightsquigarrow \begin{pmatrix}Z_{a_{d,y;x}}\\Z_{b_{d,y;x}}\end{pmatrix}\quad \text{in}\quad \ell^\infty (\tilde{\mathcal{D}}\mathcal{X}\tilde{\Y} )^2,
\end{eqnarray*}
where
\begin{align*}
Z_{a_{d,y;x}}&=\frac{B(1,x)'\pi(d) B(d,0,x)'-B(0,x)'\pi(d) B(d,1,x)'}{(B(1,x)-B(0,x))'\pi(d)}Z_{\beta}(y)\\
&+\bigg(\frac{B(d,0,x)'\beta(y) B(1,x)'-B(d,1,x)'\beta(y) B(0,x)' }{(B(1,x)-B(0,x))'\pi(d)}\\
&-\frac{\left(B(d,0,x)'\beta(y) B(1,x)'\pi(d)-B(d,1,x)'\beta(y) B(0,x)'\pi(d) \right)(B(1,x)-B(0,x))'}{((B(1,x)-B(0,x))'\pi(d))^2}\bigg)Z_{\pi}(d)
\end{align*}
and
\begin{align*}
Z_{b_{d,y;x}}&=\frac{(B(d,1,x)-B(d,0,x))'}{(B(1,x)-B(0,x))'\pi(d)}Z_{\beta}(y)+\frac{(B(d,1,x)-B(d,0,x))'\beta(y)(B(0,x)-B(1,x))'}{((B(1,x)-B(0,x))'\pi(d))^2}Z_{\pi}(d).
\end{align*}

Define $\widehat{\mu}_{d,y;x}:=\widehat{a}_{d,y;x}/\sqrt{1+\widehat{b}_{d,y;x}^2}$ and $\mu_{d,y;x}:=a_{d,y;x}/\sqrt{1+b_{d,y;x}^2}$. Then, by the functional delta method, we have
$$
\sqrt{n}(\widehat{\mu}_{d,y;x}-\mu_{d,y;x})\rightsquigarrow Z_{\mu_{d,y;x}} \quad \text{in}\quad \ell^\infty (\tilde{\mathcal{D}}\mathcal{X}\tilde{\Y} ),
$$
where $Z_{\mu_{d,y;x}}=Z_{a_{d,y;x}}/\sqrt{1+b_{d,y;x}^2}-a_{d,y;x}b_{d,y;x}Z_{b_{d,y;x}}/(1+b_{d,y;x}^2)^{3/2}$. Moreover, we have that
\begin{equation*}
\sqrt{n}\left(\widehat{\rho}_{Y_d;X}(y;x)-\rho_{Y_d;X}(y;x)\right)\rightsquigarrow Z_{\rho_{Y_d;X}(y;x)} \text{ in } \ell^\infty(\tilde\D\mathcal{X}\tilde\Y),
\end{equation*}
where $Z_{\rho_{Y_d;X}(y;x)}=-Z_{b_{d,y;x}}/(1+b_{d,y;x}^2)^{3/2}
$. Finally, another application of the functional delta method yields\footnote{If rearrangement is used in the last step of Algorithm \ref{algo:continuous}, this step additionally relies on Hadamard differentiability of the rearrangement operator established in \citet{chernozhukov2010quantile}.}
$$
\sqrt{n}\left(\widehat{F}_{Y_{d}|X}(y|x)-F_{Y_{d}|X}(y|x)\right)\rightsquigarrow \phi(\mu_{d,y;x})Z_{\mu_{d,y;x}}=: Z_{F_{Y_{d}|X}(y|x)} \quad \text{in}\quad \ell^\infty (\tilde{\mathcal{D}}\mathcal{X}\tilde{\Y} ).
$$
The validity of the bootstrap follows from the functional delta method for the bootstrap. 

\end{appendix}

\bibliographystyle{ecta}
\bibliography{LGR}

\end{document}